\documentclass[prd,aps,epsf,epsfig,preprint,nofootinbib]{revtex4-2}
\usepackage{mathrsfs}
\usepackage{dcolumn}
\usepackage{bm}
\usepackage{hyperref}
\usepackage{epsfig,graphicx,color,appendix}
\usepackage{multirow}
\usepackage{tabularx}
\usepackage{capt-of}
\usepackage{subfigure}
\usepackage[countmax]{subfloat} 
\usepackage{threeparttable}
\usepackage{amssymb}
\usepackage{array}
\usepackage{amsmath}
\usepackage{makecell}
\usepackage{booktabs}
\usepackage{longtable}
\usepackage{mathrsfs} 
\usepackage{placeins}

\newcommand{\lsp}{V_{LS}^{(+)}} 
\newcommand{\lsm}{V_{LS}^{(-)}} 
\newcommand{\lspm}{V_{LS}^{(\pm)}} 

\graphicspath{ {./images/} }

\begin{document}
\long\def\symbolfootnote[#1]#2{\begingroup%
\def\thefootnote{\fnsymbol{footnote}}\footnote[#1]{#2}\endgroup}
\def\lsim{ {\ \lower-1.2pt\vbox{\hbox{\rlap{$<$}\lower5pt\vbox{\hbox{$\sim$}
}}}\ } }
\def\gsim{ {\ \lower-1.2pt\vbox{\hbox{\rlap{$>$}\lower5pt\vbox{\hbox{$\sim$}
}}}\ } }

\font\el=cmbx10 scaled \magstep2{\obeylines\hfill \today}

\vskip 1.5 cm

\centerline{\large\bf $\boldsymbol{B_c}$ Meson Spectroscopy from Bayesian MCMC: }
 \centerline{\large\bf Probing Confinement and State Mixing}
\small
\vskip 1.0 cm

\centerline{\bf Christas \surname{Mony A.} \symbolfootnote[2]{christasmony@gmail.com} and Rohit Dhir \symbolfootnote[3]{Corresponding Author : dhir.rohit@gmail.com} }
\medskip
\centerline{\it Department of Physics and Nanotechnology, SRM Institute of Science and Technology,}
\centerline{\it Kattankulathur - 603203, Tamil Nadu, India.}
\bigskip
\bigskip
\begin{center}
{\large \bf Abstract}
\end{center}

We present a Bayesian study of the $B_c$ meson spectrum using three non-relativistic confining potentials, namely the Cornell potential, a logarithmically modified extension, and a screened Cornell form. Parameters for each potential are sampled using Markov chain Monte Carlo (MCMC), retaining correlations among the fitted parameters. The resulting parameter distributions are used to calculate the $B_c$ spectrum through the $6D$ multiplet, including masses, spin-dependent splittings, mixing angles, and wave-function observables. The predicted masses are further examined through Regge trajectories. The low-lying spectrum is relatively stable across the three potentials, while their predictions become increasingly separated with excitation as the states probe larger distances, accompanied by growing parameter-induced uncertainties. The Regge trajectories show stronger non-linearity for low-lying states and progressively approach linear behavior with excitation. These results quantify the sensitivity of the predicted excited $B_c$ spectrum to the chosen confining interaction and provide uncertainty estimates for states that remain experimentally less explored.

\medskip
\medskip

\vfill

\newpage
\section{Introduction}
\label{sec:1_intro}
The $B_c$ meson is the only known quarkonium composed of two heavy quarks of different flavors. This asymmetry has immediate physical consequences that distinguish it from charmonium and bottomonium. The explicit open flavor ($b\bar{c}$) forbids direct annihilation into gluons, so every excited state below the lowest strong-decay threshold ($BD$) decays exclusively through radiative or weak transitions, producing a sub-threshold spectrum that is both wider and populated by narrower states than in either equal-flavor system. The absence of charge conjugation symmetry permits spin-singlet and spin-triplet states of the same orbital angular momentum to mix through the antisymmetric spin-orbit interaction, an effect absent in $c\bar{c}$ and $b\bar{b}$ and directly sensitive to the interplay between one-gluon exchange and confinement components of the interquark potential. The characteristic dynamical scales of the $B_c$ system lie between those of charmonium and bottomonium, placing it in an intermediate kinematic regime where confinement effects may differ qualitatively from both limiting cases and where the system must be calibrated on its own terms.

Experimentally, the $B_c$ family remains sparsely mapped. The ground state $B_c(1S)$ at $(6274.47 \pm 0.32)$~MeV, first observed by CDF~\cite{CDF:1998ihx, CDF:1998axz} and confirmed across LHCb, CMS, and ATLAS in multiple decay channels~\cite{LHCb:2020ayi, LHCb:2019bem, CMS:2019uhm, ATLAS:2014lga, ParticleDataGroup:2024cfk}, and the first radial excitation $B_c(2S)$ at $(6871.2 \pm 1.0)$~MeV~\cite{ParticleDataGroup:2024cfk} constitute the complete PDG listing of experimentally known $B_c$ states.  The LHCb Collaboration has observed a wide structure in the $B_c^+\gamma$ mass spectrum consistent with the lowest excited $P$-wave $B_c^+$ states~\cite{LHCb:2025uce, LHCb:2025ubr}, while the ATLAS Collaboration has subsequently reported the first observation of the vector $B_c^{*+}$ state~\cite{ATLAS:2026ubk}. The individual $1P$ fine-structure components remain unresolved, however, leaving detailed masses and singlet-triplet mixing among the quantities that require improved experimental resolution. The low production rate, requiring simultaneous $c\bar{c}$ and $b\bar{b}$ pair creation, means that the bulk of the predicted spectrum, including all $D$-wave states and radial excitations beyond $2S$, awaits discovery. Theoretical predictions with quantified uncertainties are therefore essential for guiding the design and interpretation of future measurements at LHCb and other facilities.

Non-relativistic potential models with the Cornell potential have been a primary framework for $B_c$ spectroscopy~\cite{Eichten:2019gig, Li:2019tbn, Asghar:2019qjl, Devlani:2014nda, Li:2022bre, Hao:2024nqb}. A variety of refinements have broadened this landscape: quasipotential approaches~\cite{Ebert:2002pp, Ebert:2011jc} incorporate systematic relativistic corrections; relativized models~\cite{Godfrey:2004ya} use nonlocal kernels for high-momentum dynamics; instantaneous Bethe-Salpeter treatments~\cite{Wang:2022cxy} include partial-wave mixing; and coupled-channel or screened potentials~\cite{Li:2023wgq, Monteiro:2016rzi, Bokade:2025lmn, Patel:2026prg} account for unquenched effects beyond the linear confining term. 

Despite these advancements, two open questions, one physical and one methodological, remain unaddressed in a unified framework that simultaneously probes potential-shape sensitivity and performs full uncertainty propagation for $B_c$ states. The physical question is whether the strictly linear confining term remains adequate across the full radial-orbital spectrum of this intermediate regime (between charmonium and bottomonium systems), or whether a controlled modification of its intermediate- and long-distance behavior can produce distinguishable spectroscopic consequences. The methodological question concerns the uncertainty of predictions when only a limited number of independent $B_c$ spectroscopic constraints are available. Even when parameters are fixed from charmonium and bottomonium data, conventional $\chi^2$ minimization offers no rigorous criterion for selecting among potential forms or for quantifying systematic model dependence. Recent efforts have partially addressed these aspects using bootstrap methods for bottomonium~\cite{Molina:2020zao} and Monte Carlo approaches for spin-averaged $B_c$ spectra~\cite{Akan:2025nej}. However, these studies do not perform full posterior sampling of correlated parameters. A comprehensive Bayesian analysis of the $B_c$ spectrum, including credible intervals derived from the correlated parameter space, therefore remains necessary.

We address these questions by employing the Cornell potential together with a logarithmically modified extension and a screened Cornell potential within a unified Bayesian MCMC framework, constrained by PDG masses and lattice-QCD hyperfine splittings. The logarithmic and screened forms retain the short-distance Coulombic behavior while modifying the confining interaction at intermediate and large distances in different ways. The resulting parameter distributions are used to calculate the $B_c$ spectrum through the $6D$ multiplet, yielding masses, spin-dependent splittings, singlet-triplet mixing angles, and wave-function observables with asymmetric credible intervals. The calculated masses are subsequently used to examine the corresponding Regge trajectories. The flavor asymmetry permits $P$- and $D$-wave singlet-triplet mixing through the antisymmetric spin-orbit interaction, which is included consistently in the spectroscopic analysis.

The remainder of this paper is organized as follows. Section~\ref{sec:2_methodology} describes the theoretical framework and MCMC procedure, Sec.~\ref{sec:3_ND} presents the numerical results and discussion, and Sec.~\ref{sec:4_sum_conclu} summarizes our findings.

\section{Methodology}
\label{sec:2_methodology}
We investigate $B_c$ meson spectroscopy using the non-relativistic potential model. The heavy quark-antiquark interaction is taken as the Cornell potential~\cite{Eichten:1974af, Eichten:1979ms}, which combines a Lorentz-vector one-gluon-exchange (OGE) Coulomb term at short distances with a Lorentz-scalar linear confining term at long distances:
\begin{equation}\label{eqn:1_cornell}
V(r) = -\frac{4\alpha_s}{3r} + \sigma r + V_c,
\end{equation}
where $\alpha_s$ is the strong coupling constant, $\sigma$ the string tension, $V_c$ a constant energy offset, and $4/3$ the Casimir factor of the fundamental SU(3) representation. Since the relevant energy scales in the $B_c$ system lie between those of charmonium and bottomonium, the phenomenological parameters must be determined independently from those of hidden-flavor heavy quarkonia.

We also consider a phenomenological modification of the Cornell potential to probe the sensitivity of the spectrum to intermediate-distance dynamics. To explore possible departures from a strictly linear confining interaction at intermediate distances, we introduce the controlled phenomenological deformation,
\begin{equation}
\label{eqn:corplusln} 
V_{\text{Ext}}(r) = V(r) + C_0\ln(1 + \sigma'r),
\end{equation}
where $\sigma'$ and $C_0$ parametrize the onset scale and strength of the modification, respectively. The physical significance of this deformation is expressed through the resulting effective string tension
\begin{equation}
\label{eqn:sigmaeffective}
    \sigma_{\mathrm{eff}}(r) = \sigma + \frac{C_0\sigma'}{1+\sigma'r},
\end{equation}
which introduces a smooth distance dependence in the effective confining slope, while preserving the asymptotically linear behavior of the Cornell potential. The short- and long-distance limits do not uniquely determine the intermediate-distance form of the interaction. Several two-parameter deformations, including screened linear terms, power-law confining potentials ($r^\alpha$, $0<\alpha<1$), Yukawa-like corrections, and hyperbolic tangent forms, satisfy the same limiting constraints. Among these possible deformations, the logarithmic form is particularly suitable for the present study because it provides a direct interpretation in terms of a smoothly running $\sigma_{\mathrm{eff}}(r)$. Moreover, corrections to the leading confining interaction arise naturally in theoretical descriptions of the heavy-quark potential. In Effective String Theory, quantum corrections to the long-distance potential can be related to modifications of the effective string tension~\cite{Brambilla:2014eaa}, while logarithmic deformations have also been employed in holographic QCD descriptions of the heavy-quark potential and Cornell-like phenomenology~\cite{He:2010ye}. These considerations provide a theoretical motivation for using the logarithmic form as a phenomenological deformation, without implying that QCD uniquely selects this functional form. We therefore use it to probe the sensitivity of the $B_c$ spectrum to the intermediate-distance confining profile, with the dependence on $\sigma$, $\sigma'$, and $C_0$ examined in Sec.~\ref{sec:3_ND}.

The potential model provides a reliable phenomenological framework for describing heavy quark-antiquark systems. In this heavy-quark regime, a non-relativistic treatment is well justified since $m_Q \gg \Lambda_{\text{QCD}}$, allowing the interaction to be systematically expanded in powers of $1/m_Q$~\cite{Brambilla:2019esw}. The spin-independent potentials in Eqs.~\eqref{eqn:1_cornell} and \eqref{eqn:corplusln} yield degenerate masses for states with different total spin. This degeneracy is lifted by spin-dependent interactions, which arise at order $1/m_Q^2$ in the non-relativistic expansion and are treated perturbatively~\cite{Eichten:1980mw, Gromes:1984ma, DeRujula:1975qlm, Godfrey:1985xj, Kwong:1987mj, QuarkoniumWorkingGroup:2004kpm}. These corrections correspond to spin-spin, spin-orbit, and tensor interactions, originating from the underlying vector and scalar components of the potential.

The energy spectrum of the $B_c$ system is obtained by solving the Schr\"odinger equation with the potentials defined in Eqs.~\eqref{eqn:1_cornell} and \eqref{eqn:corplusln}. Expressing the total wave function as $\Psi(\vec{r}) = R_{nl}(r)Y_{lm}(\theta,\phi)$, with $R_{nl}(r) = u_{nl}(r)/r$ and normalization $\int_0^\infty |u_{nl}(r)|^2\,dr = 1$, leads to the radial Schr\"odinger equation
\begin{equation}
\label{eqn:2_radial_SE}
    u''_{nl}(r) + 2\mu\bigg[E_{nl} - V(r) - \frac{l(l+1)}{2\mu r^2}\bigg]u_{nl}(r) = 0,
\end{equation}
where $\mu$ is the reduced quark mass, and $E_{nl}$ is the energy eigenvalue. $n$ and $l$ are the principal and orbital angular momentum quantum numbers, respectively. Equation~\eqref{eqn:2_radial_SE} is solved numerically using the Runge-Kutta method to obtain $E_{nl}$ (see Appendix~\ref{app:rk} for details). The corresponding spin-independent bound-state mass is then given by 
\begin{equation}
    M_{nl} = m_1 + m_2 + E_{nl},
\end{equation}
with $m_1$ and $m_2$ denoting the constituent heavy quark masses. In contrast to hidden flavor quarkonia, for an unequal-mass system such as $B_c$ ($m_1 \neq m_2$), these spin-dependent terms take the general form~\cite{Lucha:1995zv}:
\begin{equation}
\label{eqn:VssGeneral}
    V_{SS}(r) = \frac{2}{3m_1 m_2}\,\mathbf{S}_1\cdot\mathbf{S}_2\;\Delta V_V(r),
\end{equation}
\begin{equation}
\label{eqn:VtGeneral}
    V_{T}(r) = \frac{1}{12m_1 m_2}\mathcal{S}_{12}\bigg[\frac{1}{r}\frac{d}{dr}V_V(r) - \frac{d^2}{dr^2}V_V(r)\bigg],
\end{equation}
with $\mathcal{S}_{12} \equiv 12\big[\frac{(\mathbf{S}_1\cdot\mathbf{r})(\mathbf{S}_2\cdot\mathbf{r})}{r^2} - \frac{1}{3}\mathbf{S}_1\cdot\mathbf{S}_2\big]$, and
\begin{multline}
\label{eqn:VlsGeneral}
    V_{LS}(r) = \frac{1}{4m_1^2 m_2^2 r}\bigg\{\Big[\big((m_1+m_2)^2+2m_1m_2\big)\mathbf{L}\cdot\mathbf{S}_+
    +(m_2^2-m_1^2)\mathbf{L}\cdot\mathbf{S}_-\Big]\frac{d}{dr}V_V(r) \\
    -\Big[(m_1^2+m_2^2)\mathbf{L}\cdot\mathbf{S}_+
    +(m_2^2-m_1^2)\mathbf{L}\cdot\mathbf{S}_-\Big]\frac{d}{dr}V_S(r)\bigg\},
\end{multline}
where $\mathbf{S}_\pm \equiv \mathbf{S}_1 \pm \mathbf{S}_2$. Equation~\eqref{eqn:VlsGeneral} can also be decomposed into symmetric and antisymmetric spin-orbit components expressed as, $\lsp(r)$ and $ \lsm (r)$, respectively; explicitly given as
\begin{equation}
\label{eqn:VLSPlus}
V_{LS}^{(+)}(r) = \frac{1}{4m_{1}^2 m_{2}^2 r}\bigg[\Big((m_1+m_2)^2+2m_1m_2\Big)\frac{d}{dr}V_V(r) - \big(m_1^2+m_2^2\big)\frac{d}{dr}V_S(r)\bigg]\mathbf{L}\cdot\mathbf{S}_{+},
\end{equation}
\begin{equation}
\label{eqn:VLSMinus}
V_{LS}^{(-)}(r) = \frac{(m_2^2-m_1^2)}{4m_{1}^2 m_{2}^2 r}\bigg[\frac{d}{dr}V_V(r) - \frac{d}{dr}V_S(r)\bigg]\mathbf{L}\cdot\mathbf{S}_{-}.
\end{equation}

For states that do not undergo spin mixing, the physical mass is obtained in first-order perturbation theory by adding the expectation values of the spin-dependent interactions to the spin-independent mass $M_{nl}$,
\begin{equation}
\label{eqn:massfull}
M(n^{2S+1}L_J) = M_{nl} +\langle V_{SS}\rangle +\langle V_{LS}^{(+)}\rangle +\langle V_T\rangle,
\end{equation}
where the expectation values are evaluated using the spin-independent radial wave function $u_{nl}(r)$ obtained from Eq.~\eqref{eqn:2_radial_SE}, together with the appropriate spin matrix elements.

Since charge conjugation is not a good quantum number for the $B_c$ system, the antisymmetric spin-orbit interaction mixes the $|n^3L_L\rangle$ and $|n^1L_L\rangle$ basis states. Consequently, these are not the physical mass eigenstates. Following Ref.~\cite{Godfrey:1986wj}, the effective Hamiltonian matrix in the ordered basis $\{|n^3L_L\rangle,|n^1L_L\rangle\}$ is written as
\begin{equation}
\label{eqn:mixham}
\mathcal H=
\begin{pmatrix}
M(^3L_L) & \langle V_{LS}^{(-)}\rangle\\
\langle V_{LS}^{(-)}\rangle & M(^1L_L)
\end{pmatrix},
\end{equation}
where
\begin{align*}
M(^3L_L) &= M_{nl} +\langle {}^3L_L|V_{SS}|{}^3L_L\rangle +\langle {}^3L_L|V_{LS}^{(+)}|{}^3L_L\rangle +\langle {}^3L_L|V_T|{}^3L_L\rangle,\\
M(^1L_L)&= M_{nl} +\langle {}^1L_L|V_{SS}|{}^1L_L\rangle,
\end{align*}
and
\[
\langle V_{LS}^{(-)}\rangle = \langle n^3L_L|V_{LS}^{(-)}|n^1L_L\rangle,
\]
evaluated using the common radial wave function obtained from the
spin-independent Schr\"odinger equation.
The physical $J=L$ eigenstates are defined as
\begin{align}
\label{eqn:mixing}
|nL_L\rangle' &= \cos\theta_{nL}|n^1L_L\rangle + \sin\theta_{nL}|n^3L_L\rangle, \\ 
|nL_L\rangle &= -\sin\theta_{nL}|n^1L_L\rangle + \cos\theta_{nL}|n^3L_L\rangle,
\end{align}
where diagonalization of the Hamiltonian yields\footnote{The eigenvalues of Eq.~(\ref{eqn:mixham}) are
\[
M_{\pm}= \frac{M({}^3L_L)+M({}^1L_L)}{2} \pm \frac{1}{2} \sqrt{\left[M({}^3L_L)-M({}^1L_L)\right]^2 +4\langle V_{LS}^{(-)}\rangle^2},
\]
where $M_{+}$ ($M_{-}$) denotes the heavier (lighter) physical state.}
\begin{equation}
\label{eqn:mixangle}
\tan2\theta_{nL} = \frac{2\langle V_{LS}^{(-)}\rangle} {M(^1L_L)-M(^3L_L)}.
\end{equation}
The solution for $\theta_{nL}$ is chosen such that the primed eigenvector defined above corresponds to the heavier physical eigenstate throughout this work. This convention differs only by an overall sign from some conventions adopted in the literature and therefore does not affect the physical mass spectrum or the singlet-triplet admixture probabilities.

In both potentials, $V_V(r) = -\frac{4\alpha_s}{3r}$ is taken as the sole vector component, while all confining terms (linear and logarithmic) are scalar \cite{McClary:1983xw}. The spin-spin and tensor interactions therefore have identical analytical forms for both potentials~\cite{Eichten:1979pu, Eichten:1980mw, Gromes:1984ma, Lucha:1995zv}:
\begin{equation}\label{eqn:VssOurs}
    V_{SS}(r) = \frac{32\pi\alpha_s}{9m_1 m_2}\,\mathbf{S}_1\cdot\mathbf{S}_2\;\delta(r),
    \qquad
    \langle\mathbf{S}_1\cdot\mathbf{S}_2\rangle = \frac{1}{2}s(s+1) - \frac{3}{4},
\end{equation}
where the Dirac delta is replaced by the smeared Gaussian, which incorporates relativistic corrections of $\mathcal{O}(v^2/c^2)$~\cite{Barnes:2005pb},
\begin{equation}
\label{eqn:smearfn}
    \delta(r) \to \Big(\frac{\rho}{\sqrt{\pi}}\Big)^{3}e^{-\rho^2r^2},
\end{equation}
with $\rho$ a phenomenological smearing parameter. The tensor interaction, contributing only for $L>0$, is~\cite{Eichten:1979pu, Eichten:1980mw, Gromes:1984ma, Lucha:1995zv}
\begin{equation}\label{eqn:VtOurs}
    V_{T}(r) = \frac{\alpha_s}{3m_1 m_2 r^3}\mathcal{S}_{12},
    \qquad
    \langle \mathcal{S}_{12}\rangle = \frac{4}{(2l+3)(2l-1)}\Big[s(s+1)l(l+1) - \frac{3}{2}\langle\mathbf{L}\cdot\mathbf{S}\rangle - 3\langle\mathbf{L}\cdot\mathbf{S}\rangle^2\Big].
\end{equation}

For the Cornell potential (Eq.~\eqref{eqn:1_cornell}), the spin-orbit components are
\begin{equation}
\label{eqn:VLSPlusCornell}
V_{LS}^{(+)\rm I}(r) = \frac{1}{4m_{1}^2 m_{2}^2 r}\bigg[\Big((m_1+m_2)^2+2m_1m_2\Big)\frac{4\alpha_s}{3r^2} - \big(m_1^2+m_2^2\big)\sigma\bigg]\mathbf{L}\cdot\mathbf{S}_{+},
\end{equation}
\begin{equation}
\label{eqn:VLSMinusCornell}
V_{LS}^{(-)\rm I}(r) = \frac{(m_2^2-m_1^2)}{4m_{1}^2 m_{2}^2 r}\bigg[\frac{4\alpha_s}{3r^2} - \sigma\bigg]\mathbf{L}\cdot\mathbf{S}_{-},
\end{equation}
with the spin-orbit matrix elements~\cite{Eichten:1979pu, Eichten:1980mw, Gromes:1984ma, Lucha:1995zv, Ikhdair:2004hg}
\begin{equation}
\label{eqn:LSplusop}
    \langle\mathbf{L}\cdot\mathbf{S}_{+}\rangle = \frac{1}{2}\Big[j(j+1)-l(l+1)-s(s+1)\Big],
    \qquad
    \langle\mathbf{L}\cdot\mathbf{S}_{-}\rangle = \sqrt{\frac{(2l+3)(2l-1)}{10}}\;\delta_{J,L}.
\end{equation}
In the above expressions, $l$, $s$, and $j$ denote the orbital, total spin, and total angular momentum quantum numbers of the $B_c$ state, respectively.

For the modified Cornell potential (Eq.~\eqref{eqn:corplusln}), the scalar part acquires an additional derivative $\frac{C_0\sigma'}{1+\sigma'r}$, yielding
\begin{equation}
\label{eqn:VLSPlusModCornell}
V_{LS}^{(+)\rm II}(r) = \frac{1}{4m_{1}^2 m_{2}^2 r}\bigg[\Big((m_1+m_2)^2+2m_1m_2\Big)\frac{4\alpha_s}{3r^2} - \big(m_1^2+m_2^2\big)\bigg(\sigma+\frac{C_0\sigma'}{1+\sigma'r}\bigg)\bigg]\mathbf{L}\cdot\mathbf{S}_{+},
\end{equation}
\begin{equation}
\label{eqn:VLSMinusModCornell}
V_{LS}^{(-)\rm II}(r) = \frac{(m_2^2-m_1^2)}{4m_{1}^2 m_{2}^2 r}\bigg[\frac{4\alpha_s}{3r^2} - \sigma - \frac{C_0\sigma'}{1+\sigma'r}\bigg]\mathbf{L}\cdot\mathbf{S}_{-}.
\end{equation}
Comparing Eqs.~\eqref{eqn:VLSPlusCornell}-\eqref{eqn:VLSMinusCornell} with Eqs.~\eqref{eqn:VLSPlusModCornell}-\eqref{eqn:VLSMinusModCornell}, one identifies the effective $r$-dependent string tension $\sigma_{\mathrm{eff}}(r)$ (Eq.~\eqref{eqn:sigmaeffective}) which captures the net confining slope at finite $r$ from both the linear and logarithmic contributions. The logarithmic term thereby adjusts the fine-structure splittings within $P$- and $D$-wave multiplets and modifies the mixing strength between ${}^3L_J$ and ${}^1L_J$ levels, making the $B_c$ system a sensitive probe of intermediate-distance confinement dynamics where such corrections are expected to be more significant than in equal-flavor quarkonia. The full $B_c$ mass spectra are obtained by incorporating the spin-dependent potentials in Eqs.~\eqref{eqn:VssOurs}, \eqref{eqn:VtOurs}, \eqref{eqn:VLSPlusCornell}-\eqref{eqn:VLSMinusCornell}, and \eqref{eqn:VLSPlusModCornell}-\eqref{eqn:VLSMinusModCornell}, as perturbative corrections to the spin-independent spectrum.

To determine the potential parameters and quantify their uncertainties, we employ a Bayesian MCMC analysis, a significant methodological distinction from most previous $B_c$ potential model studies that rely on $\chi^2$ minimization. As illustrated by the two deterministic solutions presented in Appendix~\ref{app:1_min}, the limited experimental information in the $B_c$ sector and the tensions that arise when bottomonium inputs are simultaneously included motivate a probabilistic exploration of the parameter space. To our knowledge, no prior work has performed a full Bayesian MCMC extraction of $B_c$ potential parameters with uncertainty propagation to the predicted spectrum.

\subsection{Alternative Deformation: Screened Cornell Potential}

\label{sec2a_ScrCorPot}

As an explicit comparison to the logarithmic deformation, we also consider the representative screened Cornell potential~\cite{Ding:1993uy}, which provides a phenomenological description of a softened confining interaction and allows us to examine the model dependence of the predicted $B_c$ spectrum. Screened potentials represent a distinct phenomenological scenario associated with the suppression of confinement at large separations, in contrast to the logarithmic deformation, which retains the asymptotically linear confining behavior relevant to the sub-threshold bound states.

\begin{equation}
\label{eqn:screenpot}
V_{\rm scr}(r) = -\frac{4\alpha_s}{3r}
+ \frac{\sigma}{\mu_{\rm scr}}\left(1 - e^{-\mu_{\rm scr} r}\right)
+ V_c,
\end{equation}
where $\mu_{\rm scr}$ controls the characteristic distance over which screening becomes significant. At short distances,
\[
\frac{\sigma}{\mu_{\rm scr}}\left(1-e^{-\mu_{\rm scr}r}\right)
\simeq \sigma r,
\]
so the screened potential reduces to the Cornell form at leading order, whereas at large separations the confining contribution saturates,
\[
\frac{\sigma}{\mu_{\rm scr}}\left(1-e^{-\mu_{\rm scr}r}\right)
\longrightarrow \frac{\sigma}{\mu_{\rm scr}},
\qquad r\rightarrow\infty,
\]
and the corresponding confining force vanishes. Thus, in contrast to the logarithmic extension, which preserves an asymptotically linear confining slope, the screened potential progressively suppresses the confining interaction at large distances.

For the screened potential, the scalar confining term $V_S(r) = \frac{\sigma}{\mu_{\rm scr}}(1 - e^{-\mu_{\rm scr} r}) + V_c$ has derivative
\begin{equation}
\frac{d}{dr}V_S(r) = \sigma e^{-\mu_{\rm scr} r},
\end{equation}
which modifies the spin-orbit components relative to the Cornell potential. The symmetric and antisymmetric spin-orbit terms become
\begin{equation}
\label{eqn:VLSPlusScrCornell}
V_{LS}^{(+)\rm III}(r) = \frac{1}{4m_1^2 m_2^2 r}
\bigg[\Big((m_1+m_2)^2+2m_1m_2\Big)\frac{4\alpha_s}{3r^2}
- \big(m_1^2+m_2^2\big)\sigma e^{-\mu_{\rm scr} r}\bigg]\mathbf{L}\cdot\mathbf{S}_+,
\end{equation}
\begin{equation}
\label{eqn:VLSMinusScrCornell}
V_{LS}^{(-)\rm III}(r) = \frac{(m_2^2-m_1^2)}{4m_1^2 m_2^2 r}
\bigg[\frac{4\alpha_s}{3r^2} - \sigma e^{-\mu_{\rm scr} r}\bigg]\mathbf{L}\cdot\mathbf{S}_-.
\end{equation}
The spin-spin and tensor interactions retain the same forms as in Eqs.~\eqref{eqn:VssOurs} and~\eqref{eqn:VtOurs}, as they originate solely from the vector (Coulomb) component. The corresponding spectroscopic predictions are compared with those of the logarithmically extended Cornell potential in the subsequent analysis to assess the sensitivity of the obtained $B_c$ spectrum to the choice of confining deformation.

\subsection{Deterministic Parameter Fits}
\label{sec:2b_det}

Before the MCMC analysis, we establish a controlled baseline through deterministic fits based primarily on the well-measured bottomonium sector. The strong coupling constant is scanned over the physically motivated interval
\begin{equation}
0.20 \le \alpha_s \le 0.56,
\end{equation}
while $\sigma$, $V_c$, and $\rho$ are adjusted to reproduce the available spectroscopic inputs.

Two representative parameter sets illustrate the range of viable solutions. Set-I is obtained by simultaneously fitting $b\bar{b}$ and $B_c$ spectroscopic data, producing a solution anchored to established heavy-quark spectroscopic data\footnote{Including $c\bar{c}$ and $b\bar{b}$ states in the fit leads to larger deviations between theoretical and experimental masses, particularly for the $B_c$ system. Fitting hidden heavy-flavor states together is already known to be challenging~\cite{A:2023bxv}.}. The joint constraints, however, introduce tension between the fitted sectors, reflected in 
\begin{equation} \chi^2_{\text{Set-I}} = 1.28 \times 10^{5}. 
\end{equation} 
Set-II is fitted exclusively to the available $B_c$ states. With fewer constraints, the parameters adjust without cross-sector tension, achieving 
\begin{equation} 
\chi^2_{\text{Set-II}} = 3.3 \times 10^{-9}. 
\end{equation} 
The near-vanishing $\chi^2$ of Set-II reflects parametric flexibility rather than superior predictive power: the $B_c$ sector alone is insufficiently constraining to uniquely determine the potential parameters. Set-I therefore represents the empirically guided optimization, while Set-II demonstrates the parametric freedom available when only $B_c$ inputs are used. Explicit parameter values and the resulting spectra are given in Appendix~\ref{app:1_min}.

These examples illustrate that deterministic $\chi^2$ minimization with limited $B_c$ inputs yields multiple acceptable parameter solutions whose excited-state predictions diverge appreciably. This is also evident from the recent study by Mateu et al.~\cite{Mateu:2018zym}, which employed an extensive bottomonium dataset and revealed strong correlations among the parameters of the Cornell potential fit. These considerations motivate the Bayesian MCMC framework employed in the main analysis, in which the parameter space is explored systematically to identify statistically preferred regions and quantify inter-parameter correlations.

\subsection{MCMC Exploration of Parameter Space}
\label{sec:2c_MCMC}
The $B_c$ meson contains two heavy quarks of different flavors, so neither the bottomonium nor the charmonium parameters apply directly. The relevant energy scales lie between those of the two systems; empirical constraints on excited states are limited, and the unequal-mass configuration naturally introduces strong correlations among potential parameters. These characteristics render deterministic fitting susceptible to local minima and parameter degeneracies.

We use the \texttt{emcee} Python package (version $3.1.6$)~\cite{Foreman-Mackey:2012any}, which implements the affine-invariant ensemble sampler of Ref.~\cite{Goodman:2010dyf}. The affine invariance greatly improves sampling efficiency in the highly correlated parameter spaces characteristic of quarkonium potential models. The free parameters of the Cornell potential are
\[
\Theta=\{\alpha_s,\sigma,\rho,V_c\},
\]
where we extend $\Theta$ by $\{C_0,\sigma'\}$ for the modified Cornell potential and by $\{\mu_{\rm scr}\}$ for the screened Cornell potential, respectively. The number of walkers is chosen as
\begin{equation*}
N_{\text{walkers}}=8\times N_{\text{parameters}},
\end{equation*}
resulting in 32, 48, and 40 walkers for the Cornell, modified Cornell, and screened Cornell potentials, respectively. For each parameter $\theta \in \Theta$, we adopt a uniform prior within physically motivated bounds:
\begin{equation}
\label{eqn:uniprior}
\log p(\theta) = 
\begin{cases}
0, & \theta_{\min} \le \theta \le \theta_{\max},\\
-\infty, & \text{otherwise},
\end{cases}
\end{equation}
where the bounds are 
\begin{table}[h!]
\centering
\begin{tabular}{@{}l@{\hspace{2.0cm}}l@{}}
$\alpha_s \in [0.20,0.56]$ & $\sigma' \in [0.05,0.50]~\text{GeV}$ \\
$\sigma \in [0.10,0.23]~\text{GeV}^2$ & $C_0 \in [0.05,0.50]~\text{GeV}$ \\
$\rho \in [1.00,4.00]~\text{GeV}$ & $\mu_{\rm scr} \in [0.01,0.20]~\text{GeV}$ \\
$V_c \in [-0.250,0]~\text{GeV}$ &  
\end{tabular}
\end{table}

The constituent quark masses are treated as effective phenomenological parameters within the non-relativistic potential model. They are not identified with pole, kinetic, or $\overline{\rm MS}$ running masses, but represent effective constituent masses appropriate for the description of heavy-quark bound states. We adopt
\begin{equation*}
m_c = 1.5~\text{GeV}, \qquad m_b = 4.8~\text{GeV},
\end{equation*}
following our previous heavy-quarkonium analysis~\cite{A:2023bxv}, where these effective heavy-quark masses were used within the Cornell potential framework for the charmonium and bottomonium spectra. We retain the same values in the present $B_c$ analysis to maintain consistency in the heavy-quark parameterization across the charmonium, bottomonium, and mixed-flavor $c\bar b$ systems.

The likelihood is defined through a $\chi^2$ form,
\begin{equation*}
\chi^2(\Theta) = \sum_i\left(\frac{X_i^{\text{theo}}(\Theta) - X_i^{\text{expt}}}{\Delta X_i^{\text{expt}}}\right)^2,
\qquad
\log\mathcal{L}(\Theta) = -\tfrac{1}{2}\chi^2(\Theta),
\end{equation*}
where $\Theta$ denotes the set of free potential parameters. $X_i^{\text{theo}}$ are theoretically computed masses including spin-dependent contributions, and $X_i^{\text{expt}}$ and $\Delta X_i^{\text{expt}}$ are the experimental data and uncertainties. The experimental inputs are $M_{B_c(1S)}$, $M_{B_c(2S)}$, the mass difference $M_{B_c(2S)}-M_{B_c(1S)}$ from the PDG~\cite{ParticleDataGroup:2024cfk}, and the LQCD hyperfine splitting $M_{B_c^*(1S)}-M_{B_c(1S)}$ from Ref.~\cite{Mathur:2018epb}. The mass difference, though not strictly independent, is included as an additional constraint to stabilize the determination of level splittings. For each walker position, the theoretical spectrum is computed by solving Eq.~\eqref{eqn:2_radial_SE} and applying all perturbative spin-dependent corrections described earlier, giving the posterior
\begin{equation*}
\log P(\Theta) = \log p(\Theta) + \log\mathcal{L}(\Theta).
\end{equation*}

Since the available experimental information for the $B_c$ system is limited, we also investigate the impact of incorporating prior information from heavy quarkonium. Specifically, Gaussian priors are adopted for the parameters $\alpha_s$, $\sigma$, and $V_c$, with prior means taken from the bottomonium-based fit (Set-I in Appendix~\ref{app:1_min}). This provides a complementary Bayesian inference in which well-constrained heavy-quarkonium information is propagated to the $B_c$ sector, allowing us to examine the stability of the extracted parameters.
The prior means are
\[
\mu_{\alpha_s}=0.373,\qquad
\mu_{\sigma}=0.182~{\rm GeV}^2,\qquad
\mu_{V_c}=-0.067~{\rm GeV},
\]
while the prior widths are chosen as one quarter of the corresponding uniform prior ranges,
\[
\sigma_i=\frac{\theta_{\max}-\theta_{\min}}{4},
\]
yielding moderately informative priors that retain sufficient flexibility for the $B_c$ data to influence the posterior. Gaussian priors are applied only to $\alpha_s$, $\sigma$, and $V_c$, since these parameters are already constrained by the heavy-quarkonium analysis. The smearing parameter $\rho$ and the additional parameters introduced in the modified Cornell and screened potentials retain uniform priors because no independent external information is available for them. In particular, the bottomonium-based fit value of $\rho$ lies at the boundary of the adopted uniform interval, making a Gaussian prior unsuitable. The corresponding prior is therefore written as
\begin{equation}
\log p(\Theta) = \sum_{i\in\{\alpha_s,\sigma,V_c\}} \log \mathcal N(\theta_i;\mu_i,\sigma_i) + \sum_{j\in U} \log p_U(\theta_j),
\end{equation}
where $\mathcal N(\theta_i;\mu_i,\sigma_i)$ denotes the Gaussian prior for parameter $\theta_i$ with mean $\mu_i$ and standard deviation $\sigma_i$, while $p_U(\theta_j)$ denotes the uniform prior assigned to the remaining free parameters.
The resulting posterior distributions are compared with those obtained using uniform priors to evaluate the robustness of the inferred parameters with respect to the adopted prior assumptions.

For each prior choice, three independent MCMC chains with different random seeds are generated. Convergence is assessed using the integrated autocorrelation time ($\tau_{\mathrm{int}}$), effective sample size (ESS), acceptance fraction, and Gelman--Rubin ($\hat{R}$) statistic. All chains satisfy the \texttt{emcee} criterion $N_{\rm steps}>50~\tau_{\rm int}$ and yield statistically consistent posterior distributions. After discarding the initial $20\%$ of samples as burn-in, one representative converged chain is used for the subsequent posterior inference and spectrum calculations, while the other chains serve as independent convergence checks. The posterior distributions are visualized using \texttt{corner} plots~\cite{corner}. Detailed sampler settings, convergence diagnostics, thinning strategy, prior-sensitivity tests, and WAIC and LOO cross-validation results are given in Appendix~\ref{app:2_MCMC}.

From the retained samples, we extract the median and $1\sigma$ credible intervals for each parameter (Table~\ref{tab:Bc_parameters}). The median is adopted as the central estimator in preference to the mean, since the posterior distributions are generally asymmetric owing to the nonlinear dependence of the spectrum on the potential-model parameters. Spectral uncertainties are propagated using $\sim\!5000$ posterior samples from the representative chain after burn-in and thinning based on $\tau_{\mathrm{int}}$. The corresponding ESS values are $\sim2300-2500$ for the Cornell potential, $\sim2070-4350$ for the modified Cornell potential, and $\sim3331-8702$ for the screened Cornell potential. For each posterior draw, we compute the masses of all states up to $6D$, including ${}^3\!P_1-{}^1\!P_1$ and ${}^3\!D_2-{}^1\!D_2$ mixing. The resulting posterior spectra provide the median predictions and credible bands reported in Sec.~\ref{sec:3_ND}.

\subsection{Wave Function at the Origin and Related Quantities}
\label{sec:wfo}

The square of the radial wave function at the origin, $|R_{nl}^{(l)}(0)|^2$, characterizes the short-distance structure of the $B_c$ meson and enters directly into physical observables such as decay constants and production cross sections. It is extracted from the solution of the radial Schr\"odinger equation as
\begin{equation}
    \label{eqn:wfo}
    R_{nl}^{(l)}(0) = \frac{d^l R_{nl}(r)}{dr^l}\Bigg|_{r=0},
\end{equation}
giving $|R_{nl}(0)|^2$, $|R_{nl}'(0)|^2$, and $|R_{nl}''(0)|^2$ for $S$-, $P$-, and $D$-wave states, respectively (for details see Appendix~\ref{app:rk}).

The root-mean-squared (RMS) radius is computed from the normalized reduced radial wave function $u_{nl}(r) = rR_{nl}(r)$ as
\begin{equation}
    \sqrt{\langle r^2 \rangle} = \sqrt{\int_0^\infty u_{nl}^*(r)\, r^2\, u_{nl}(r)\, dr},
\end{equation}
and provides a measure of the spatial extent of the bound state, with particular sensitivity to the long-range confining potential.

The kinetic energy expectation value $\langle T \rangle$ is evaluated via the virial theorem~\cite{Lucha:1991vn},
\begin{equation}
    \langle T \rangle = \frac{1}{2}\left\langle r\frac{dV(r)}{dr}\right\rangle,
\end{equation}
from which the squared momentum follows as
\begin{equation}
    \langle p^2 \rangle = 2\mu\langle T \rangle, \qquad \mu = \frac{m_c m_b}{m_c + m_b}.
\end{equation}
The individual heavy-quark velocity expectations are then given by
\begin{equation}
    \langle v_c^2 \rangle = \frac{\langle p^2 \rangle}{m_c^2}, \qquad \langle v_b^2 \rangle = \frac{\langle p^2 \rangle}{m_b^2},
\end{equation}
and serve as a diagnostic of the validity of the non-relativistic approximation.

All quantities above are evaluated for each of the $\sim 5000$ posterior samples drawn from the MCMC chains of all three potentials. The quoted central values correspond to the median of the resulting distributions, with uncertainties reflecting the $1\sigma$ credible intervals, thereby consistently propagating inter-parameter correlations to all derived observables.

\subsection{Regge Trajectories}
\label{sec:regge}

Regge trajectories correlate the squared masses of bound states with their radial ($n_r$) or orbital quantum numbers ($l$), providing a phenomenological window onto the global structure of the spectrum~\cite{Regge:1959mz, Regge:1960zc}. For mesons, the trajectories are generally expected to be linear, with the hadron mass expressed as
\begin{equation}\label{eqn:22_regge_linear}
    M^2 = \alpha_{lin}\, l + \beta_{lin}\, n_r + C_{lin},
\end{equation}
where $\alpha_{lin}$ and $\beta_{lin}$ are the Regge slopes and $C_{lin}$ is a constant~\cite{Chew:1961ev, Chew:1962eu, Chen:2016spr}.

The linear form, however, was from the outset an approximation~\cite{Chew:1961ev, Chew:1962eu}, and several theoretical studies have shown that trajectories in heavy quarkonium systems can deviate from strict linearity~\cite{Martin:1986rtr, Lucha:1991vn, Ebert:2013iya, Chen:2018hnx}, a point of particular relevance for the $B_c$ meson. Indeed, the linear confining potential commonly used in non-relativistic treatments does not necessarily produce linear Regge trajectories~\cite{Martin:1986rtr, Lucha:1991vn}. To account for these possible departures, especially among low-lying states, we also consider the non-linear Regge form proposed in Refs.~\cite{Feng:2023txx, Chen:2023djq},
\begin{equation} \label{eqn:23_regge_NL}
    M^2 = m_R^2 + \beta_x^2\,(x + c_{0x})^{4/3} + 2m_R\beta_x\,(x + c_{0x})^{2/3}, \qquad (x = n_r,\, l),
\end{equation}
with parameters defined as
\begin{equation} \label{eqn:24}
    \beta_x = c_{fx}\, c_x\, c_c, \qquad m_R = m_1 + m_2 + V_c,
\end{equation}
\begin{equation} \label{eqn:25}
    c_{n_r} = \frac{(3\pi)^{2/3}}{2}, \qquad c_{l} = \frac{3}{2}, \qquad c_c = \left(\frac{\sigma^2}{\mu}\right)^{1/3}.
\end{equation}
Here $c_{fn_r}$ and $c_{fl}$ are fit parameters expected to be close to unity, while $c_{0n_r}$ and $c_{0l}$ vary across individual trajectories.

We apply this non-linear form unchanged to the mass spectra from all three potentials enabling a direct comparison of the resulting Regge behavior between them. The parameters of both the linear (Eq.~\eqref{eqn:22_regge_linear}) and non-linear (Eq.~\eqref{eqn:23_regge_NL}) trajectories are extracted by fitting the MCMC-derived mass spectra using the \texttt{iminuit} package (version $2.18.0$)~\cite{iminuit}, based on the MINUIT optimization algorithm~\cite{James:1975dr}. The resulting trajectories are discussed in Sec.~\ref{sec:3C_Regge}.

\section{Numerical Results and Discussion}
\label{sec:3_ND}

We present the numerical results for the $B_c$ mass spectrum obtained from the Cornell potential (Eq.~\eqref{eqn:1_cornell}), its logarithmic extension (Eq.~\eqref{eqn:corplusln}), and screened Cornell potential (Eq.~\eqref{eqn:screenpot}), hereafter Potential~I, Potential~II, and Potential~III, respectively. The role of the logarithmic term is quantified through the $r$-dependent effective string tension $\sigma_{\rm eff}(r)$ (Eq.~\eqref{eqn:sigmaeffective}). All parameters and uncertainties are obtained from the Bayesian MCMC analysis described in Sec.~\ref{sec:2_methodology}, with asymmetric credible intervals quoted at $1\sigma$.

The fitted parameters for all three potentials are listed in Table~\ref{tab:Bc_parameters}. The short-range parameters $\alpha_s$, $\rho$, and $V_c$ are consistent between Potentials I and II within uncertainties. In Potential~II, the logarithmic term, governed by $\sigma' = 0.22~\text{GeV}$ and $C_0 = 0.225~\text{GeV}$, assumes part of the confining role, allowing the linear string tension to relax from $\sigma = 0.187~\text{GeV}^2$ (Potential~I) to $0.15~\text{GeV}^2$. The three confinement-sector parameters collectively generate the effective $r$-dependent tension $\sigma_{\rm eff}(r)$, and correspondingly carry larger, asymmetric uncertainties that reflect their mutual correlations. Interestingly, Potential~III favors a slightly larger $\alpha_s$ and $\sigma$ and a smaller $\rho$ than Potentials I and II, with $\mu_{\text{scr}} = 0.058~\text{GeV}$ and a comparatively broad credible interval, indicating that screening becomes relevant mainly at larger separations.

\begin{figure}
    \centering
    \includegraphics[width=0.9\linewidth]{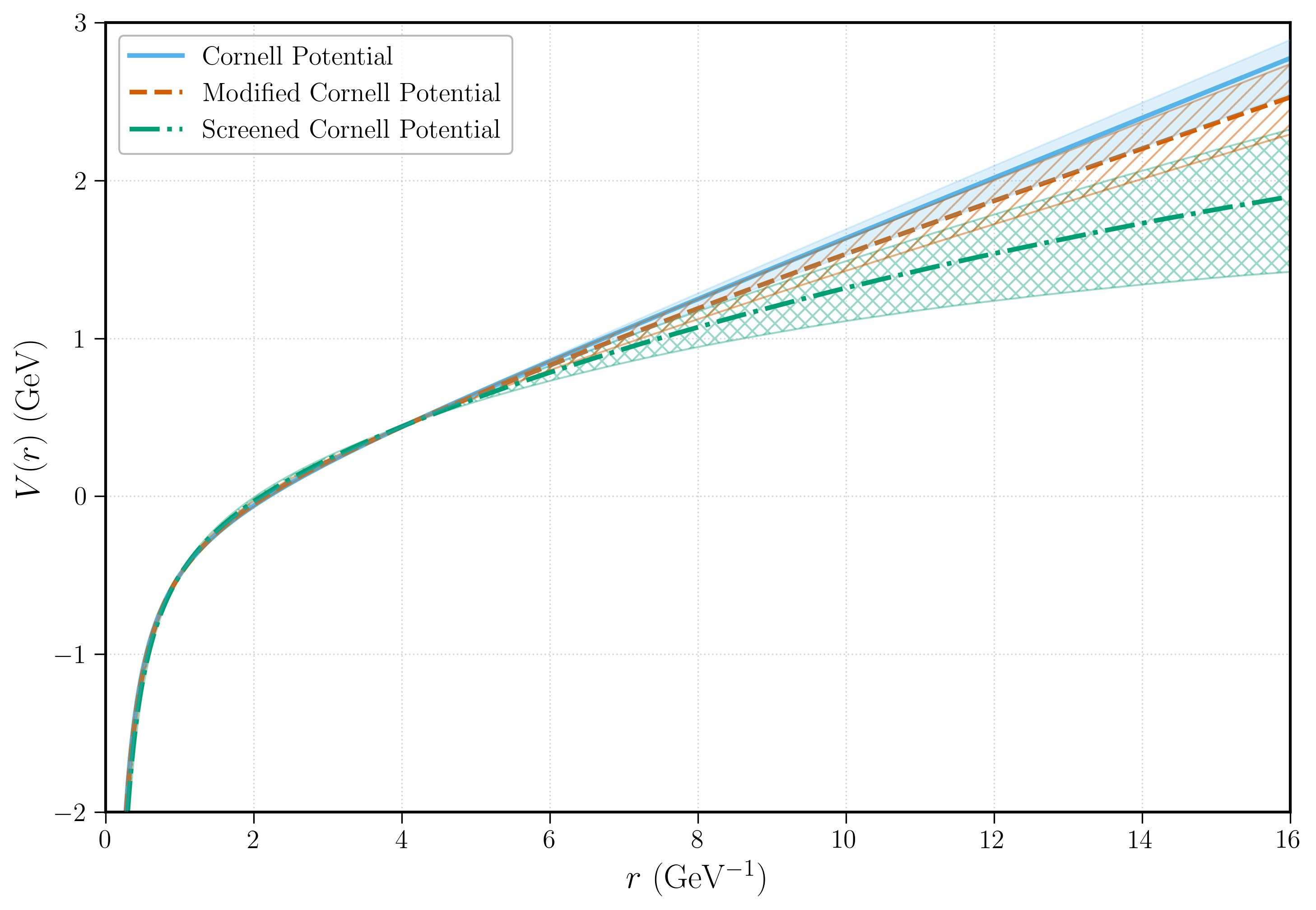}
    \caption{Comparison of the Cornell (Potential I, Eq.~\eqref{eqn:1_cornell}), Modified Cornell (Potential II, Eq.~\eqref{eqn:corplusln}), and Screened Cornell (Potential III, Eq.~\eqref{eqn:screenpot}) potentials, using median parameter values Table~\ref{tab:Bc_parameters}. Shaded bands denote credible regions based on $\sim 5000$ MCMC samples.}
    \label{fig:BcPots}
\end{figure}

Figure~\ref{fig:BcPots} presents a direct comparison of the potentials constructed from $\sim\!5000$ posterior samples.\footnote{See Appendix~\ref{app:2_MCMC} for sampler details.} At short distances, all three potential forms coincide, reflecting the common short-range structure. At intermediate and large $r$, Potential~II develops a systematically shallower slope due to the logarithmic correction while retaining its asymptotically linear behavior, whereas Potential~III shows a stronger suppression at large separations from the screening term. The credible bands widen with increasing $r$, most prominently at large separations, indicating that the confinement-sector parameters are less constrained by the available low-lying data.

Comparing the MCMC-extracted parameters of Potential~I with the deterministic fits in Table~\ref{tab:min_params} (Appendix~\ref{app:1_min}), the values of $\alpha_s$ and $\sigma$ lie between those of Set-I and Set-II but remain closer to Set-I, consistent with the stronger kinematic affinity of the $B_c$ system to $b\bar{b}$. The parameter $\rho$ remains relatively broad across the three potentials, reflecting the limited number of $B_c$ states constraining the spin-spin interaction.

\begin{figure}
    \centering
    \includegraphics[width=0.8\linewidth]{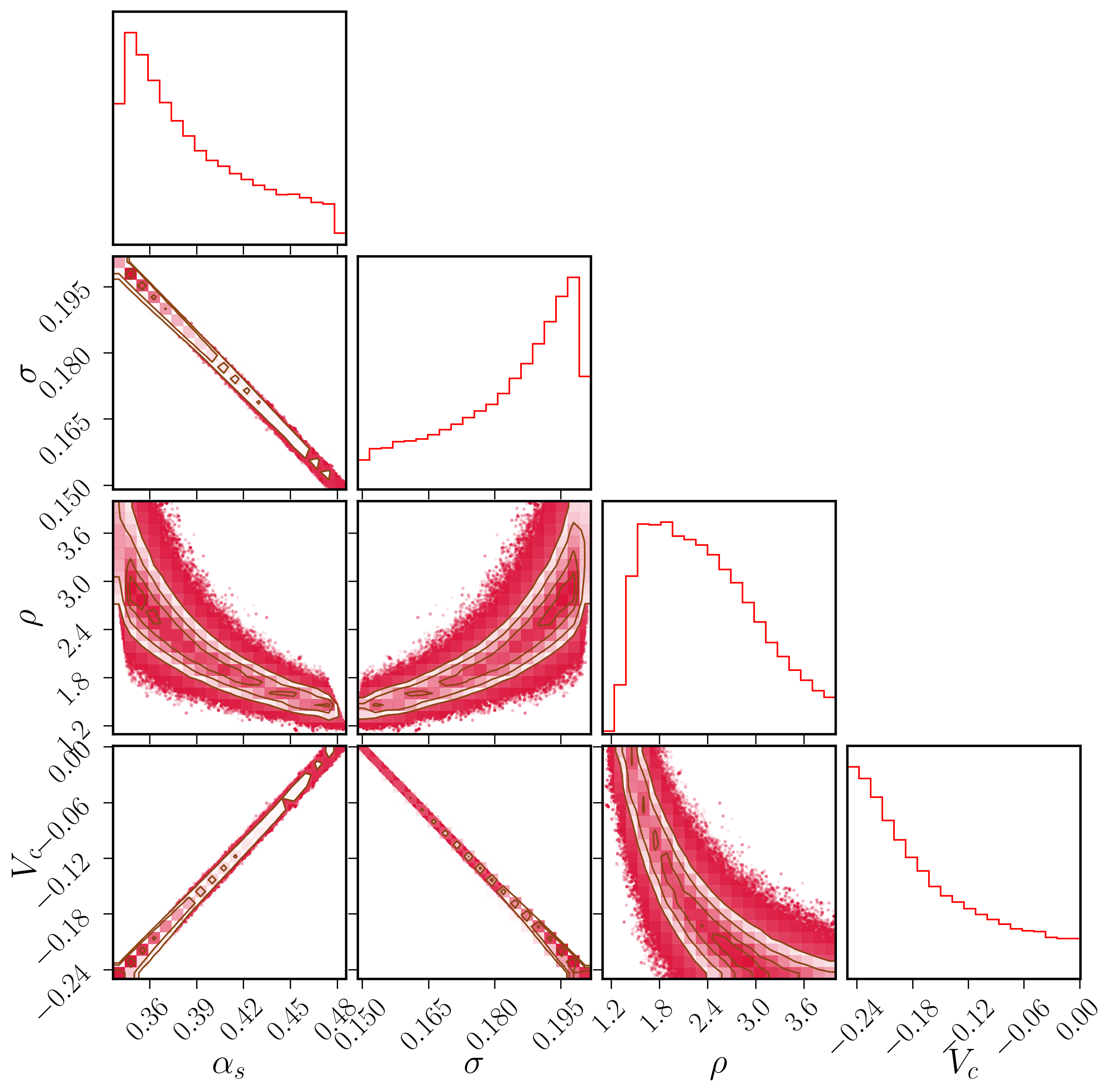}
    \caption{Corner plot for Cornell Potential (Potential I).}
    \label{fig:CornerCornel}
\end{figure}

\begin{figure}
    \centering
    \includegraphics[width=\linewidth]{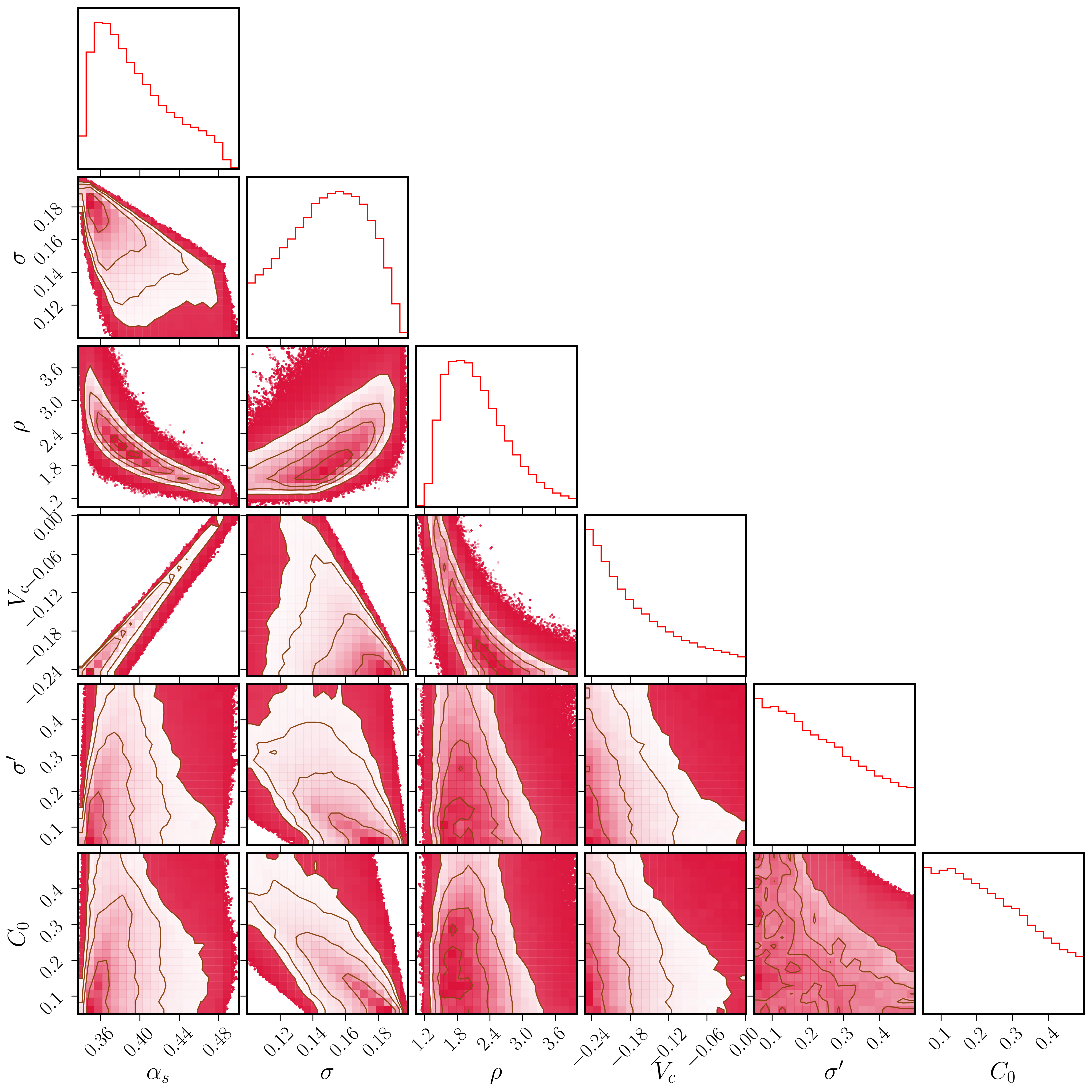}
    \caption{Corner plot for Modified Cornell Potential (Potential II).}
    \label{fig:CornerModCornel}
\end{figure}

\begin{figure}
    \centering
    \includegraphics[width=\linewidth]{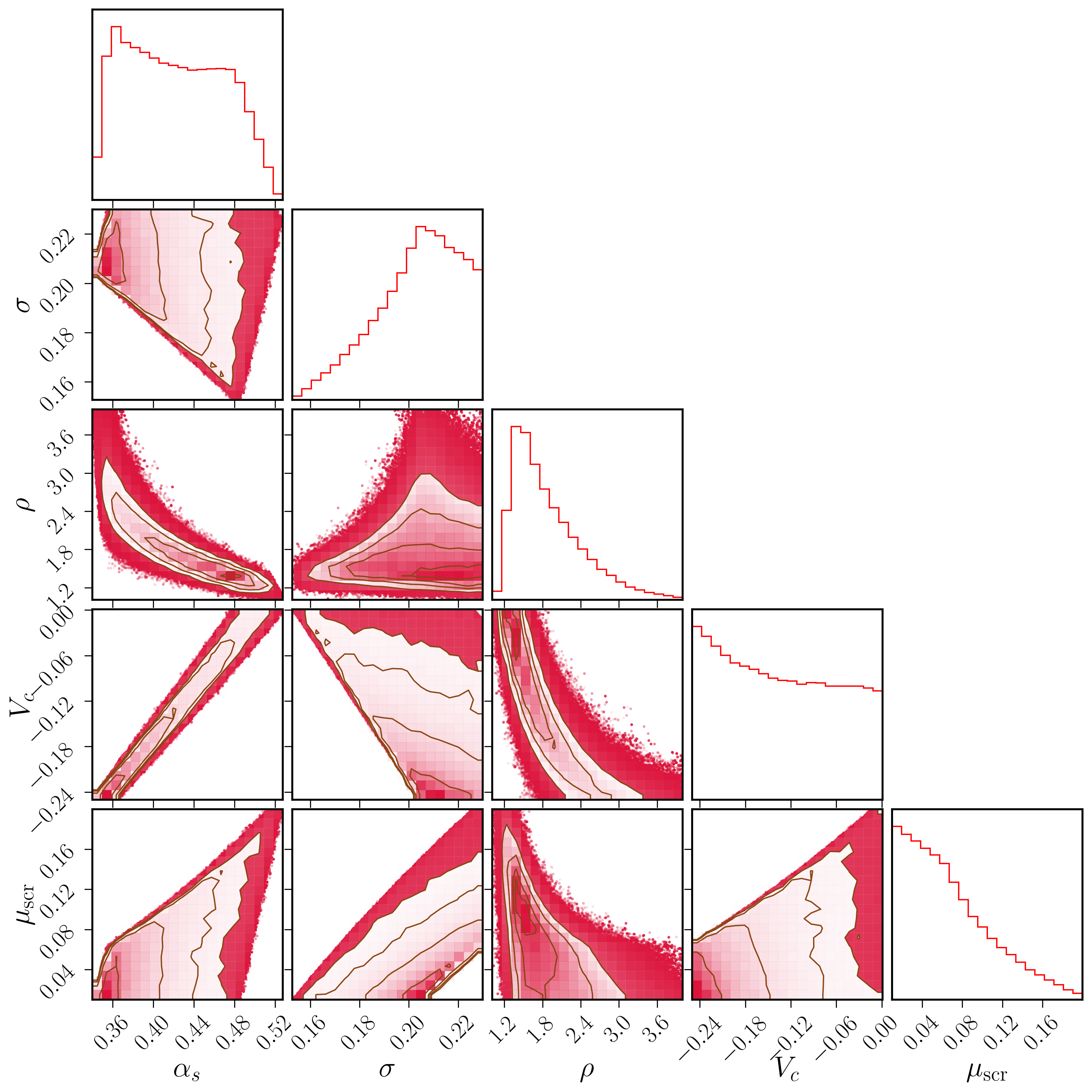}
    \caption{Corner plot for Screened Cornell Potential (Potential III).}
    \label{fig:CornerScrCornel}
\end{figure}

The parameter correlations are displayed in the corner plots of Figs.~\ref{fig:CornerCornel}-\ref{fig:CornerScrCornel}. For Potential~I, $\alpha_s$ and $\sigma$ are well-peaked, while $\rho$ and $V_c$ are broader and skewed. The joint distributions reveal a tight negative $\alpha_s-\sigma$ correlation and a compensatory positive $\alpha_s-V_c$ correlation, with the $\rho-\alpha_s$ and $\rho-\sigma$ contours exhibiting characteristic banana-shaped curvature. Interestingly, the observed negative $\alpha_s-\sigma$ correlation is consistent with that reported by Mateu et al.~\cite{Mateu:2018zym}. For Potential~II, the $\alpha_s-V_c$ correlation persists, but the addition of $\sigma'$ and $C_0$ introduces extended, diffuse contours in the $\sigma-\sigma'$, $\sigma-C_0$, and $\sigma'-C_0$ planes, confirming that these three parameters collectively parametrize the effective confinement and cannot be individually resolved without additional experimental input.\footnote{We have also applied the modified Cornell potential to bottomonium, where the larger number of observed states enables a more robust assessment; a systematic improvement in higher excited states is found and will be detailed in a forthcoming publication.} For Potential~III, the $\alpha_s-V_c$ and curved $\rho-\alpha_s$ correlations observed in the other potentials persist, while the introduction of $\mu_{\rm scr}$ leads to broader and more asymmetric marginal distributions. In particular, $\alpha_s$ is comparatively diffuse, $\sigma$ is skewed toward the upper prior boundary, and $\mu_{\rm scr}$ peaks at its lower boundary, indicating a posterior preference for weaker screening within the adopted prior range.

The present $B_c$ dataset provides three effectively independent experimental constraints, since $M_{B_c(2S)}-M_{B_c(1S)}$ is algebraically related to the two individual masses. Thus, with four, six, and five free parameters, Potentials~I, II, and III are formally under-determined. The corresponding WAIC effective parameter counts, $p_{\rm WAIC}\simeq 1.23$, $1.27$, and $1.27$, further indicate that the data constrain only a limited number of parameter combinations, consistent with the extended posterior degeneracies observed in the corner plots. We assess prior sensitivity by repeating the inference with moderately informative Gaussian priors centered on the bottomonium-based Set-I values for $\alpha_s$, $\sigma$, and $V_c$, while retaining uniform priors for $\rho$ and the additional parameters of Potentials~II and III. Although these priors generally narrow the posterior parameter distributions, their effect on the predicted spectrum is small (as discussed in Appendix~\ref{app:2_MCMC}).

Model comparison using WAIC and PSIS-LOO yields no statistically significant preference among the three potentials, including between Potentials~I and II, and the same conclusion holds under the Gaussian prior. Thus, while the logarithmic extension provides a useful phenomenological framework for probing intermediate-distance confinement, the present $B_c$ data do not support a statistically preferred confinement model. We therefore retain the uniform-prior results as the baseline inference for the subsequent spectrum analysis, with the Gaussian-prior results serving as a prior-sensitivity check. The resulting credible intervals should be interpreted in light of the limited experimental constraints and associated parametric freedom. With this understanding, we now turn to the sector-wise predictions for the $B_c$ spectrum.

\subsection{Mass Spectra}
\label{sec:3A_Mass}

The mass spectra from the potentials are presented in Tables~\ref{tab:Bcmass_S}, \ref{tab:Bcmass_P}, and \ref{tab:Bcmass_D} for the $S$-, $P$-, and $D$-wave $B_c$ states, respectively, with Potential~I-III in columns~2-4. Each entry reports the median with its asymmetric $1\sigma$ credible interval. We compare against experimental data~\cite{ParticleDataGroup:2024cfk} and a range of theoretical results: LQCD~\cite{Mathur:2018epb, Dowdall:2012ab, Davies:1996gi}, the Godfrey-Isgur relativized quark model (GI)~\cite{Godfrey:2004ya}, the relativistic quark model of Ebert et al.\ (EFG)~\cite{Ebert:2002pp, Ebert:2011jc}, the modified GI model with screening (LLWL)~\cite{Li:2023wgq}, the full Salpeter equation solution (WWLC)~\cite{Wang:2022cxy}, the phenomenological relativistic model with coupled-channel effects (MBK)~\cite{Monteiro:2016rzi}, relativistic screened potential model (BB)~\cite{Bokade:2025lmn}, semi-relativistic constituent-quark framework with screened potential (PLCR)~\cite{Patel:2026prg} and various non-relativistic potential models (EQ~\cite{Eichten:2019gig}, LZ~\cite{Li:2019tbn}, AAMS~\cite{Asghar:2019qjl}, DKR~\cite{Devlani:2014nda}, LTFWP~\cite{Li:2022bre}, HZ~ \cite{Hao:2024nqb}).

\begin{enumerate}

\item The pseudoscalar states $B_c(1S)$ and $B_c(2S)$ are reproduced in excellent agreement with the PDG~\cite{ParticleDataGroup:2024cfk} by all three potentials,
\[
M_{\rm I}(1^1S_0) = 6274.48^{+0.31}_{-0.30}~\text{MeV},\quad M_{\rm II}(1^1S_0) = 6274.47^{+0.31}_{-0.31}~\text{MeV},
\]
\[\text{and }M_{\rm III}(1^1S_0) = 6274.47_{-0.31}^{+0.32}~\text{MeV,}\]
with sub-MeV uncertainties and negligible inter-potential differences, confirming robust short-range calibration. The predictions also agree with LQCD estimates~\cite{Mathur:2018epb, Dowdall:2012ab}, as do the $2S$ masses within uncertainties~\cite{Dowdall:2012ab}. Among the thirteen independent reference models, all the results cluster within a $9$~MeV band around the experimental value, except for AAMS~\cite{Asghar:2019qjl}, which lies $\sim 44$~MeV above the ground state.

\item The $1S$ hyperfine splitting,
\begin{align*}
\Delta M_{1S}^{\rm I} &= 55.71^{+3.63}_{-3.97}~\text{MeV},\\ 
\Delta M_{1S}^{\rm II} &= 55.65^{+3.90}_{-3.91}~\text{MeV},\\ 
\Delta M_{1S}^{\rm III} &= 55.27_{-3.9}^{+3.98}~\text{MeV},
\end{align*}
agrees with the LQCD results of $55(3)$~MeV~\cite{Mathur:2018epb} and $54(3)$~MeV~\cite{Dowdall:2012ab}, validating the spin-contact interaction parametrization. Among the reference models the spread is $39-67$~MeV, with MBK's value of $39$~MeV~\cite{Monteiro:2016rzi} indicating a more pronounced suppression of the contact interaction. The $2S$ splitting,
\begin{align*}
\Delta M_{2S}^{\rm I} &= 29.66^{+4.38}_{-5.22}~\text{MeV},\\ 
\Delta M_{2S}^{\rm II} &= 27.76^{+4.64}_{-4.68}~\text{MeV},\\ 
\Delta M_{2S}^{\rm III} &= 23.6_{-5.08}^{+5.5}~\text{MeV},
\end{align*}
are reduced by approximately $47\%$, $50\%$, and $57\%$ relative to their respective $1S$ values for Potentials~I, II, and III. For Potentials~I and II, the $\sim47\%$ reduction follows from the smeared contact-overlap ratio $\langle\delta\rangle_{2S}/\langle\delta\rangle_{1S}\approx0.53$ (Eq.~\eqref{eqn:smearfn}), substantially weaker than the pure-Coulomb $1/n^3$ expectation because of the confining enhancement of the near-origin wave function and the finite range of the Gaussian-smeared spin-spin operator. The stronger suppression in Potential~III reflects the additional modification of the excited-state radial wave function induced by screening, leading to the smallest $2S$ hyperfine splitting among the three potentials.

\item Recently, the ATLAS Collaboration reported the first observation of the $B_c^{*+}$ meson, measuring the hyperfine splitting as $\Delta M_{1S}=64.5\pm1.4~\text{(stat.)}_{-1.4}^{+1.0}~\text{(syst.)}~\text{MeV}$ \cite{ATLAS:2026ubk}. This value is higher than the posterior medians obtained in the present analysis, $\Delta M_{1S}=55.71$, $55.65$, and $55.27$ MeV for Potentials~I, II, and III, respectively. The present inference was performed using the lattice input $55(3)$ MeV available when the analysis was undertaken~\cite{Mathur:2018epb}, which is consistent with the independent lattice determination $54(3)$ MeV \cite{Dowdall:2012ab}. To assess the sensitivity of the inferred parameters to the precision assigned to the lattice constraint, we further broadened its uncertainty from $3$ to $10$~MeV through posterior importance reweighting using $3\times10^5$ uniformly thinned posterior samples. The posterior median of $\rho$ shifts by approximately $+0.21$, $+0.24$, and $+0.18$~GeV for Potentials~I, II, and III, respectively, while remaining within the corresponding original $68\%$ credible intervals, and the shifts in the other parameter posteriors are modest. The stability of the $\rho$ posterior under this uncertainty broadening indicates that its width is not an artifact of the assumed $3$~MeV lattice precision. The hyperfine splitting constrains a correlated combination of $\alpha_s$, $\rho$, and the radial wave function through the smeared contact matrix element, rather than fixing $\rho$ independently. 

\item The $S$-wave predictions from Potentials~I and II agree closely through $2S$, beyond which a systematic inter-potential separation develops and grows nearly linearly with $n$. The mass gap $\delta M_n \equiv M_{\rm I}(n^1S_0) - M_{\rm II}(n^1S_0)$ accumulates at $\sim 14$~MeV per unit $n$, reaching $\sim 70$~MeV by $6S$. This divergence originates in the long-range confining sector where the two potentials differ, as shown in Fig.~\ref{fig:BcPots}, making $B_c$ radial excitations with $n \geq 4$ effective discriminators between confinement models. The MCMC-propagated uncertainties widen monotonically with $n$. The low-$n$ states carry symmetric or mildly asymmetric uncertainties and the high-$n$ states develop lower-skewed bands (e.g., $M_{\rm I}(6^3S_1) = 8232.47^{+39.46}_{-80.35}$~MeV), reflecting the absence of experimental constraints on higher excitations. Overall, both potentials yield comparable $S$-wave spectra for low-lying states while diverging progressively for higher excitations, with Potential~II clustering excited levels more tightly due to the shallower effective slope. For comparison, Potential~III remains close to Potentials~I and~II for the $1S$ and $2S$ states, but its $6S$ prediction is lower by $\sim260$~MeV relative to Potential~I and by $\sim188$~MeV relative to Potential~II. Its $6^3S_1$ uncertainty, $^{+150}_{-241}$~MeV, is also substantially larger than those of Potentials~I and~II, providing a quantitative measure of the increased uncertainty at higher radial excitation.

\item For low-lying states, Potentials I and II agree with all reference models except AAMS~\cite{Asghar:2019qjl}, which consistently underestimates the excited-state masses. For $n \geq 5$, LLWL~\cite{Li:2023wgq}, LZ~\cite{Li:2019tbn}, and LTFWP~\cite{Li:2022bre} give lower masses than our median predictions, while the remaining models show similar trends.
Potential~III changes the comparison at higher radial excitation. With its stronger screening of the confining interaction, the predicted masses for $n\geq3$ become progressively lower than those from Potentials~I and II and are closer to several other screened-potential calculations. In particular, the recent BB~\cite{Bokade:2025lmn} and PLCR~\cite{Patel:2026prg} results are close to Potential~III for the higher $S$ states, while LLWL remains lower. LZ~\cite{Li:2019tbn} and LTFWP~\cite{Li:2022bre}, which give lower masses than Potentials~I and II at higher excitation, are generally above Potential~III from $4S$ onward. The HZ predictions~\cite{Hao:2024nqb}, available up to $3S$, are also above Potential~III in this sector. Recent Bayesian and Monte-Carlo studies provide additional points of comparison. Molina et al.~\cite{Molina:2020zao} used bootstrap-based uncertainty propagation for bottomonium, whereas our Bayesian MCMC framework resolves parameter correlations and propagates asymmetric posterior uncertainties in the experimentally sparse and unequal-mass $B_c$ system, including spin-dependent interactions and the $^3P_1-^1P_1$ and $^3D_2-^1D_2$ mixing. Akan~\cite{Akan:2025nej} considered a spin-averaged Cornell analysis; here we include the full fine structure and introduce a logarithmic deformation to examine the sensitivity to the intermediate-distance confinement profile. Our results thus combine uncertainty-quantified spectroscopic predictions with a comparison of the dependence on the confinement potential.

\item For $P$-wave states (Table~\ref{tab:Bcmass_P}), the $1^3P_0$ and axial-vector $1P_1$ medians from Potentials I and II agree excellently with LQCD~\cite{Mathur:2018epb, Dowdall:2012ab, Davies:1996gi}, while the $1P_1'$ and $1^3P_2$ states are reproduced within the LQCD uncertainties~\cite{Davies:1996gi}. The $1P$ fine-structure spread,
\begin{align*}
\Delta M_{\rm fine}^{\rm I}(1P) &= M(1^3P_2) - M(1^3P_0) = 49.95~\text{MeV},\\
\Delta M_{\rm fine}^{\rm II}(1P) &= 52.56~\text{MeV},\\
\Delta M_{\rm fine}^{\rm III}(1P) &= 62.22~\text{MeV},
\end{align*}
is determined by the spin-orbit and tensor interactions, where Potential~III gives the largest separation among the three potentials. The $1^3P_0$ predictions are
$6699$--$6714$~MeV for Potentials I and II and
$6714.64_{-5.74}^{+8.31}$~MeV for Potential III, placing the three predictions within the range covered by the reference models, with the Potential III result also consistent with the available LQCD determinations within uncertainties. 

\item The $P$-wave inter-potential agreement holds through $2P$, after which the spectra separate progressively, with the mass difference between Potentials~I and II reaching $\sim 80$~MeV at $6^3P_2$. Potential~III shows a stronger compression of the higher $P$-wave spectrum, with its $6^3P_2$ mass lying $\sim 308$~MeV below Potential~I. The mixed axial-vector states remain closely spaced across all three potentials, with $M_{nP_1'}-M_{nP_1}$ varying from $\sim 5$ to $\sim 7$~MeV for Potentials~I and II and from $\sim 8$ to $\sim 6$~MeV for Potential~III between $1P$ and $6P$. The corresponding $P_1-P_1'$ mixing angles increase with radial excitation. For Potentials~I and II, $\theta_{1P}=4.82^\circ$ and $6.62^\circ$, respectively, with large uncertainties of about $\pm9^\circ$--$10^\circ$ at $1P$, while both reach $\theta_{6P}\approx19^\circ$ with smaller uncertainties. Potential~III gives $\theta_{1P}=(12.85_{-9.87}^{+4.32})^\circ$ and increases gradually to $\sim20^\circ$ at $6P$. 

\item The low-lying $P$-wave predictions agree with the reference models except AAMS~\cite{Asghar:2019qjl}. For higher excitations, EFG \cite{Ebert:2002pp, Ebert:2011jc} gives slightly larger masses than our medians for Potentials~I and II, while LLWL~\cite{Li:2023wgq}, LZ~\cite{Li:2019tbn}, and LTFWP~\cite{Li:2022bre} give comparatively smaller masses. Potential~III is closer to the screened-potential calculations, including LLWL~\cite{Li:2023wgq}, LTFWP~\cite{Li:2022bre}, BB~\cite{Bokade:2025lmn}, and PLCR~\cite{Patel:2026prg}, particularly for the higher excitations. The separation between the screened and unscreened potentials becomes more pronounced with increasing excitation, as reflected in the predicted spectra. For the mixing angles, the rotation convention adopted in the present work is consistent with those used in GI~\cite{Godfrey:2004ya}, LLWL~\cite{Li:2023wgq}, and MBK~\cite{Monteiro:2016rzi}. The numerical values should nevertheless be compared with care across different models. Although the rotation convention is common to these works, the mixing angles remain sensitive to the adopted spin-dependent interaction, fitted parameters, and phase or charge-conjugation conventions. In addition, the association of the primed label with a particular mass eigenstate is not universal across the literature.

\item The recent LHCb observation of the orbitally excited $B_c(1P)$ structure in the $B_c^+\gamma$ invariant-mass spectrum~\cite{LHCb:2025ubr,LHCb:2025uce} provides direct experimental evidence for the $P$-wave $B_c$ sector. At the current experimental resolution, however, the observed enhancement is described by an effective two-peak fit, with masses $M_1=6704.8$~MeV and $M_2=6752.4$~MeV, rather than by individually resolved $1^3P_0$, $1P_1$, $1P_1'$, and $1^3P_2$ states. Consequently, these two peak positions cannot be unambiguously assigned to individual members of the $1P$ multiplet and are not used as state-specific likelihood inputs in the present analysis. The $1P$ masses predicted by Potentials~I--III, obtained from fits that do not use these LHCb measurements, populate the mass region covered by the observed structure, providing a qualitative posterior-predictive consistency check for the $P$-wave spectrum. A quantitative test of the predicted fine structure and $P_1-P_1'$ mixing pattern will require experimentally resolved measurements of the individual $1P$ states.

\item The $D$-wave predictions (Table~\ref{tab:Bcmass_D}) agree between the Potentials I and II for low-lying states and diverge steadily at high $n$, mirroring the $S$- and $P$-wave behavior. The gap at $6^3D_3$ reaches $\sim 90$~MeV, larger than for $S$- or $P$-waves at the same $n$, because higher-$l$ states are more spatially extended and thus more sensitive to long-range potential differences. Potential~III follows the same qualitative trend for the lowest $D$-wave multiplet, while its stronger screening leads to a more pronounced downward shift for higher excitations. Unlike the $P$-wave case, the mass splitting of the mixed $D$-wave states $M_{nD_2'} - M_{nD_2}$ decreases with $n$ ($\sim 10$~MeV at $1D$; $\sim 3$~MeV at $6D$). The $D_2-D_2'$ mixing angle is large and negative throughout, $\theta_{1D} = -52.52^\circ$ ($-52.29^\circ$) for Potential~I (II). The magnitude decreases slowly with $n$ ($\theta_{6D} = -47.58^\circ$ and $-47.38^\circ$ for Potential~I and II, respectively), with uncertainties of $\sim 1.5^\circ$ for the lowest-lying multiplets, in stark contrast to the $P$-wave case, where $\theta_{nP}$ uncertainties reach $\pm 10^\circ$ at low $n$. Potential~III shows the same behavior, with slightly smaller magnitudes, varying from $\theta_{1D}=-51.22^\circ$ to $\theta_{6D}=-44.35^\circ$; however, the uncertainties on the angles are slightly larger than the other two potentials.

\item The $1D$ predictions from Potentials I and II agree with the majority of reference models, with MBK~\cite{Monteiro:2016rzi} and AAMS~\cite{Asghar:2019qjl} as outliers. For higher excitations, LLWL~\cite{Li:2023wgq}, AAMS~\cite{Asghar:2019qjl}, and LTFWP~\cite{Li:2022bre} fall systematically below our medians. Potential~III predictions follow a similar pattern compared to other screened-potential calculations, including LLWL~\cite{Li:2023wgq}, BB~\cite{Bokade:2025lmn}, PLCR~\cite{Patel:2026prg}, and LTFWP~\cite{Li:2022bre}.

\item The mass ordering of the low-lying $D$-wave multiplets follows $1D_2 < 1^3D_3 < 1^3D_1 < 1D_2'$ and $2D_2 < 2^3D_1 < 2^3D_3 < 2D_2'$, which deviates from the naive expectation $n^3D_1 < nD_2 < nD_2' < n^3D_3$. This non-standard ordering persists through $4D$ for Potentials~I and II, while for Potential~III it persists through $2D$ before reverting to the conventional hierarchy at higher excitations. This level-dependent sign reversal is not a numerical artifact; similar observations are made by other models including MBK~\cite{Monteiro:2016rzi}, EQ~\cite{Eichten:2019gig}, LZ~\cite{Li:2019tbn}, DKR~\cite{Devlani:2014nda}, and LTFWP~\cite{Li:2022bre}.
\end{enumerate}

\begin{figure}
    \centering
    \includegraphics[width=\linewidth]{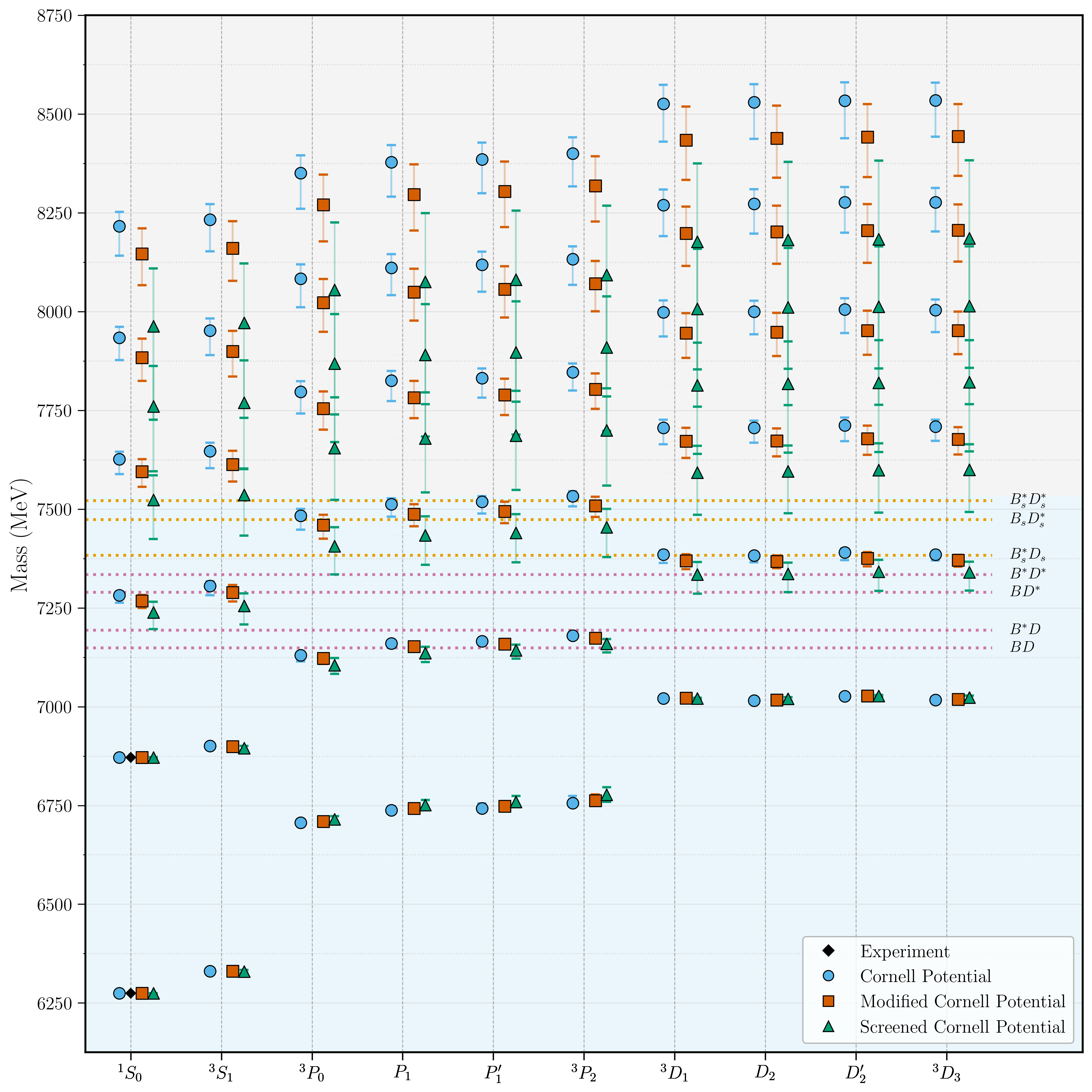}
    \caption{The $B_c$ mass spectrum (in MeV) computed using Cornell (Potential I), modified Cornell (Potential II), and screened Cornell (Potential III) potentials in solid blue circle, solid red square, and solid green triangle, respectively. The black diamonds are experimental values \cite{ParticleDataGroup:2024cfk}. The dotted lines indicate the thresholds for open-flavor $B_{(s)}^{(*)} D_{(s)}^{(*)}$ channels. The grey shaded region hosts states with $\langle v_c^2\rangle\sim0.5$ and above.}
    \label{fig:MassSpectraBc}
\end{figure}

The $B_c$ mass spectra predicted by the three potentials are presented as a Grotrian diagram in Fig.~\ref{fig:MassSpectraBc}, with PDG values~\cite{ParticleDataGroup:2024cfk} indicated by black diamonds and open-flavor $B_{(s)}^{(*)}D_{(s)}^{(*)}$ thresholds by dotted horizontal lines. States below the lowest threshold ($BD$), namely $1S$, $2S$, $1P$, $2P$, and $1D$, are expected to be narrow and decay predominantly via electromagnetic or weak interactions, while higher excitations above threshold will decay strongly into heavy-light meson pairs. The logarithmic term in Potential~II provides additional confinement at intermediate distances, shifting excited levels downward and clustering them more tightly relative to Potential~I; this results in a more compact arrangement of the higher states with respect to the open-flavor thresholds. Potential~III produces a more pronounced compression of the higher spectrum, with the screening-induced downward shift becoming increasingly significant for larger $n$ and $l$. Consequently, the locations of highly excited states relative to the nearby open-flavor thresholds become more sensitive to the choice of confinement model. The growth of mass uncertainties above threshold, most pronounced for high-$n$, high-$l$ states, confirms that these levels probe the intermediate- and long-range potential more deeply, where the confinement parameters are less constrained by existing data. The corresponding spectra obtained with the alternative Gaussian priors, presented in Appendix~\ref{app:2_MCMC}, show the same qualitative level ordering and threshold pattern, indicating that the principal spectroscopic conclusions are robust against the prior specification.

For states above the lowest open-flavor threshold, the quoted MCMC uncertainties are conditional on the adopted single-channel potential and do not include coupled-channel effects, which can be sizeable and are strongly state- and model-dependent. A coupled-channel calculation~\cite{Hao:2024nqb} finds mass shifts of $-60$ to $-64$~MeV ($1S$), $-91$ to $-99$~MeV ($1P$), $-108$ to $-110$~MeV ($2S$), $-115$ to $-117$~MeV ($1D$), $-122$ to $-138$~MeV ($2P$), and $-123$ to $-170$~MeV ($2D$). An earlier calculation with a different Hamiltonian and channel set~\cite{Monteiro:2016rzi} gives smaller $S$-wave shifts ($-27$ to $-30$~MeV for $1S$, $-7$ to $-8$~MeV for $2S$) but much larger $P$/$D$-wave shifts ($-76$ to $-211$~MeV for $1P$, $-148$ to $-540$~MeV for $2P$, $-61$ to $-231$~MeV for $1D$). A screened-potential treatment~\cite{Li:2023wgq}, which mimics unquenching via a modified Godfrey-Isgur model rather than explicit channel coupling, instead gives shifts that grow gradually with $n$: $-30$ to $-32$~MeV ($3S$), $-76$~MeV ($4S$), $-125$ to $-132$~MeV ($5S$). Given this strong state- and model-dependence, we do not interpret these shifts as a universal uncertainty to be added to the MCMC credible intervals, which quantify only the parameter-induced uncertainty of the single-channel potentials adopted here. Instead, we regard the spread among these descriptions as an additional model-dependent theoretical systematic for the above-threshold spectrum, to be considered when comparing with future data.

The median energy eigenvalues $E_{nl}$ and individual spin-dependent contributions ($\langle V_{SS}\rangle$, $\langle \lsp\rangle$, $\langle \lsm\rangle$, and $\langle V_T\rangle$), with their $1\sigma$ credible intervals, are presented in Tables~\ref{tab:Bcmass_S_contri}-\ref{tab:Bcmass_D_contri}. The eigenvalues increase monotonically with $n$ in all wave sectors, while the low-lying states are nearly identical across Potentials~I-III. At high excitation, Potential~I exceeds Potential~II by approximately $70$, $80$, and $90$~MeV for the $6S$, $6P$, and $6D$ states, respectively, whereas Potential~III lies approximately $187$, $223$, and $258$~MeV below Potential~II and $258$, $305$, and $349$~MeV below Potential~I. Thus, the growing inter-potential mass divergence and the stronger high-$n$ spectral compression of Potential~III are driven predominantly by the spin-independent eigenvalue $E_{nl}$ rather than by the spin-dependent corrections.

For $S$-wave states (Table~\ref{tab:Bcmass_S_contri}), the only spin-dependent contribution is $\langle V_{SS}\rangle$, whose magnitude decreases from approximately $41.78$ and $13.93$~MeV for the singlet and triplet $1S$ states to $11.91$ and $3.97$~MeV at $6S$ for Potential~I. Agreement among the potentials is better than $0.15\%$ at $1S$ but reaches $\sim15\%$ at $6S$. Potential~III exhibits the strongest suppression, with $6S$ contributions of $-6.29$~MeV (singlet) and $2.10$~MeV (triplet), compared with $-11.91$ and $3.97$~MeV for Potential~I, consistent with its smaller high-$n$ hyperfine splittings.

The $P$-wave states (Table~\ref{tab:Bcmass_P_contri}) receive contributions from all four spin-dependent terms. The symmetric spin-orbit contribution $\langle \lsp\rangle$ dominates the fine structure and remains nearly $n$-independent, varying by only $\sim4.7\%$ from $1P$ to $6P$, owing to partial cancellation between the decreasing OGE and compensating confinement contributions to the spin-orbit radial integral. The antisymmetric contribution $\langle\lsm\rangle$ increases from $0.44$ to $2.28$~MeV for Potential~I, driving the increase of $\theta_{nP}$, while $\langle V_T\rangle$ decreases by $\sim21\%$ from $1P$ to $6P$ while retaining the expected angular-momentum structure across the $^3P_J$ multiplet. The $\lspm$ contributions show larger inter-potential differences than $\langle V_{SS}\rangle$ and $\langle V_T\rangle$, reflecting their sensitivity to both scalar and vector components of the potential and hence to the logarithmic correction in Potential~II. Potential~II gives larger $\lspm$ contributions at low $n$ but approaches Potential~I at high $n$. Potential~III exhibits a distinct radial behavior, with $\lsp$ decreasing from $\sim16.5$~MeV for $1^3P_2$ to $\sim9.8$~MeV for $6^3P_2$, while $\lsm$ remains approximately $1.6-1.9$~MeV. Nevertheless, $\theta_{nP}$ increases from $12.85^\circ$ at $1P$ to $20.58^\circ$ at $6P$, indicating that the different radial behavior of $\lsm$ in Potential~III leads to a distinct $n$-dependence of the $P_1$-$P_1'$ mixing angle.

For $D$-wave states (Table~\ref{tab:Bcmass_D_contri}), $\langle V_{SS}\rangle$ is negligible ($\lesssim0.1$~MeV), while $\langle \lsp\rangle$ and $\langle V_T\rangle$ are reduced relative to their $P$-wave counterparts. The dominant spin-dependent contribution is $\langle\lsm\rangle$, which remains negative and decreases in magnitude from $-5.16$~MeV at $1D$ to $-1.72$~MeV at $6D$ for Potential~I, opposite to the $n$-dependence found for the $P$ waves. This behavior drives the monotonic shift of $\theta_{nD}$ away from $-54.7^\circ$. The dominance of scalar confinement over OGE in the symmetric spin-orbit interaction at low $n$ also produces a sign change in $\lsp(^3D_1)$ between $1D$ ($+3.24$~MeV) and $3D$ ($-1.43$~MeV), whose zero crossing leads to the non-standard level ordering. The $\lspm$ contributions are generally larger for Potential~I than for Potential~II, reflecting the suppression induced by the logarithmic term. Potential~III exhibits a weaker antisymmetric spin-orbit contribution, with $\langle\lsm\rangle$ decreasing from $-3.74$~MeV at $1D$ to $-0.67$~MeV at $6D$, while $\theta_{nD}$ shifts from $-51.22^\circ$ to $-44.35^\circ$. Its $\lsp(^3D_1)$ contribution is already negative at $1D$ ($-0.41$~MeV) and remains negative, so the zero crossing present for Potentials~I and II is absent. Consequently, the non-standard $D$-wave ordering persists only through $2D$ for Potential~III. The uncertainties of the individual spin-dependent matrix elements are comparable across the potentials, while at high $n$ the dominant contribution to the mass uncertainty comes from the eigenvalue $E_{nl}$ rather than the spin-dependent corrections.

\subsection{Wave Function at the Origin and Related Quantities}
\label{sec:3B_WFO}

We provide median values with $1\sigma$ credible intervals for $|R_{nl}^{(l)}(0)|^2$, RMS radii $\sqrt{\langle r^2\rangle}$, kinetic energy $\langle T\rangle$, squared momentum $\langle p^2\rangle$, and heavy-quark velocities from the potentials in Table~\ref{tab:Bc_otherobs}. The squared radial wave function at the origin encodes the short-distance quark-antiquark overlap and governs leptonic decays and production cross sections. For $S$-waves, $|R_{nS}(0)|^2$ decreases monotonically from $\approx 2.16~\text{GeV}^3$ ($1S$) to $0.89~\text{GeV}^3$ ($6S$) under Potential~I, a $\sim 59\%$ suppression for higher states, the increasing number of nodes and larger spatial extent reduce the relative uncertainty. The relative uncertainty is largest at $1S$ ($^{+18.5\%}_{-8.1\%}$, due to the combined $\alpha_s$ and $\sigma$ sensitivity) and becomes small and symmetric by $6S$, as the wavefunction normalization becomes less sensitive to the short-range $\alpha_s$. Potential~III follows the same trend but with stronger suppression, from $2.57$ to $0.639~\mathrm{GeV}^3$, corresponding to $\sim75\%$ reduction. It therefore gives the largest $1S$ value but increasingly smaller values at higher excitation.

For $P$- and $D$-waves, $|R_{nl}^{(l)}(0)|^2$ instead increases with $n$, from $\approx 0.250~\text{GeV}^5$ to $0.578~\text{GeV}^5$ for $P$-waves and from $\approx 0.066~\text{GeV}^7$ to $0.482~\text{GeV}^7$ for $D$-waves under Potential~I. This inversion arises from the centrifugal barrier, which develops a secondary probability lobe at short range in higher-$n$ states with $l > 0$, enhancing the derivative of the wavefunction near the origin. At fixed $n$, the ordering $|R_{nS}(0)|^2 > |R_{nP}^{(1)}(0)|^2 > |R_{nD}^{(2)}(0)|^2$ holds, with $|R_{1S}(0)|^2 > |R_{1P}^{(1)}(0)|^2$ and $|R_{1P}^{(1)}(0)|^2 > |R_{1D}^{(2)}(0)|^2$ by a factor of $\sim8$ and $\sim3.5$, respectively. This trend reflects the increasing centrifugal barrier, which progressively suppresses the wavefunction and its derivatives at the origin for higher $l$. Potential~III shows weaker growth, with $P$-wave values increasing from $0.261$ to $0.343~\mathrm{GeV}^5$ and $D$-wave values from $0.058$ to $0.222~\mathrm{GeV}^7$ between the lowest and highest states. Thus, the screening primarily suppresses the short-distance overlap of highly excited states.

The RMS radii $\sqrt{\langle r^2\rangle}$ grow systematically with both $n$ and $l$, from $1.71~\text{GeV}^{-1}$ ($1S$) to $8.47~\text{GeV}^{-1}$ ($6S$) under Potential~I. The inter-potential discrepancy in $\sqrt{\langle r^2\rangle}$ grows from $\sim 0.9\%$ at $1S$ to $\sim 4\%$ at $6D$, with Potential~II consistently predicting more spatially extended states which are consistent with its softer long-range confining slope. The MCMC uncertainties on $\sqrt{\langle r^2\rangle}$ gradually grow with radial and orbital excitation under both potentials, identifying confinement dynamics as the fundamental limit on predictive precision for excited states. Potential~III gives the smallest $1S$ radius, $1.66~\mathrm{GeV}^{-1}$, but becomes increasingly more extended at high excitation, reaching $10.01$, $10.90$, and $11.75~\mathrm{GeV}^{-1}$ for $6S$, $6P$, and $6D$, respectively. Thus, the screening produces a crossover from a more compact ground state to more extended highly excited states, accompanied by increasing uncertainties with excitation.

The kinetic energy $\langle T\rangle$ increases with excitation under Potentials~I and II, from $0.37$ and $0.38~\text{GeV}$ at $1S$ to $0.79$ and $0.73~\text{GeV}$, respectively, at $6S$. This trend indicates the increasing importance of the linear confinement term relative to the Coulomb component at larger radii. Its dependence on orbital excitation also exhibits a crossover. At $n=1$, the $P$-wave has a smaller kinetic energy than the corresponding $S$-wave, as the centrifugal barrier shifts the wave function towards larger radii. By $n=3$, the ordering reverses to $\langle T\rangle_{3S}<\langle T\rangle_{3P}<\langle T\rangle_{3D}$, with values $0.524$, $0.562$, and $0.608~\text{GeV}$, respectively, reflecting the increasing influence of the confining interaction on higher-$l$ states. In contrast, Potential~III gives $\langle T\rangle=0.402~\text{GeV}$ at $1S$ but only $0.551~\text{GeV}$ at $6S$, with the corresponding values at $6P$ and $6D$ being $0.554$ and $0.559~\text{GeV}$. Thus, screening produces a pronounced reduction of the kinetic scale for highly excited states, consistent with their increasingly extended wave functions. The MCMC uncertainties on $\langle T\rangle$ also show an inversion of asymmetry between $1S$ and $2S$, reflecting the competing sensitivity to the short- and long-distance potential
parameters.

The heavy-quark squared velocities $\langle v_Q^2\rangle$ provide an internal diagnostic of the applicability of the non-relativistic approximation. For the ground state, $\langle v_b^2\rangle \approx 0.037$ and $\langle v_c^2\rangle \approx 0.378$, indicating a strongly non-relativistic $b$-quark but appreciable relativistic effects for the $c$-quark. The ratio $\langle v_b^2\rangle/\langle v_c^2\rangle \approx m_c^2/m_b^2$ follows the expected mass-squared scaling, since $\langle p^2\rangle$ is a state-level property shared between the two quarks. With increasing excitation, $\langle v_c^2\rangle$ grows steadily, reaching $\sim 0.88$ for the $6D$ state in Potential~I, whereas $\langle v_b^2\rangle$ remains below $\sim 0.09$ throughout the spectrum. Thus, the increasing relativistic character of the excited spectrum is driven predominantly by the charm quark, while the bottom quark remains comparatively non-relativistic. Taking $\langle v_c^2\rangle\sim0.5$ as a diagnostic benchmark, the median $\langle v_c^2\rangle$ exceeds this threshold from approximately the $3S$, $3P$, and $3D$ states onward in Potentials~I and II, indicating the progressively increasing importance of relativistic effects for higher excitations. The spin-spin, spin-orbit, and tensor contributions considered here are the leading spin-dependent terms within the same $1/m_Q^2$ expansion. Although their absolute magnitudes generally decrease for highly excited states, the increasing value of $\langle v_c^2\rangle$ implies that higher-order relativistic contributions may become relatively more relevant to the detailed spin-dependent structure; conversely, the decreasing size of the leading splittings may limit their absolute impact. The present decomposition therefore provides a systematic leading-order description whose reliability is expected to be strongest for the low-lying spectrum and progressively more sensitive to higher-order corrections for highly excited states. For the screened Potential~III, the relativistic diagnostic shifts toward smaller values at higher excitation, with $\langle v_c^2\rangle$ reaching $\sim0.56$ for the $6D$ state and the $\langle v_c^2\rangle\sim0.5$ benchmark being exceeded from the $4S$, $4P$, and $4D$ levels onward. This behavior follows from the larger spatial extent and lower momentum scale of the highly excited states generated by the screened interaction, illustrating the sensitivity of the relativistic diagnostic to the long-distance form of the potential. Nevertheless, $\langle v_c^2\rangle\sim0.56$ at $6D$ shows that screening does not eliminate the growing relativistic content of the upper spectrum. Our results for the higher excitations should therefore be regarded as indicative rather than precision benchmarks, particularly once $\langle v_c^2\rangle$ becomes large. The $6D$ inter-potential separation of $\sim90$~MeV is comparable to the quoted credible intervals, limiting the present discriminating power to at most $\sim1\sigma$. The higher-$n$ spectrum is therefore most useful as a guide for future experimental studies, whose benchmarking value will ultimately depend on the feasibility of producing and resolving these states.

We emphasize that the posterior for the potential constant $V_c$ is centered at $-182$~MeV with asymmetric uncertainties ($-50$~MeV, $+99$~MeV) for Potential~I (Table~\ref{tab:Bc_parameters}). The observed asymmetry reflects nonlinear correlations among $V_c$, $\alpha_s$, and $\sigma$ in the Bayesian inference. Within the present non-relativistic parametrization, $V_c$ is a phenomenological additive constant and should not be assigned a unique dynamical interpretation. Its fitted value can depend on the chosen potential parametrization and on correlations with the other model parameters. Relativistic heavy-quarkonium studies have also shown that relativistic corrections can depend on the Lorentz structure assigned to the potential; McClary and Byers~\cite{McClary:1983xw}, for example, investigated such corrections and their implications for heavy-quarkonium spectroscopy. This, however, does not establish a direct connection between the negative posterior value found here and relativistic corrections. Since our Hamiltonian does not include explicit relativistic corrections, the origin of the fitted $V_c$ cannot be disentangled from the phenomenological parametrization and parameter correlations within the present analysis. A consistent relativistic treatment would therefore be required to determine whether and to what extent the fitted potential parameters, including $V_c$, are modified by such corrections.

Comparing Potentials I and II, the differences in $|R_{nl}^{(l)}(0)|^2$ change trend with $n$. Potential~II predicts $\sim 5\%$ larger values at $1S$ but progressively smaller values at higher excitations, reaching $\sim -15\%$ at $6D$. Simultaneously, Potential~II yields uniformly larger RMS radii except for $1S$, indicating a softer short-range core and a more extended long-range tail relative to Potential~I. These differences can affect leptonic widths $\Gamma(B_c\to\ell\nu)\propto |R_{1S}(0)|^2$, $E1$ radiative transitions sensitive to the spatial extent of the states, and $gg$-fusion production rates $\propto |R_{nS}(0)|^2$, for which the two potentials increasingly diverge at higher $n$. Potential~III strengthens this contrast, giving the largest $1S$ overlap, $|R_{1S}(0)|^2=2.57~\text{GeV}^3$, about $13\%$ above Potential~II, but progressively smaller values at higher excitation, reaching approximately $24\%$, $35\%$, and $46\%$ below Potential~II for $6S$, $6P$, and $6D$, respectively. At the same time, its RMS radii exceed those of Potential~II by about $14\%$, $16\%$, and $18\%$ for $6S$, $6P$, and $6D$. Thus, the stronger screening enhances the model dependence of short-distance observables and transition matrix elements in the highly excited spectrum.

\begin{figure}
     \centering
     \subfigure[Potential I, $S$-wave.]{\includegraphics[width=.325\textwidth]{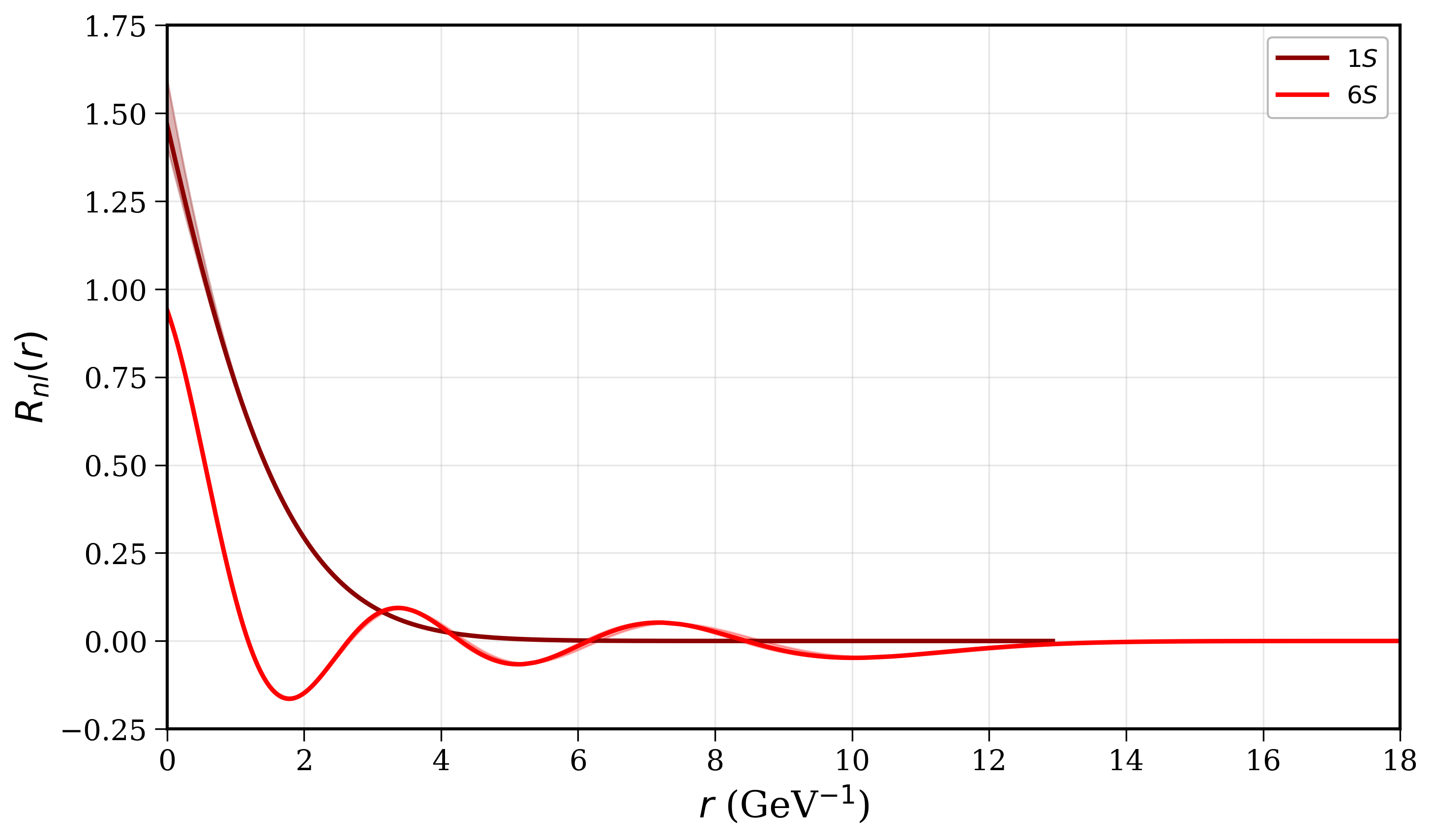}\label{fig:Cornell_WF_S}}
     \hfill
     \subfigure[Potential II, $S$-wave.]{\includegraphics[width=.325\textwidth]{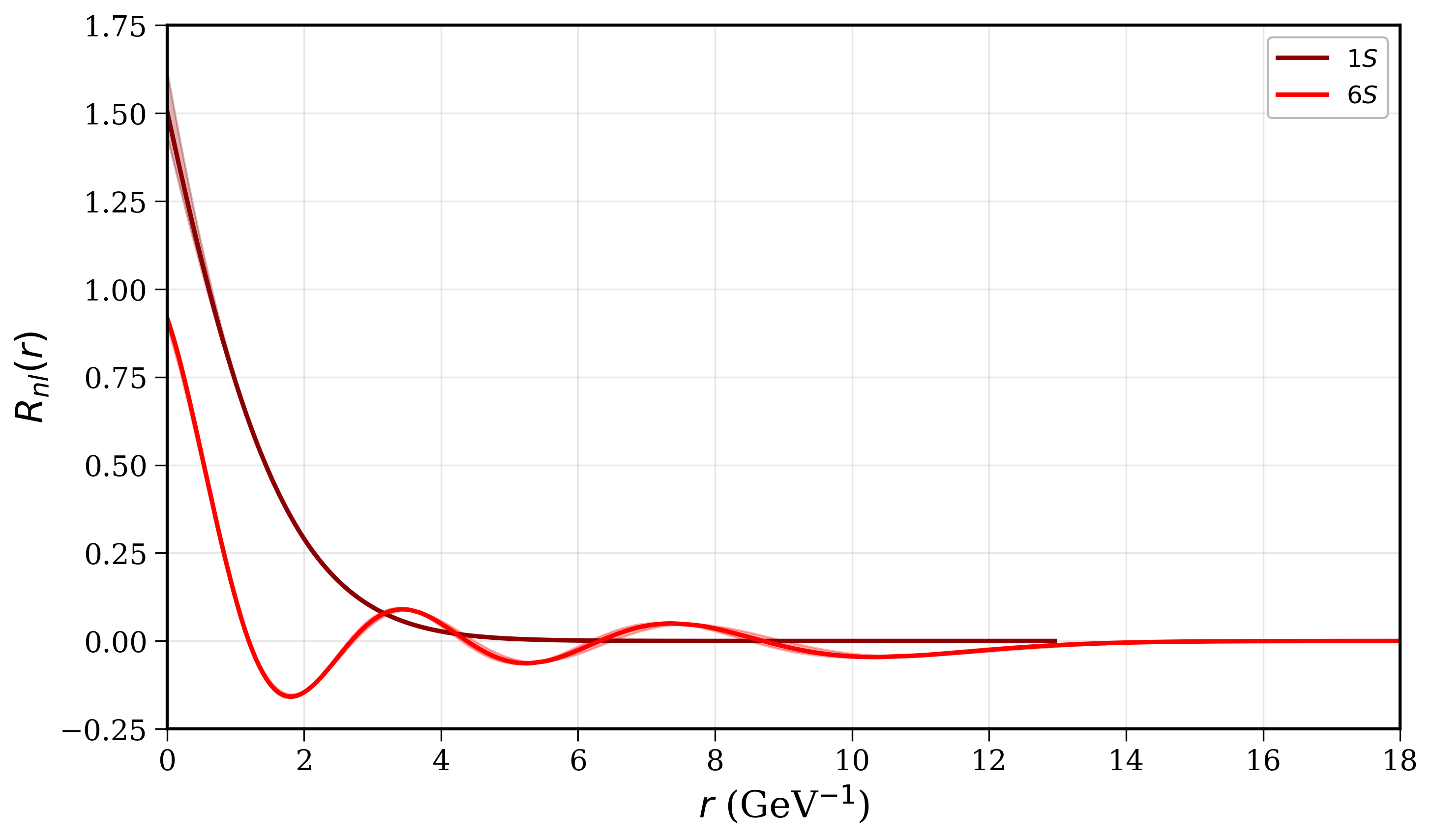}\label{fig:Modcornell_WF_S}}
     \hfill
     \subfigure[Potential III, $S$-wave.]{\includegraphics[width=.325\textwidth]{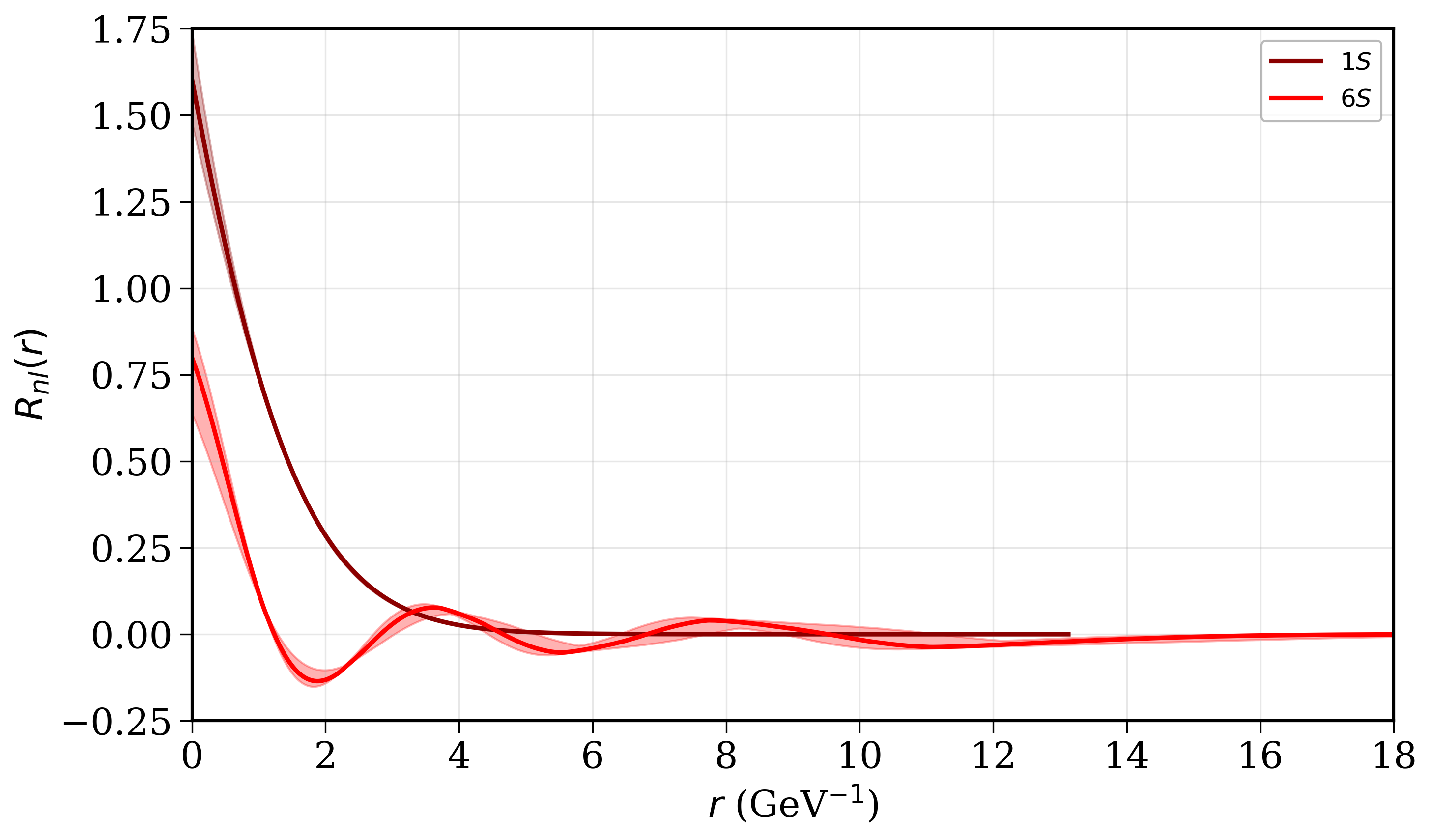}\label{fig:Screencornell_WF_S}}
     \vspace{2mm}
     \subfigure[Potential I, $P$-wave.]{\includegraphics[width=.325\textwidth]{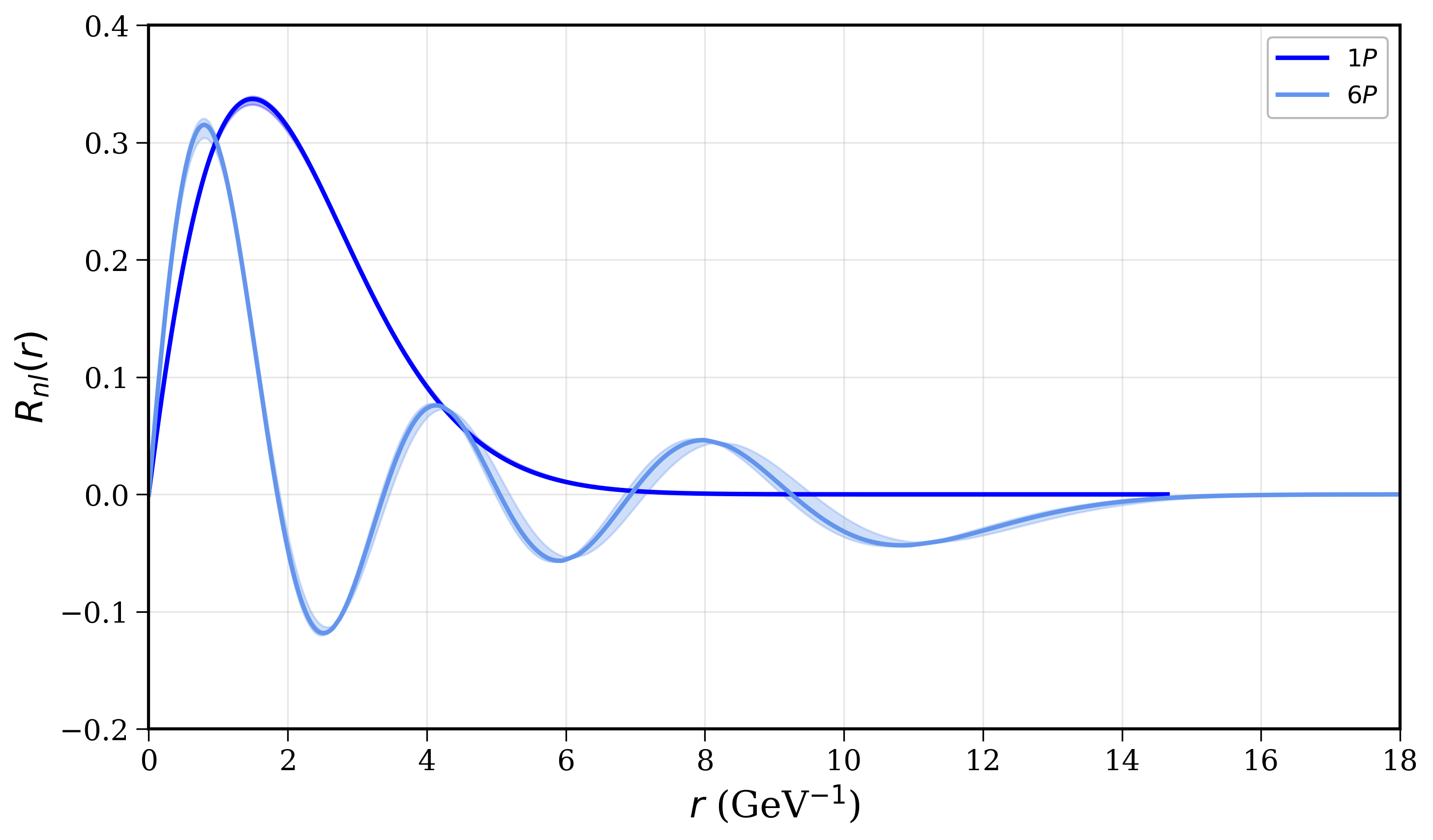}\label{fig:Cornell_WF_P}}
     \hfill
     \subfigure[Potential II, $P$-wave.]{\includegraphics[width=.325\textwidth]{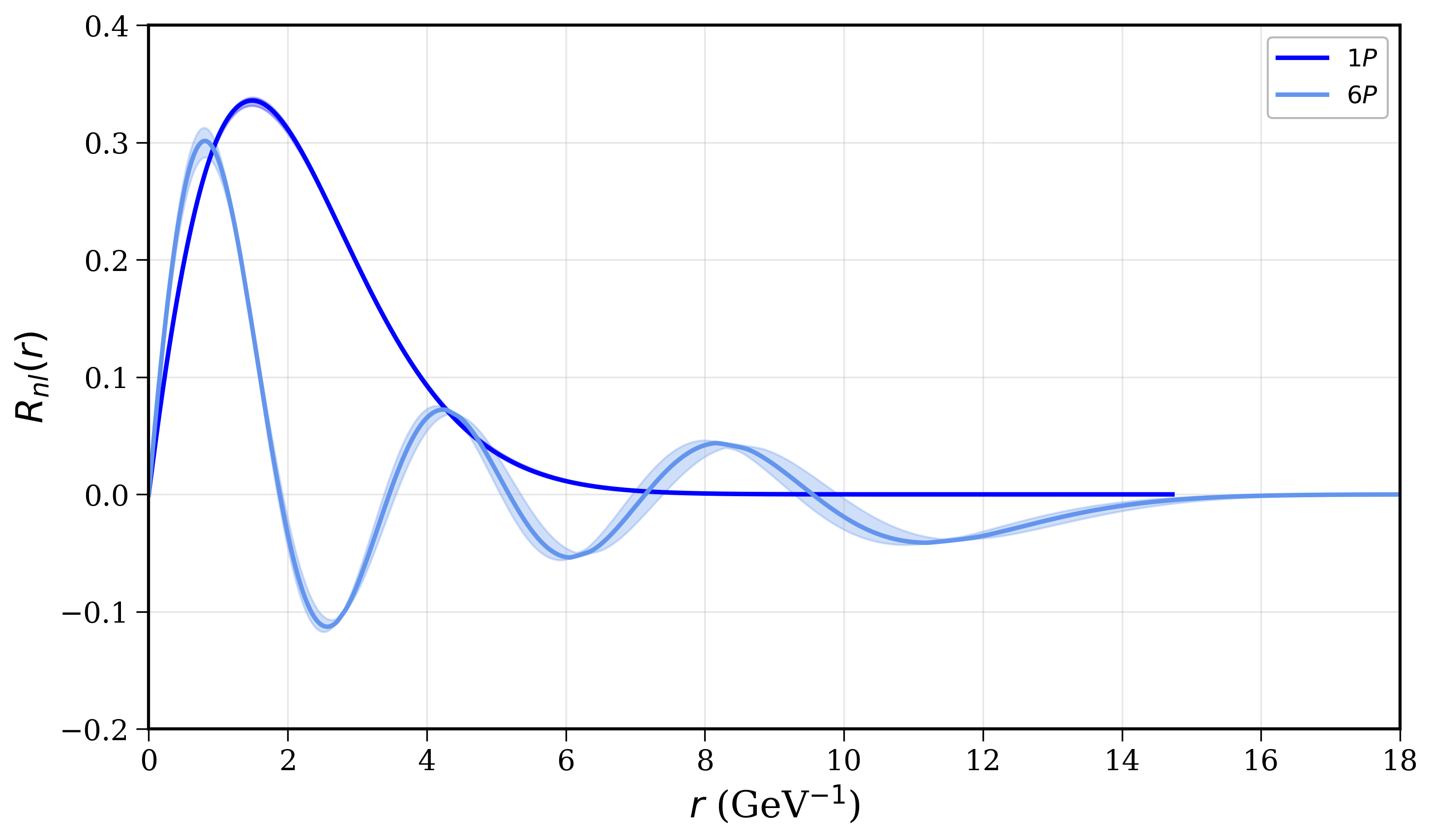}\label{fig:Modcornell_WF_P}}
     \hfill
     \subfigure[Potential III, $P$-wave.]{\includegraphics[width=.325\textwidth]{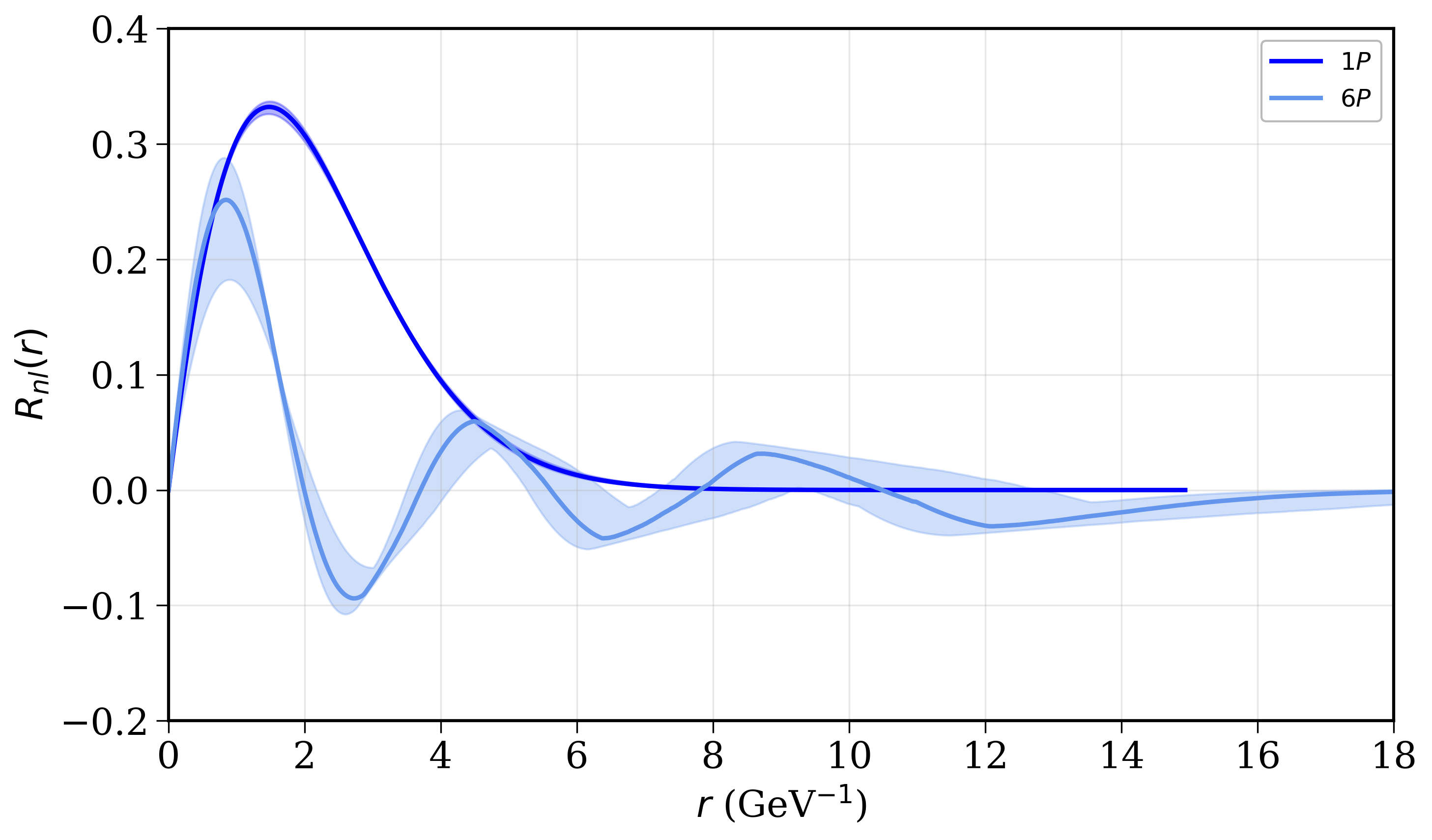}\label{fig:Screencornell_WF_P}}
     \vspace{2mm}
     \subfigure[Potential I, $D$-wave.]{\includegraphics[width=.325\textwidth]{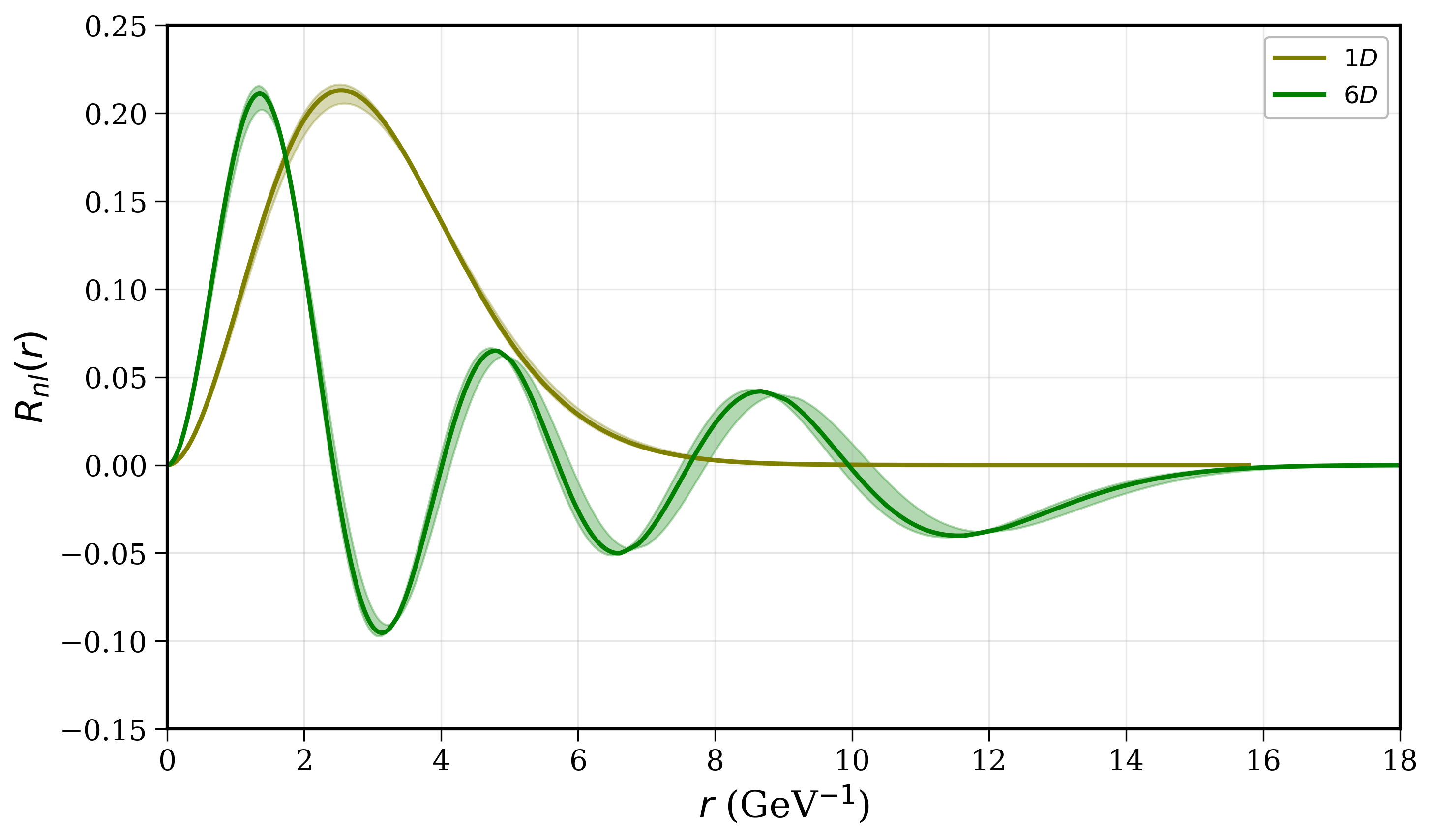}\label{fig:Cornell_WF_D}}
     \hfill
     \subfigure[Potential II, $D$-wave.]{\includegraphics[width=.325\textwidth]{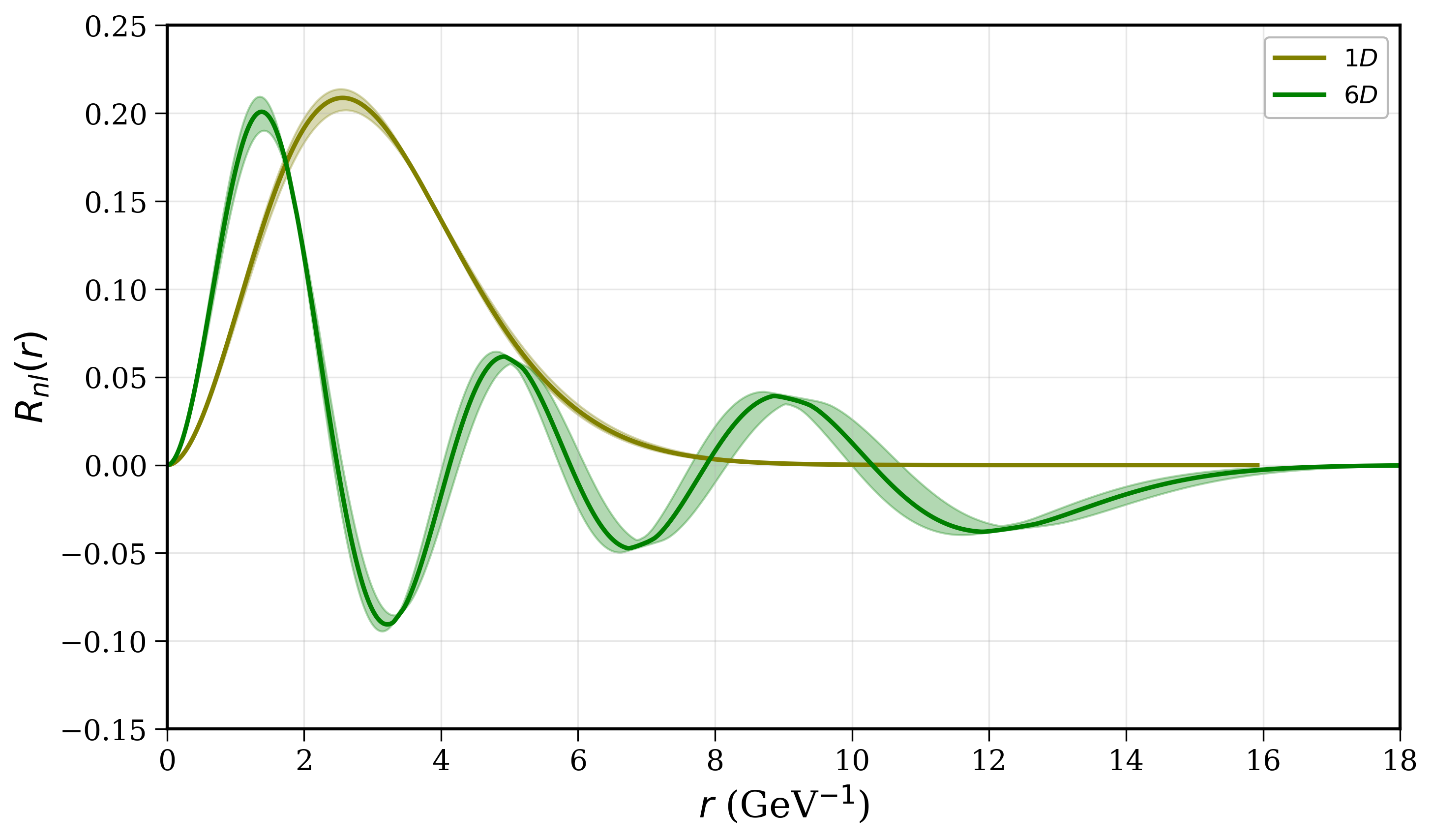}\label{fig:Modcornell_WF_D}}
     \hfill
     \subfigure[Potential III, $D$-wave.]{\includegraphics[width=.325\textwidth]{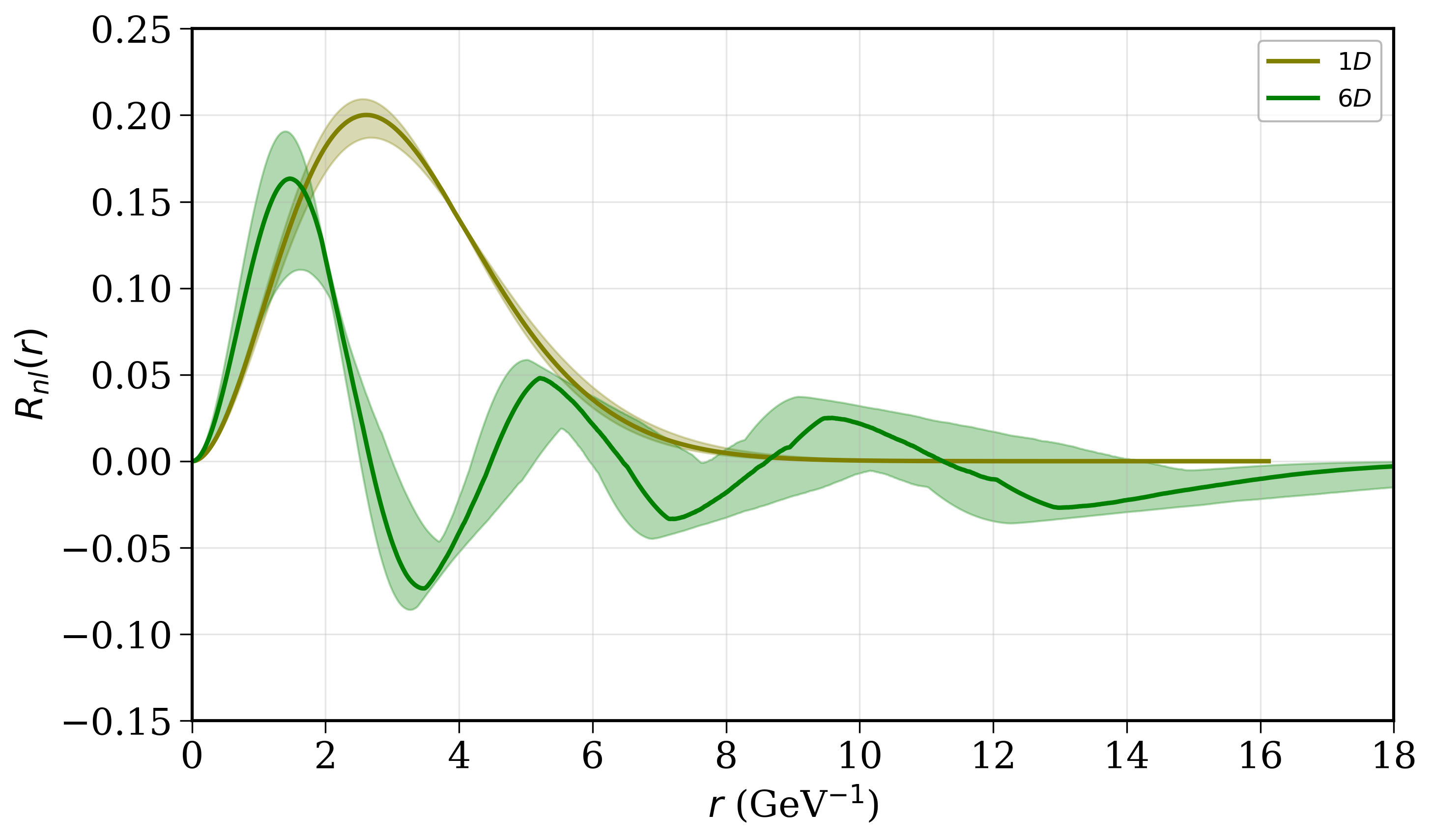}\label{fig:Screencornell_WF_D}}
     
        \caption{Radial wave functions of the $B_c$ system obtained using Cornell (Potential I, Eq.~\eqref{eqn:1_cornell}, left), Modified Cornell (Potential II, Eq.~\eqref{eqn:corplusln}, middle), and Screened Cornell (Potential III, Eq.~\eqref{eqn:screenpot}, right) potentials. Shaded bands represent the propagated uncertainties from the MCMC analysis ($\sim 5000$ samples).}
        \label{fig:Bc_WF}
\end{figure}

The radial wave functions for representative states, constructed from $\sim5000$ posterior samples, are shown in Fig.~\ref{fig:Bc_WF}. Potentials I and II give nearly identical ground-state wave functions, while at higher excitation Potential~II shows slightly outward-shifted nodes and more extended tails, consistent with its softer long-range confinement and the corresponding reduction in average momentum and kinetic energy. The credible bands broaden with orbital excitation and are largest for the $6D$ state, where a small inter-potential shift in node positions is discernible. Potential~III produces a more pronounced modification of the excited-state wave functions, with the $6P$ and $6D$ states showing longer tails, broader credible bands, and visible changes in nodal structure relative to Potentials I and II. The near-coincidence of the low-lying wave functions is retained, while the enhanced spatial spreading at high excitation is consistent with the larger RMS radii in Table~\ref{tab:Bc_otherobs}.

\subsection{Effects of Modification in the Cornell Potential}
\label{sec:3C_ModPot}

\begin{figure}
     \centering
     \subfigure[$V(r)$ vs $r$]{\includegraphics[width=.473\textwidth]{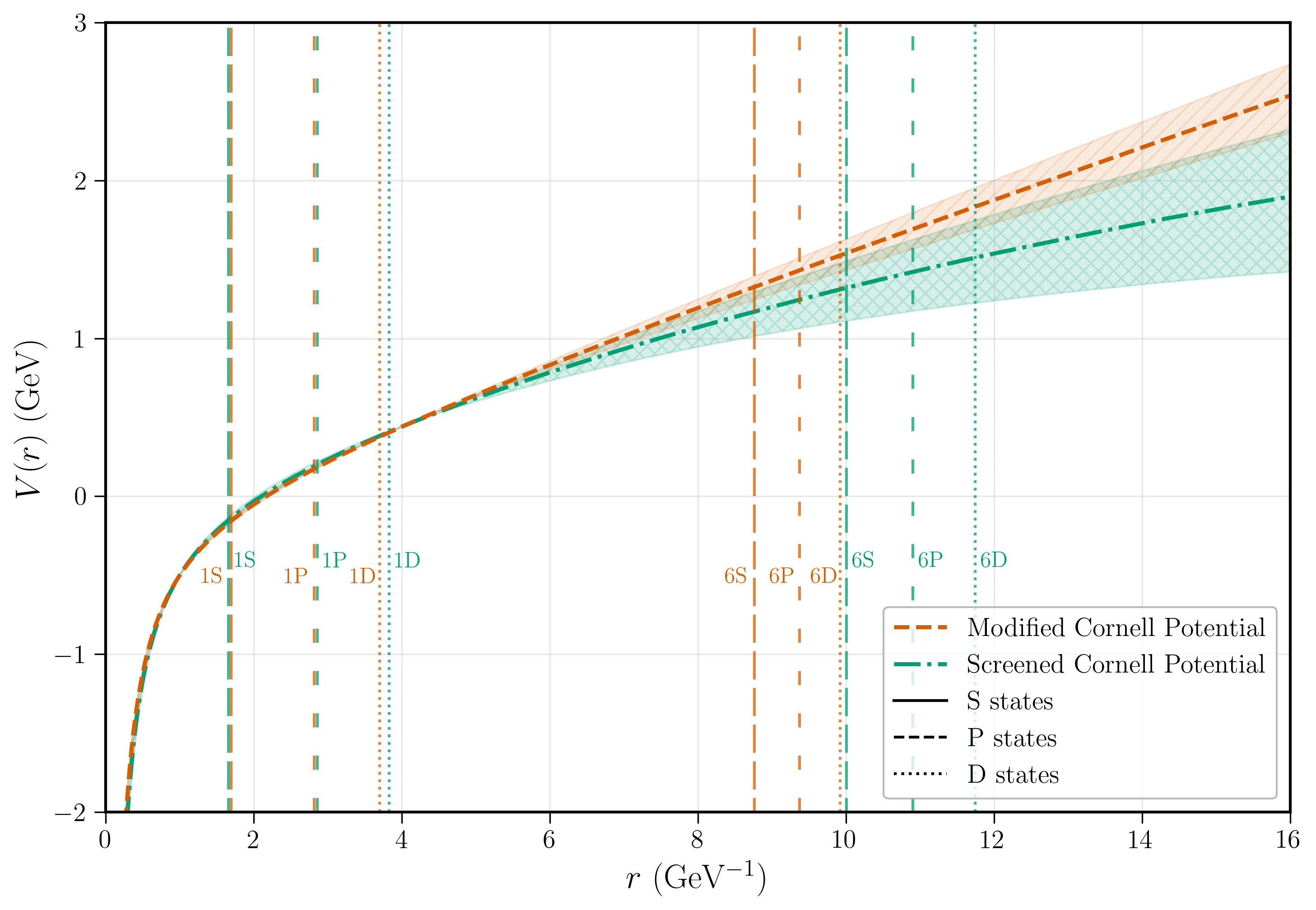}\label{fig:comparison_V_r}}
     \hfill
     \subfigure[$V'(r)$ vs $r$]{\includegraphics[width=.473\textwidth]{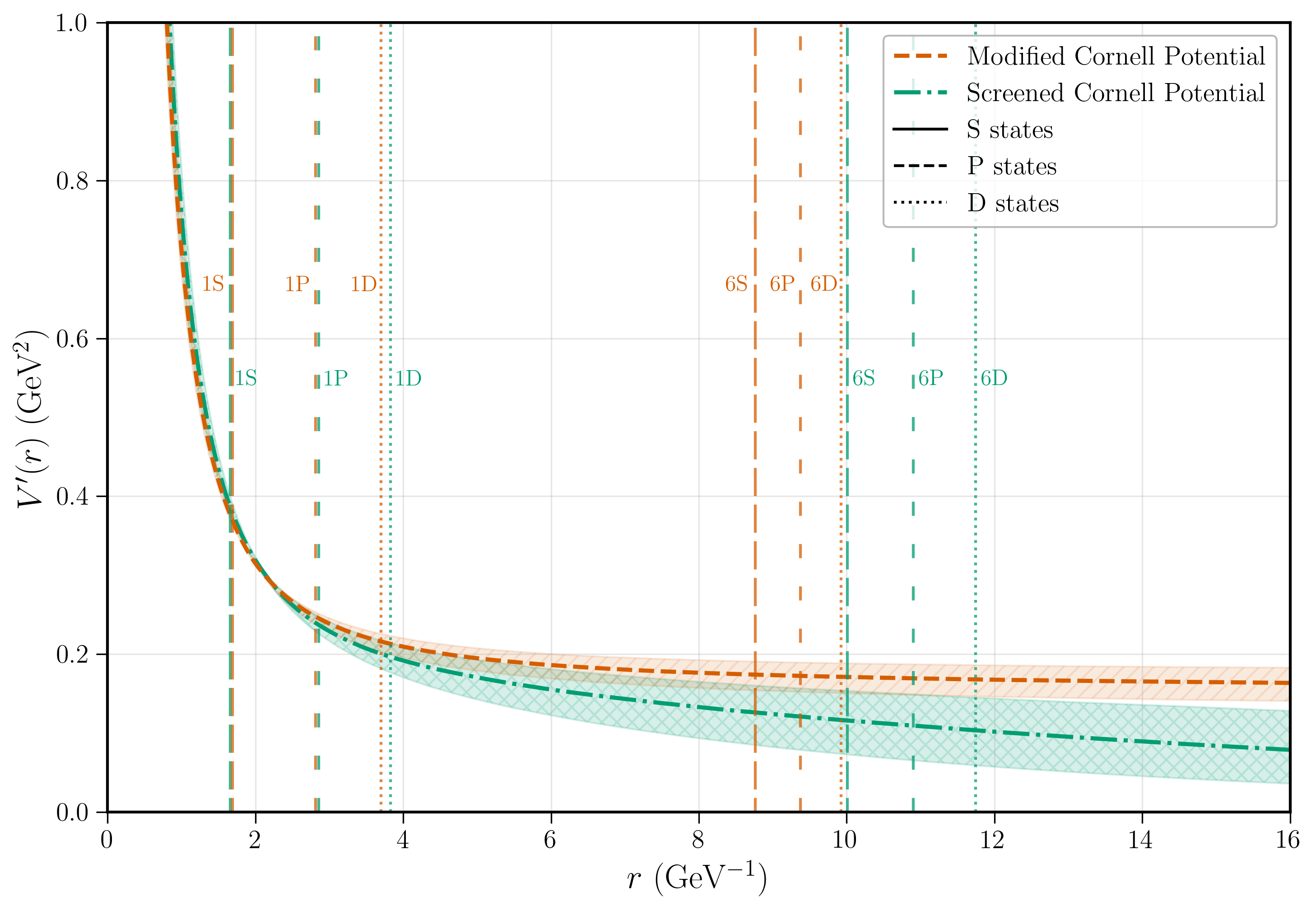}\label{fig:comparison_Vprime_r}}
     \subfigure[$V''(r)$ vs $r$]{\includegraphics[width=.473\textwidth]{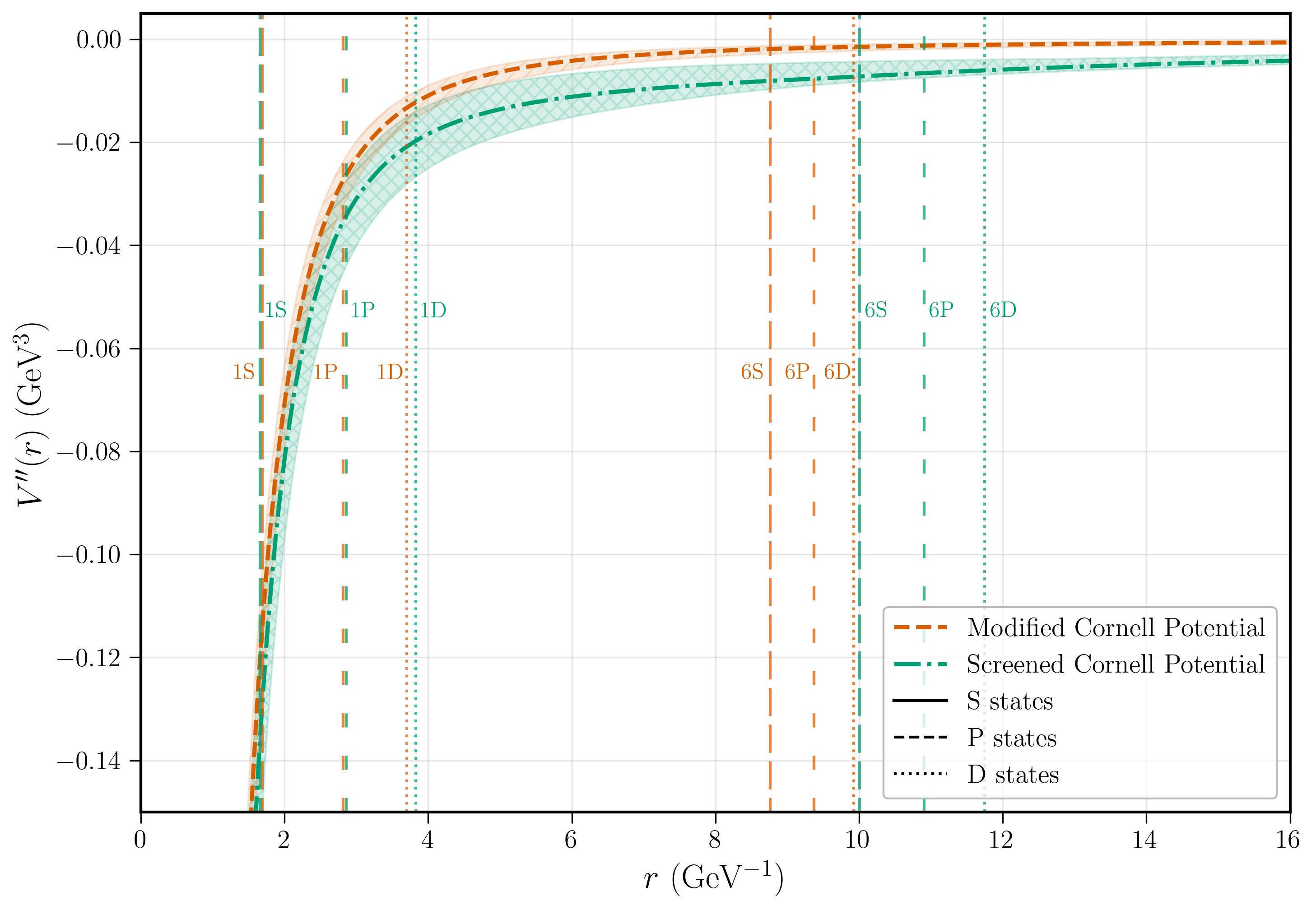}\label{fig:comparison_Vdoubleprime_r}}
     \hfill
     \subfigure[$\sigma_{\text{eff}}(r)$ vs $r$]{\includegraphics[width=.473\textwidth]{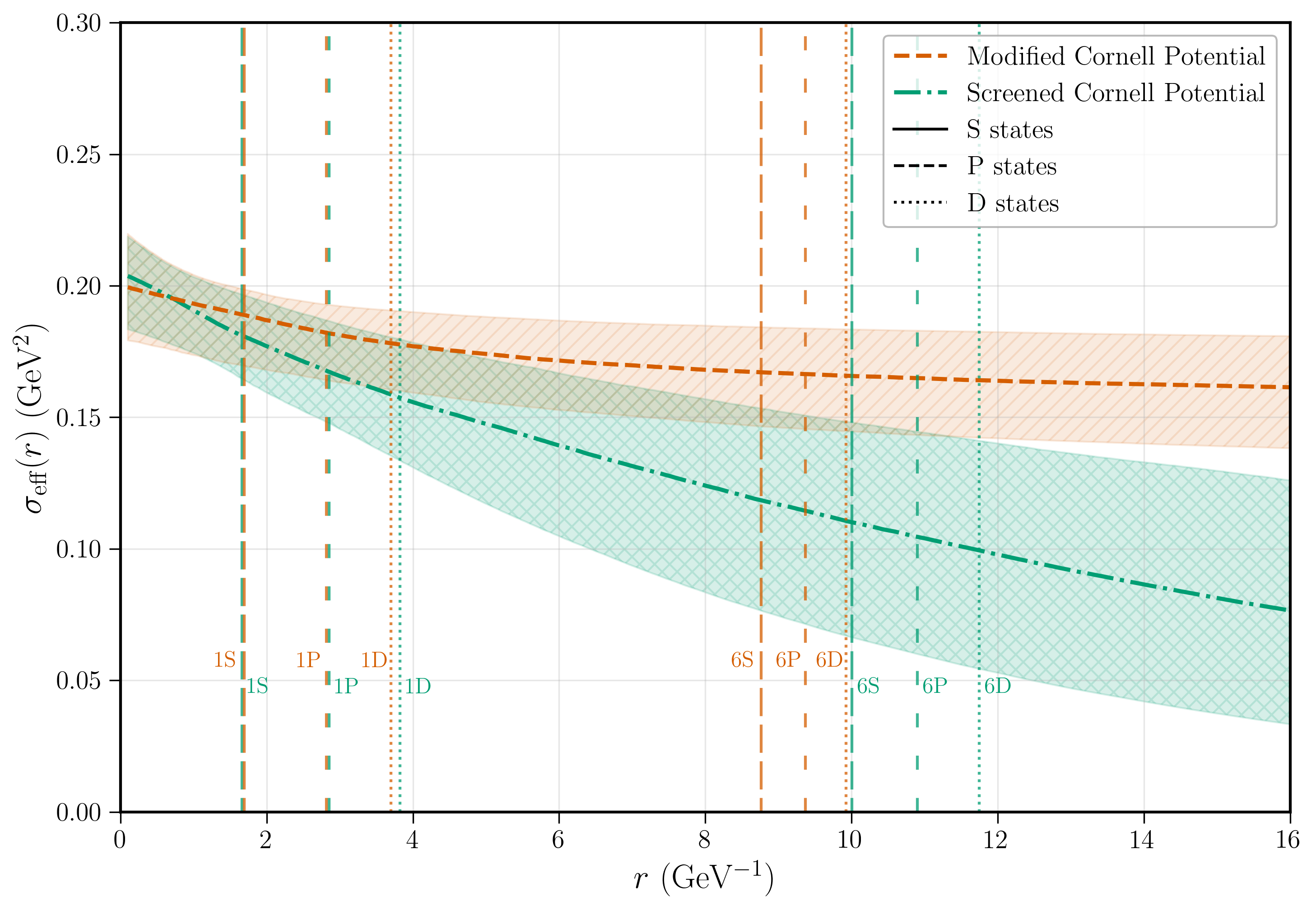}\label{fig:comparison_sigmaeff_r}}
     
        \caption{Radial dependence of the modified Cornell potential $V_{\rm II}(r)$ (Eq.~\eqref{eqn:corplusln}) and screened Cornell potential $V_{\rm III}(r)$ (Eq.~\eqref{eqn:screenpot}), their first and second derivatives, and the effective string tension $\sigma_{\text{eff}}(r)$. The vertical dashed lines indicate the RMS radii of select $B_c$ states, illustrating the regions of the potential probed by ground and excited states. Shaded bands represent the propagated uncertainties from the MCMC analysis ($\sim 5000$ samples).}
        \label{fig:CombinedCornellRMSr}
\end{figure}

Figure~\ref{fig:CombinedCornellRMSr} shows the radial dependence of the modified Cornell potential $V_{\rm II}(r)$ and the screened Cornell potential $V_{\rm III}(r)$, together with their first and second derivatives and the corresponding effective string tension $\sigma_{\rm eff}(r)$. The RMS radii of representative $B_c$ states are superimposed as vertical markers, indicating the radial regions sampled by different spectroscopic levels and providing a direct connection between the long-distance behavior of the potentials and the calculated spectrum. At short distances, $V_{\rm II}(r)$ remains essentially indistinguishable from the Cornell potential, since the logarithmic correction $C_0\ln(1+\sigma' r)$ is small compared with the Coulomb term. Its modification becomes increasingly visible at larger $r$, where the derivative $V'_{\rm II}(r)$ approaches a smaller confining slope than the linear Cornell form. The corresponding curvature $V''_{\rm II}(r)$ remains negative and approaches zero, indicating a gradual softening of the confining interaction. This behavior is quantified by the effective string tension (Eq.~\eqref{eqn:sigmaeffective})
\begin{equation*}
\sigma_{\rm eff}(r)=\sigma+\frac{C_0\sigma'}{1+\sigma'r},
\end{equation*}
which decreases with $r$ toward the asymptotic value $\sigma$. The MCMC credible bands of $\sigma_{\rm eff}(r)$ are relatively narrow at small $r$ and broaden toward larger distances, reflecting the stronger constraints from the low-lying spectrum and the reduced sensitivity of the fitted parameters to the long-distance interaction. Potential~III exhibits a substantially stronger modification at large $r$. Its derivative decreases continuously toward zero, while $V''_{\rm III}(r)$ remains negative and approaches zero, indicating progressive suppression of the confining force. The effective string tension likewise decreases with $r$, in contrast to the finite asymptotic confining slope retained by Potential~II. The difference between the two modified potentials is small in the region sampled by the low-lying states but becomes pronounced for the highly excited states, whose larger RMS radii extend into the region where the potentials separate most strongly.

The RMS-radius markers provide a direct geometric connection to the spectroscopy. The $1S$, $1P$, and $1D$ states probe relatively short distances, where Potentials~I and II remain close and consequently give similar eigenvalues, wavefunction overlaps $|R_{nl}^{(l)}(0)|^2$, and masses. With increasing excitation, the RMS radii move into the region where the confining slopes differ, and Potential~II produces systematically lower masses, smaller kinetic energies, and more extended spatial distributions than Potential~I. The corresponding inter-potential mass differences increase to approximately $70$~MeV, $80$~MeV, and $90$~MeV for the $6S$, $6P$, and $6D$ states, respectively. The same radial separation is reflected in the spatial observables discussed in Sec.~\ref{sec:3A_Mass} and Sec.~\ref{sec:3B_WFO}. For Potential~III, the RMS markers of the highly excited states extend further into the screened region, and the separation from the unscreened and logarithmically modified potentials becomes substantially larger. This is consistent with the stronger changes in the high-excitation masses, radii, kinetic energies, and wavefunction overlaps obtained with Potential~III.

The figure therefore provides a radial interpretation of the model dependence observed in the spectrum. Potential~II introduces a gradual reduction of the confining slope while retaining a finite asymptotic value, whereas Potential~III progressively suppresses the confining interaction at large distances. The increasing RMS radii of the excited states move their probability distributions into precisely the region where these differences become significant. Thus, the growing inter-potential variation at high excitation is not uniform across the spectrum but follows the increasing sensitivity of the states to the long-distance form of the interaction. The widening MCMC credible bands at large $r$ further indicate that this region is less tightly constrained by the available low-lying input observables, providing a statistical counterpart to the increasing model dependence of the upper spectrum.

\subsection{Regge Trajectories}
\label{sec:3C_Regge}

The MCMC mass spectra from all three potentials are fitted to the linear and non-linear Regge forms of Eqs.~\eqref{eqn:22_regge_linear} and~\eqref{eqn:23_regge_NL} using \texttt{iminuit}~\cite{iminuit, James:1975dr}. For each trajectory, the median MCMC-derived masses are used as the input points, with $M^2$ taken as the fitting variable. To define a common effective weighting prescription for the Regge fits, a mass scale of $\Delta M=10$~MeV is adopted for all states. The corresponding effective uncertainty in the fitting variable is taken as $\sigma_{M_i^2}=2M_i\Delta M$, where $\Delta M=0.010$~GeV. The fit then minimizes
\begin{equation}
\chi^2 =
\sum_i
\frac{\left[M_i^2-M_{\mathrm{Regge}}^2(x_i;\boldsymbol{\theta})\right]^2}
{\sigma_{M_i^2}^2},
\label{eqn:chi2_regge}
\end{equation}
where $M_i$ denotes the MCMC-derived mass for the state labelled by $x_i$, $M_{\mathrm{Regge}}$ is the corresponding mass obtained from the fitted Regge parametrization, and $\boldsymbol{\theta}$ denotes the set of trajectory parameters. Thus, although the same effective mass scale is adopted for all states, the resulting weights in $M^2$ remain state dependent through the factor $2M_i$. This prescription is used solely to define the effective weights in the independent Regge fits and is not related to, nor derived from, the state-dependent posterior uncertainties of the MCMC mass analysis. The latter are therefore not propagated into the Regge fits, and the effective $\Delta M=10$~MeV should not be interpreted as a statistical or model uncertainty of the predicted masses. The uncertainties quoted for the Regge parameters are the corresponding \texttt{iminuit} fit uncertainties conditional on this weighting prescription. The numerical values in Tables~\ref{tab:Regge_radial} and~\ref{tab:Regge_orbital} provide the resulting linear and non-linear trajectory parameters, together with their Regge-fit uncertainties, for the different radial and orbital families.

\begin{figure}
     \centering
     \subfigure[$n^1S_0$, $nP_1'$, and $nD_2'$ states in Potential~I]{\includegraphics[width=.325\textwidth]{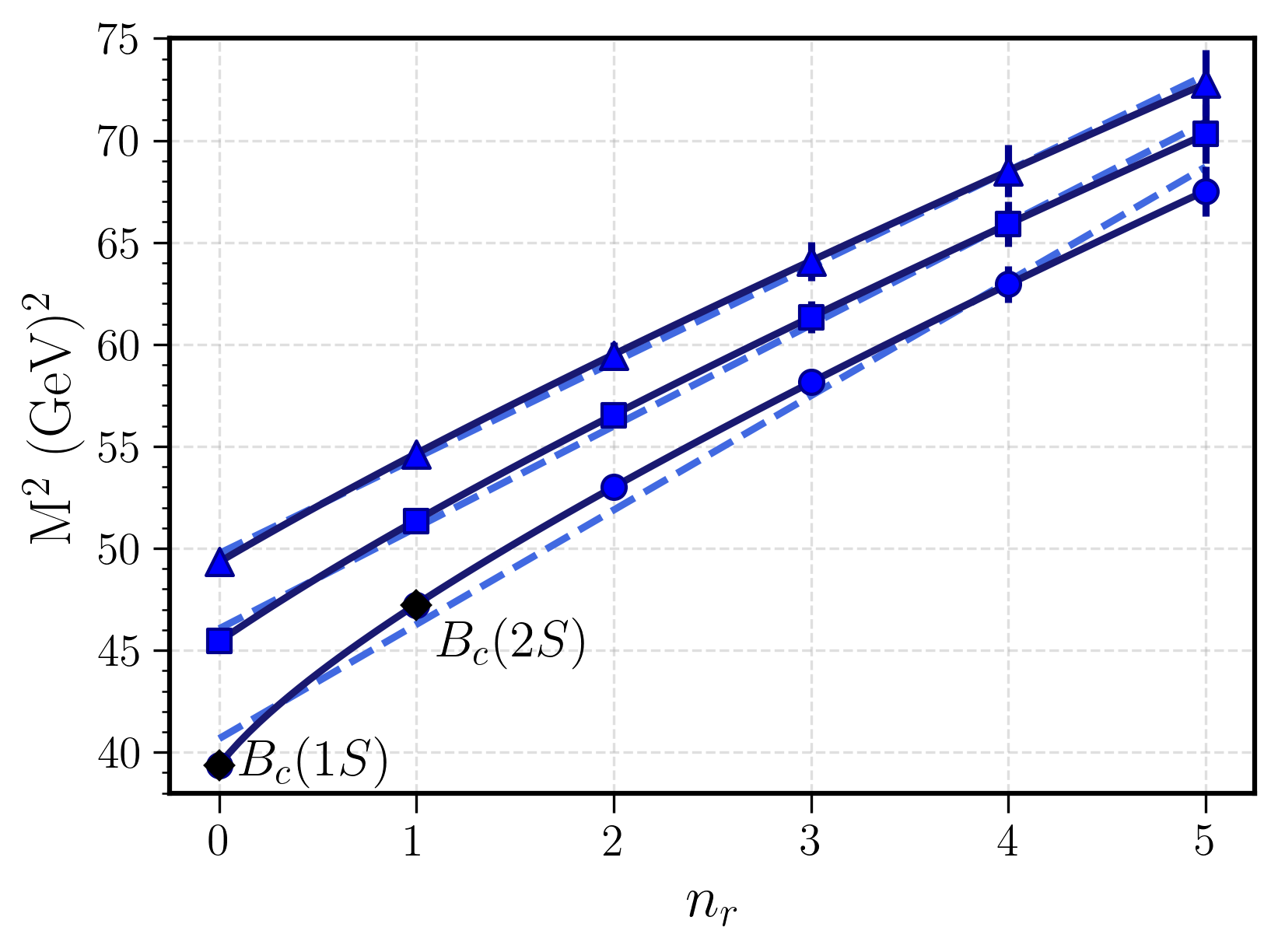}\label{fig:Regge_Bc_Cornell_nMsqerr_singlets}}
     \hfill
     \subfigure[$n^1S_0$, $nP_1'$, and $nD_2'$ states in Potential~II]{\includegraphics[width=.325\textwidth]{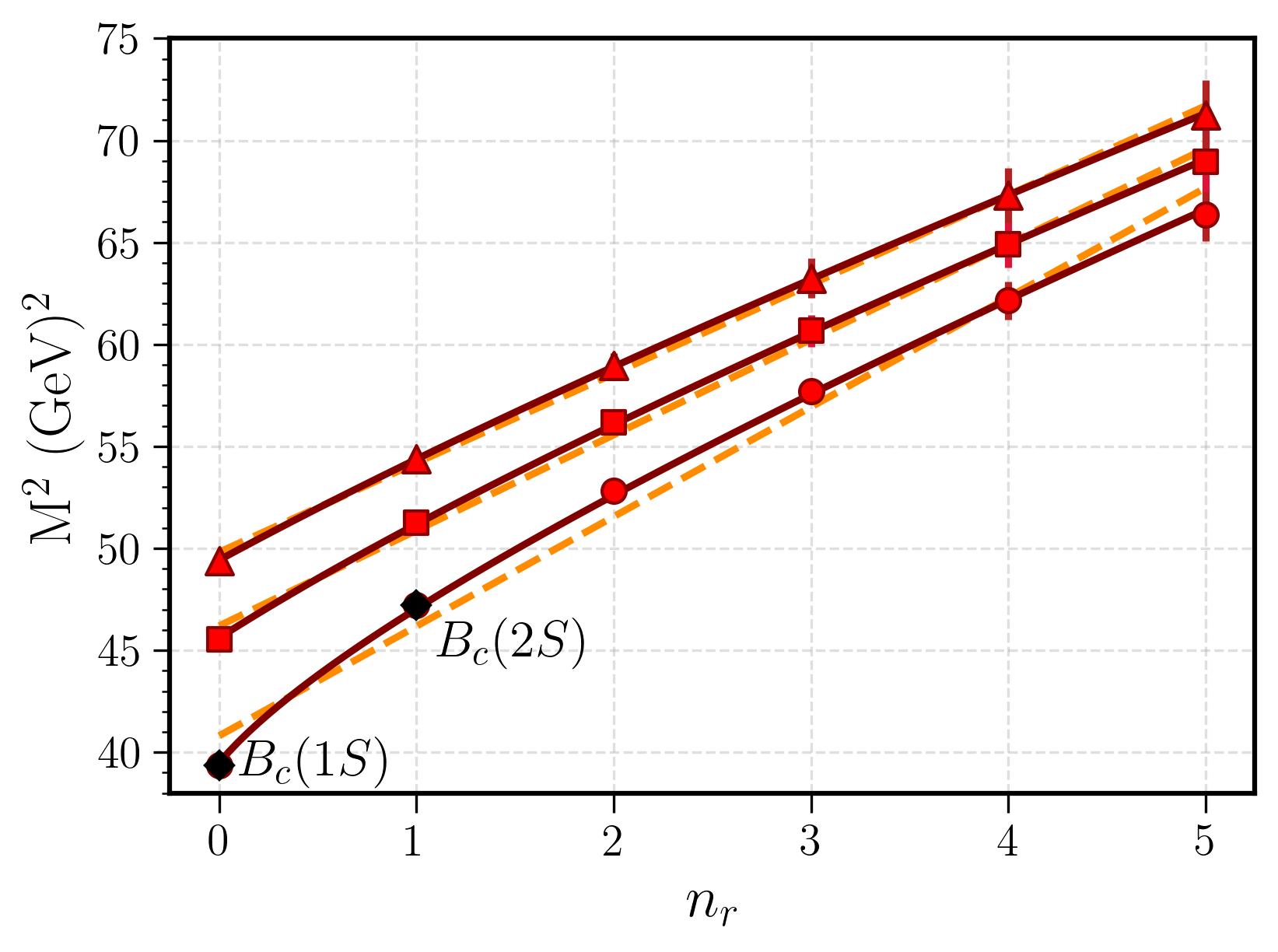}\label{fig:Regge_Bc_ModCornellpln_nMsqerr_singlets}}
     \hfill
     \subfigure[$n^1S_0$, $nP_1'$, and $nD_2'$ states in Potential~III]{\includegraphics[width=.325\textwidth]{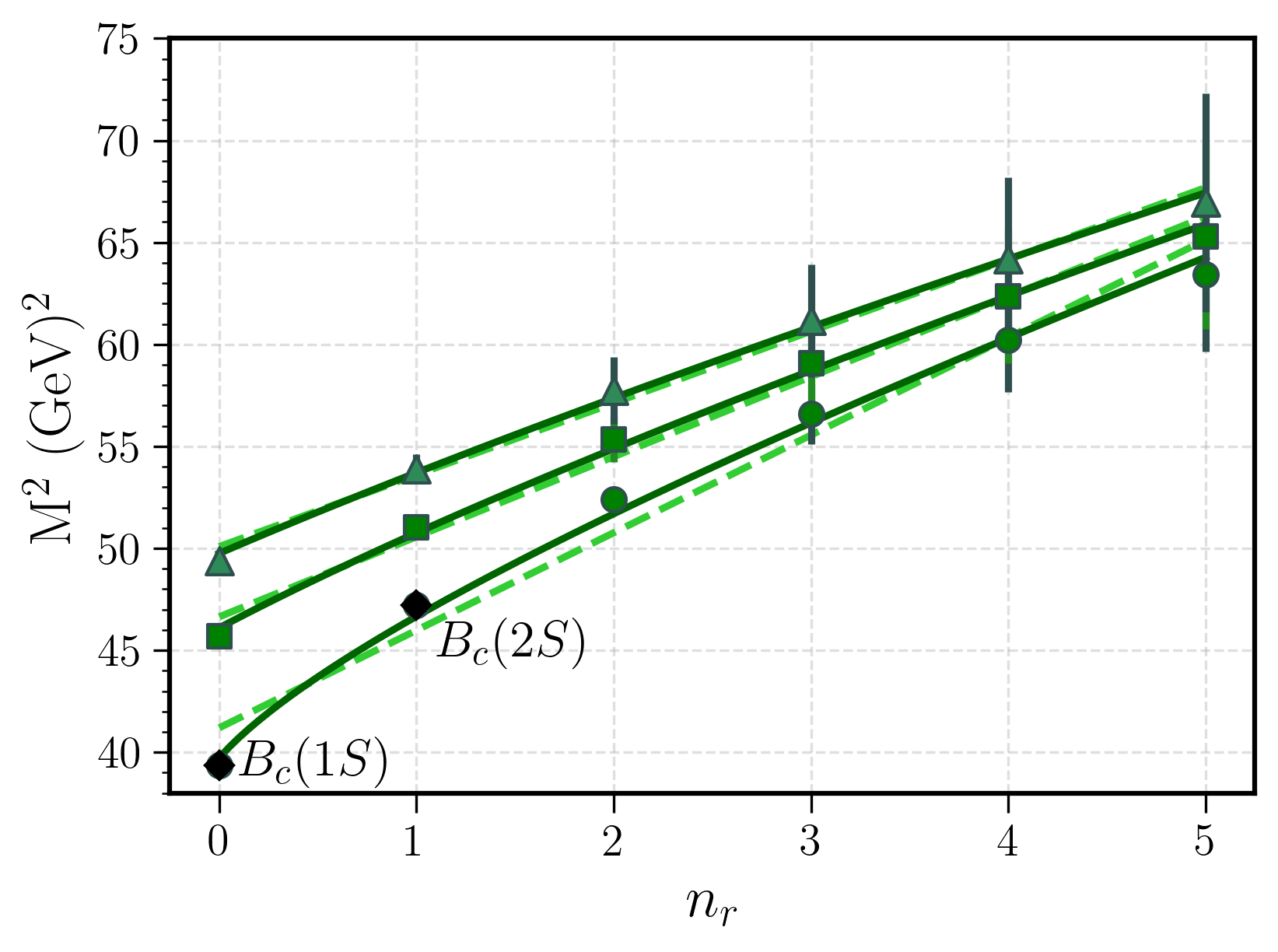}\label{fig:Regge_Bc_ScreenedCornell_nMsqerr_singlets}}
     \vspace{2mm}
     \subfigure[$n^3S_1$, $n^3P_2$, and $n^3D_3$ states in Potential~I]{\includegraphics[width=.325\textwidth]{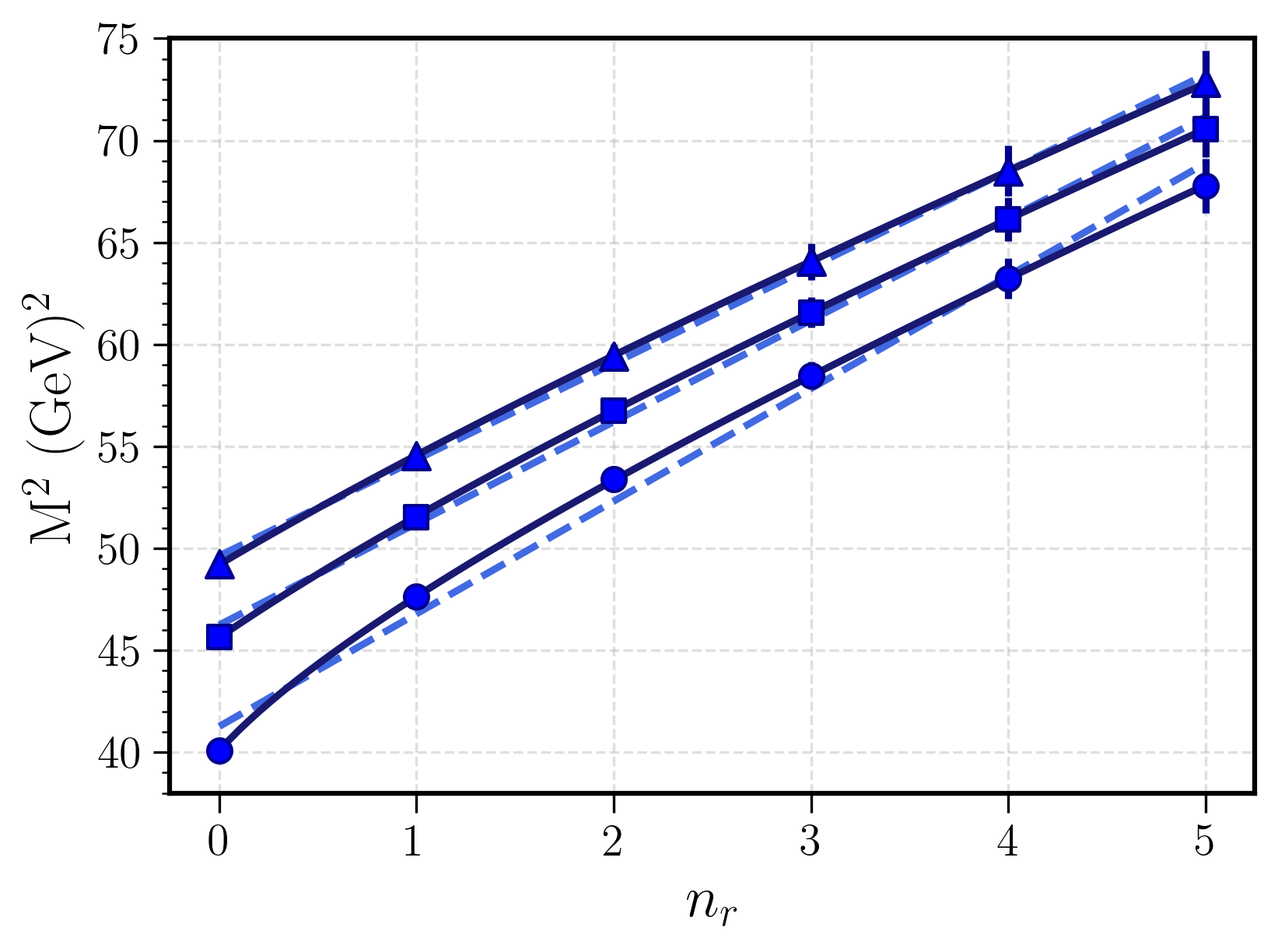}\label{fig:Regge_Bc_Cornell_nMsqerr_triplets}}
     \hfill
     \subfigure[$n^3S_1$, $n^3P_2$, and $n^3D_3$ states in Potential~II]{\includegraphics[width=.325\textwidth]{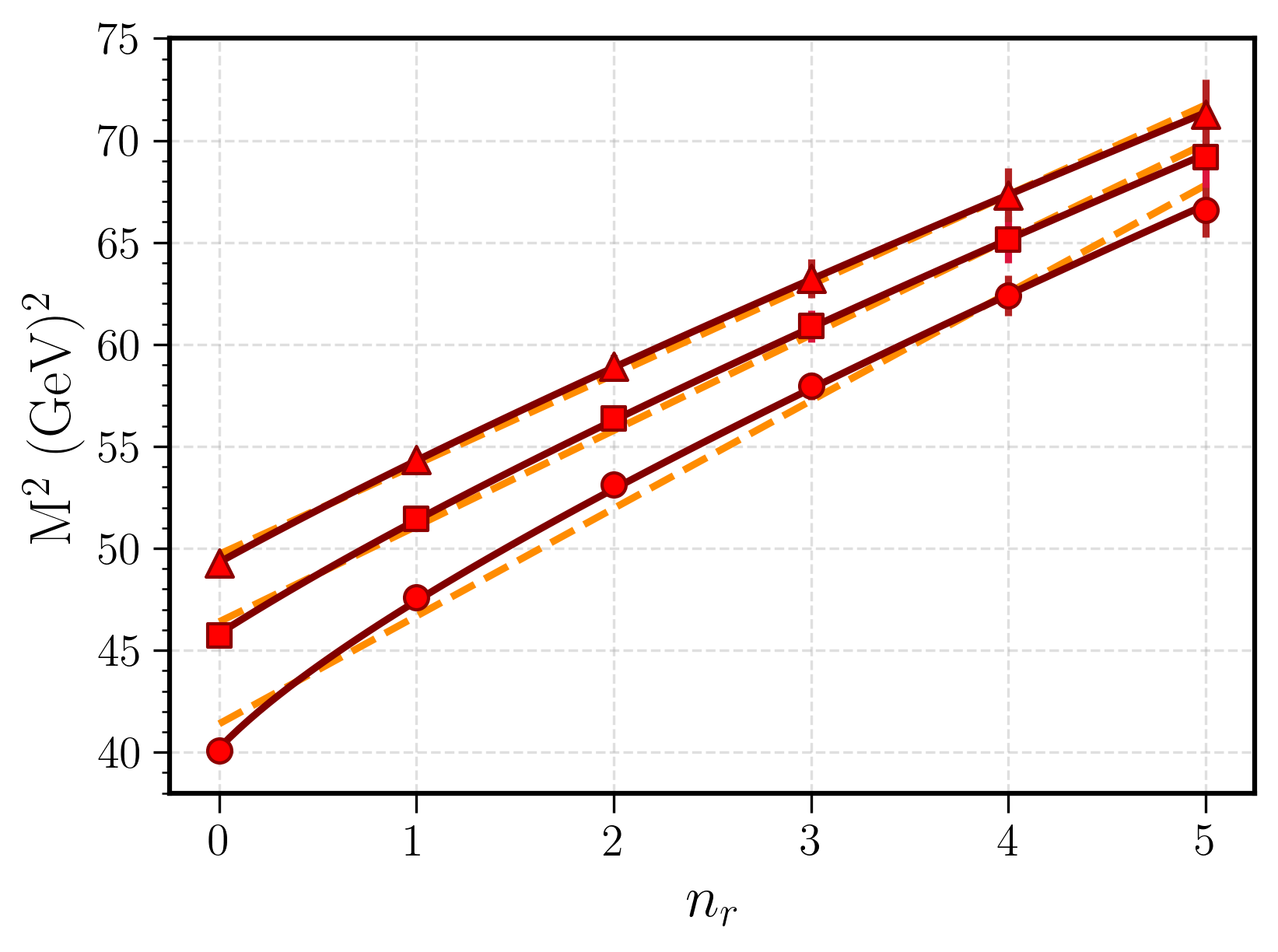}\label{fig:Regge_Bc_ModCornellpln_nMsqerr_triplets}}
     \hfill
     \subfigure[$n^3S_1$, $n^3P_2$, and $n^3D_3$ states in Potential~III]{\includegraphics[width=.325\textwidth]{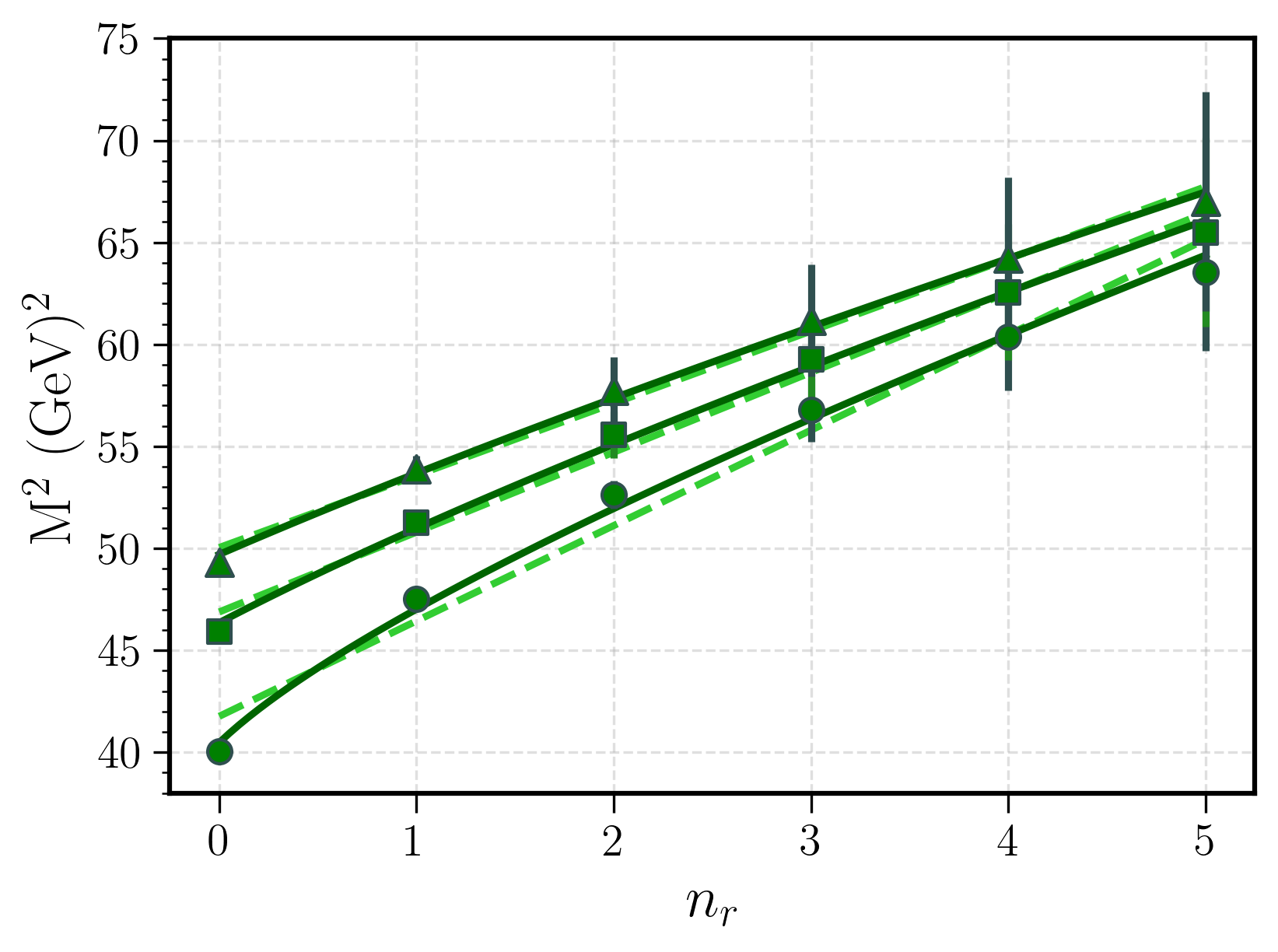}\label{fig:Regge_Bc_ScreenedCornell_nMsqerr_triplets}}
        \caption{Radial Regge trajectories of $B_c$ states. From top to bottom, the panels correspond to the $n^1S_0$, $nP_1'$, and $nD_2'$ states, and the $n^3S_1$, $n^3P_2$, and $n^3D_3$ states, shown for Potentials I (blue, left), II (red, middle), and III (green, right), respectively. Within each panel, the lowest, middle, and highest trajectories correspond to the $S$-, $P$-, and $D$-wave states, respectively. Experimental values are indicated by solid black diamonds, while the calculated masses are shown by solid circles ($S$), squares ($P$), and triangles ($D$), with uncertainty bars denoting the 16th–84th percentile intervals of the MCMC posterior distributions. The Regge fits use the effective weighting prescription described in the text. The dashed lines and solid curves denote the linear and non-linear Regge fits, respectively. The same legend is used throughout.}
        \label{fig:5_Bc_radial_regge}
\end{figure}

\begin{figure}
     \centering
     \subfigure[States with $J^P = 0^-, 1^+$, and $2^-$ in Potential~I]{\includegraphics[width=.325\textwidth]{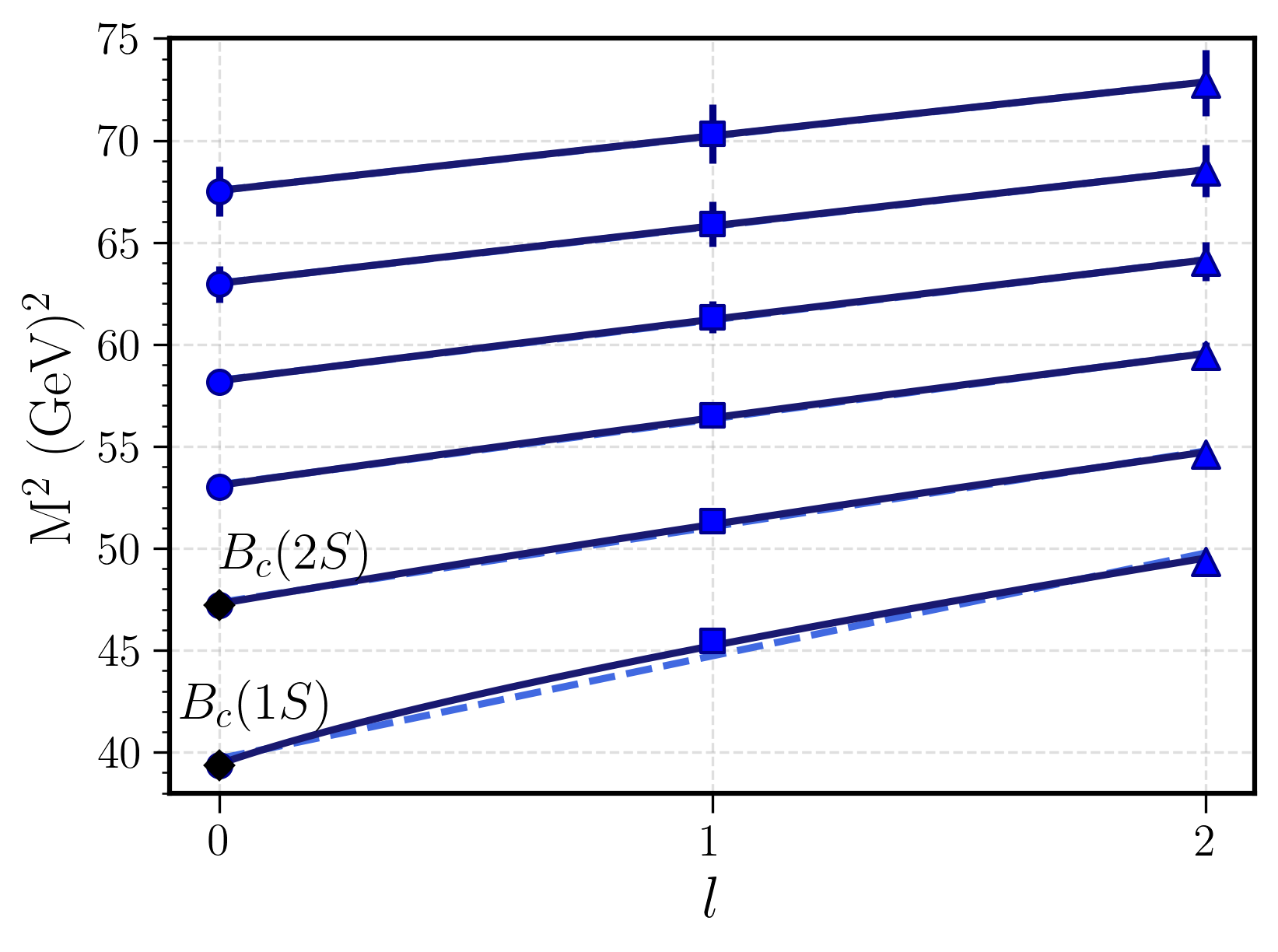}\label{fig:Regge_Bc_Cornell_lMsqerr_singlets}}
     \hfill
     \subfigure[States with $J^P = 0^-, 1^+$, and $2^-$ in Potential~II]{\includegraphics[width=.325\textwidth]{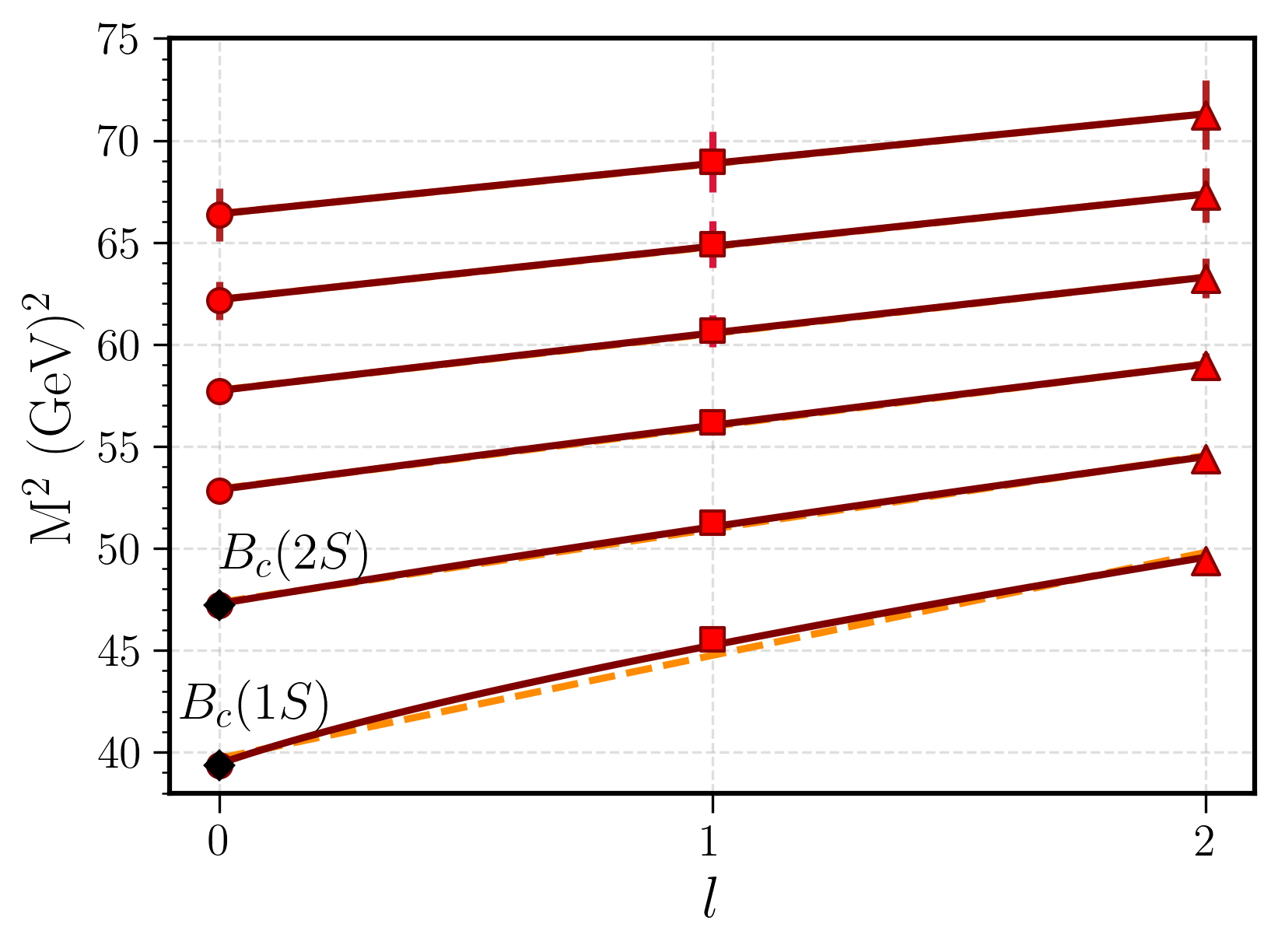}\label{fig:Regge_Bc_ModCornellpln_lMsqerr_singlets}}
     \hfill
     \subfigure[States with $J^P = 0^-, 1^+$, and $2^-$ in Potential~III]{\includegraphics[width=.325\textwidth]{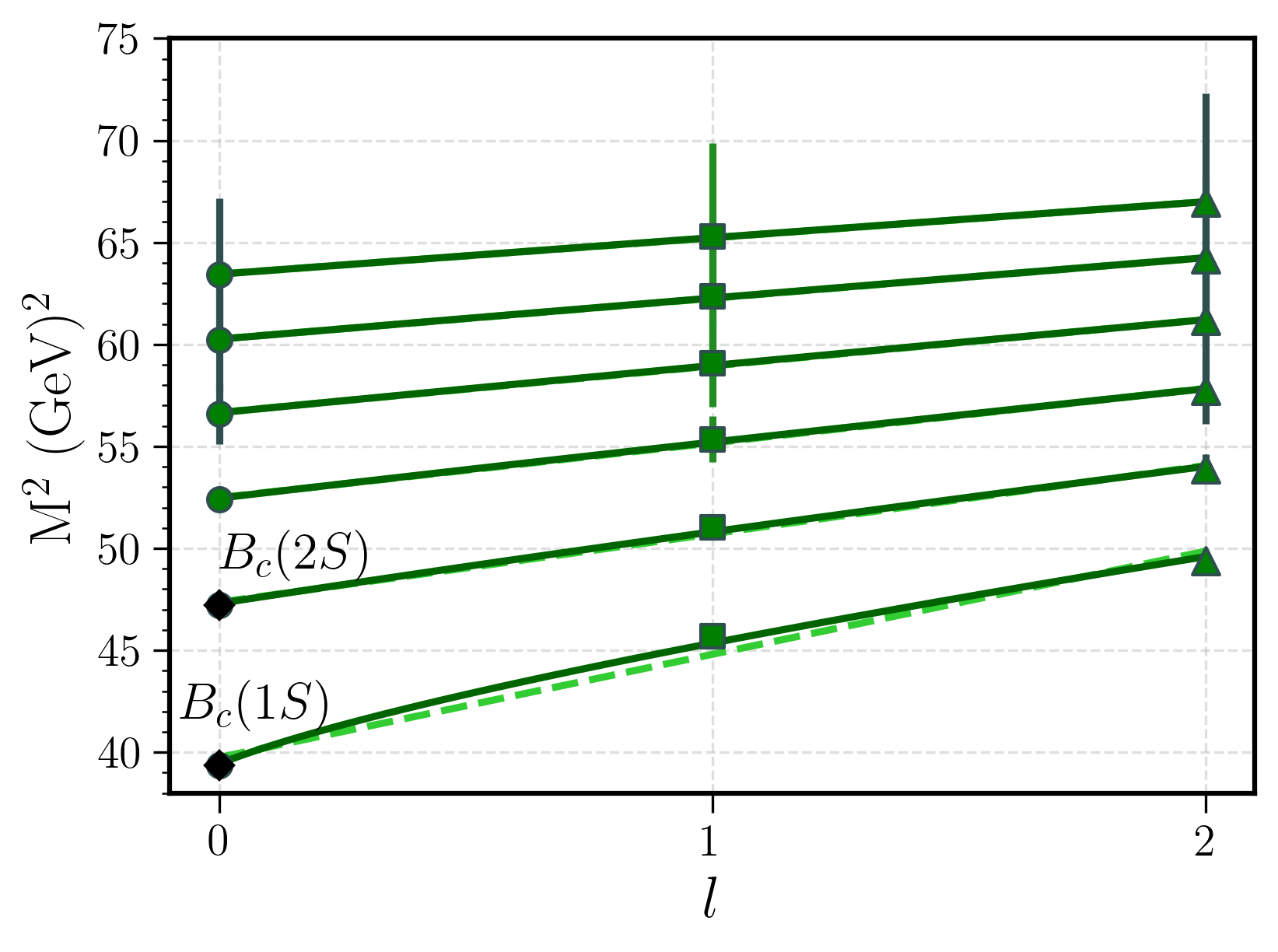}\label{fig:Regge_Bc_ScreenedCornell_lMsqerr_singlets}}
     \vspace{2mm}
     \subfigure[States with $J^P = 1^-, 2^+$, and $3^-$ in Potential~I]{\includegraphics[width=.325\textwidth]{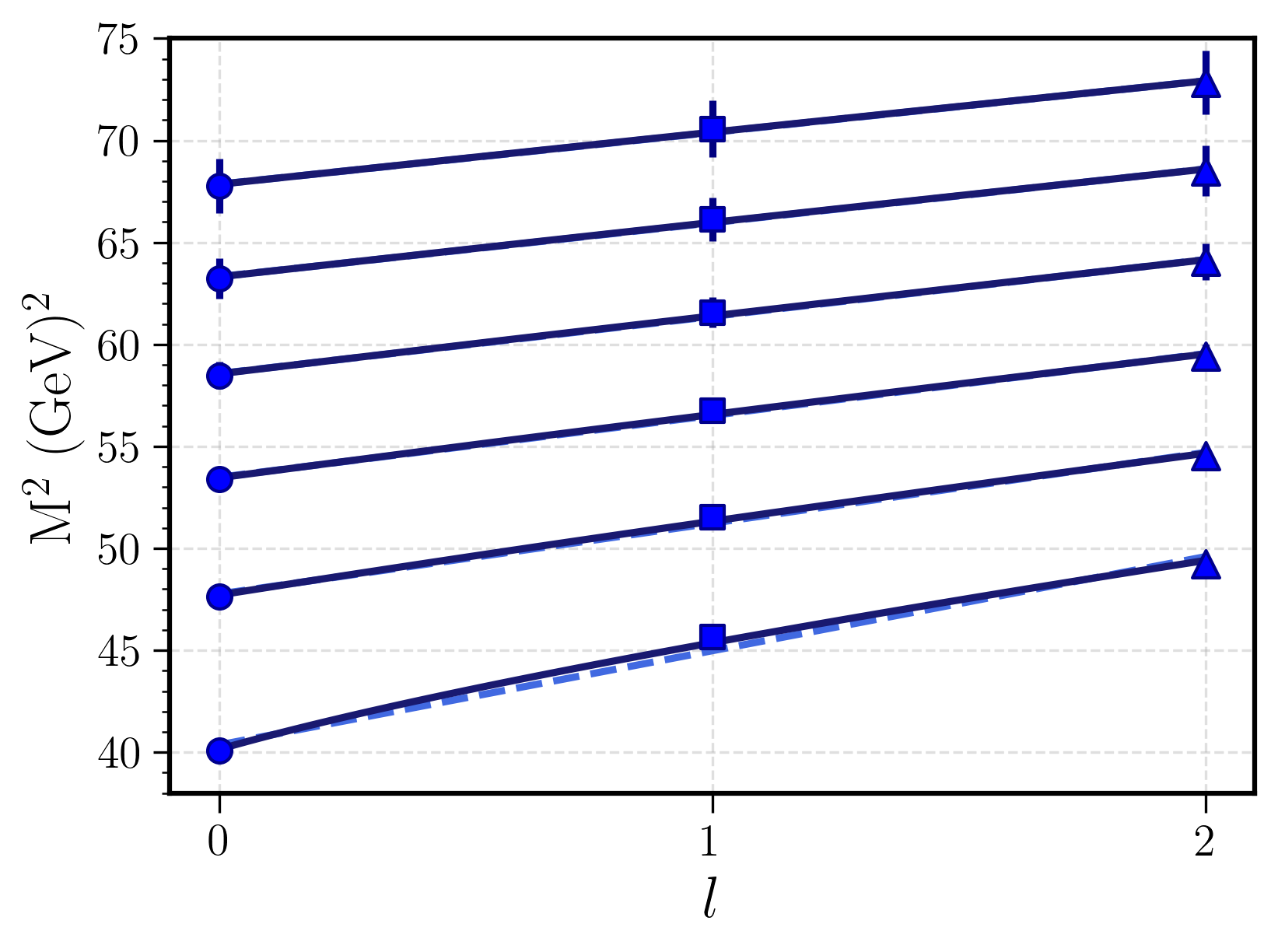}\label{fig:Regge_Bc_Cornell_lMsqerr_triplets}}
     \hfill
     \subfigure[States with $J^P = 1^-, 2^+$, and $3^-$ in Potential~II]{\includegraphics[width=.325\textwidth]{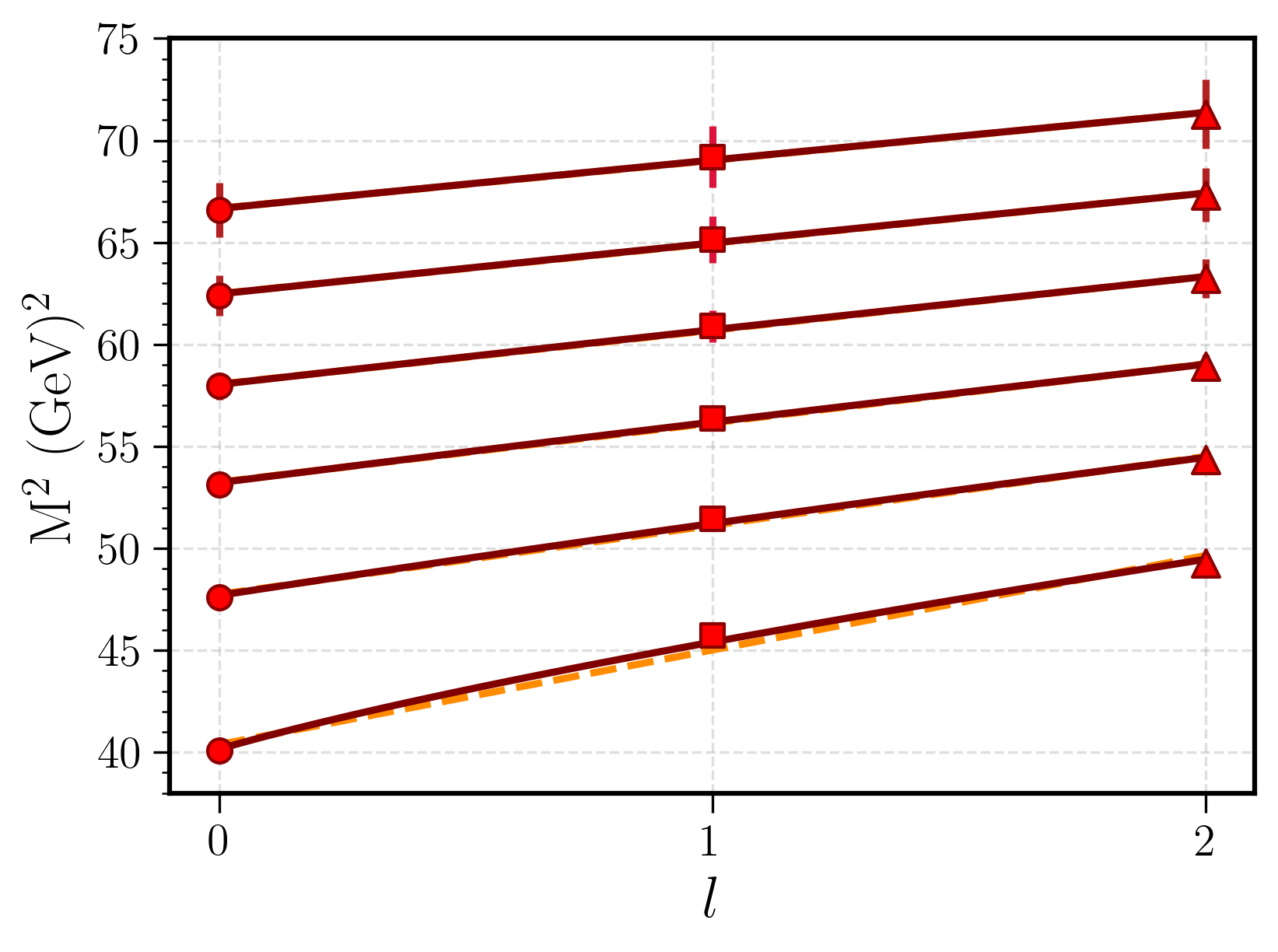}\label{fig:Regge_Bc_ModCornellpln_lMsqerr_triplets}}
     \hfill
     \subfigure[States with $J^P = 1^-, 2^+$, and $3^-$ in Potential~III]{\includegraphics[width=.325\textwidth]{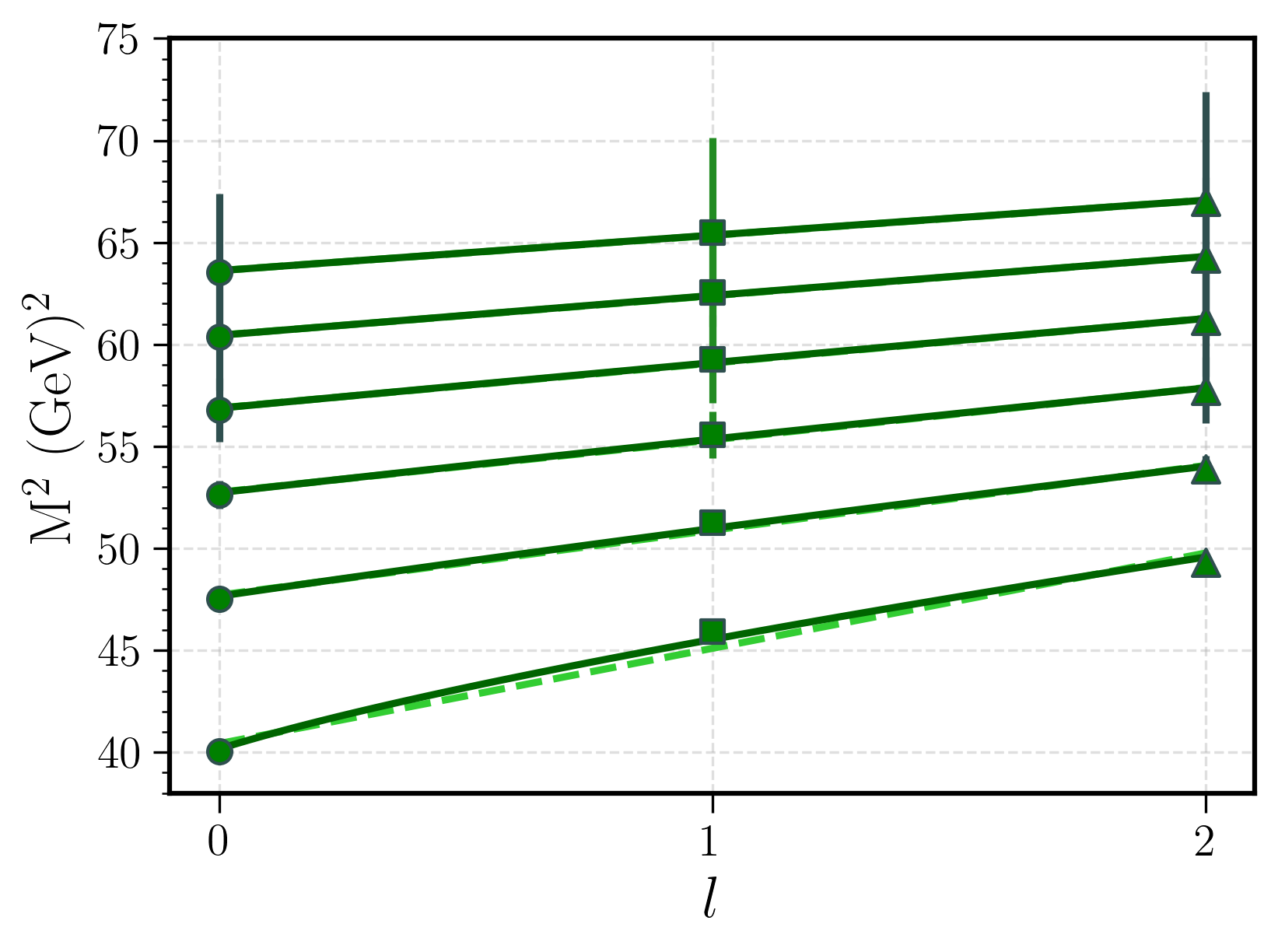}\label{fig:Regge_Bc_ScreenedCornell_lMsqerr_triplets}}
     
        \caption{Parent and daughter orbital Regge trajectories of $B_c$ mesons. The top panels correspond to states with $J^P = 0^-, 1^+$, and $2^-$, while the bottom panels show states with $J^P = 1^-, 2^+$, and $3^-$. The trajectories are shown for Potentials~I (blue, left), II (red, middle), and III (green, right), respectively. The calculated masses are shown with uncertainty bars denoting the 16th–84th percentile intervals of the MCMC posterior distributions. The Regge fits use the effective weighting prescription described in the text. The legend is the same as in Fig.~\ref{fig:5_Bc_radial_regge}.}
        \label{fig:6_Bc_orbital_regge}
\end{figure}

For the non-linear Regge parametrization, the coefficients $c_{fn_r}$ and $c_{fl}$ and the corresponding offset parameters $c_{0n_r}$ and $c_{0l}$ are fitted independently as free parameters for each individual sub-trajectory, rather than being fixed from a reference or lowest-lying trajectory. For the linear form, the coefficients $\alpha_{\rm lin}$, $\beta_{\rm lin}$, and $C_{\rm lin}$ are likewise fitted independently for each trajectory. This allows the degree of non-linearity and its excitation dependence to be examined directly across the different spin, parity, radial, and orbital families, as shown in Figs.~\ref{fig:5_Bc_radial_regge} and~\ref{fig:6_Bc_orbital_regge}.

The independent fits reveal a clear hierarchy in the radial curvature coefficients. For Potential~I, $c_{fn_r}$ remains close to unity across both singlet and triplet trajectories. For the singlet $n^1S_0$, $nP_1'$, and $nD_2'$ families, the values range from $0.994$ to $1.015$, while the corresponding triplet trajectories give values between $0.998$ and $1.015$. The small variation across the different $l$ values indicates that the radial curvature is comparatively stable for the Cornell potential. Potential~II gives systematically larger coefficients, with $c_{fn_r}$ decreasing from approximately $1.14$ for the $S$-wave trajectories to about $1.10$ for the $D$-wave trajectories. In contrast, Potential~III gives values below unity, decreasing from approximately $0.84$ for the $S$-wave trajectories to about $0.75$ for the $D$-wave trajectories. Thus, the independently fitted radial coefficients are not constrained to a prescribed universal value, while their systematic displacement above or below unity retains information on the modification of the long-range interaction.

A similar but more pronounced excitation dependence is observed for the orbital trajectories. For Potential~I, $c_{fl}$ decreases smoothly with increasing $n_r$, from $1.173$ to $1.075$ for the $J^P=0^-,1^+,2^-$ family and from $1.126$ to $1.042$ for the $J^P=1^-,2^+,3^-$ family. This gradual approach toward unity accompanies the progressively weaker curvature of the daughter trajectories. Potential~II maintains systematically larger coefficients, with $c_{fl}$ decreasing from $1.366$ to $1.172$ and from $1.312$ to $1.143$ for the two orbital families, respectively. The screened Potential~III shows the strongest modification, with $c_{fl}$ decreasing from $1.072$ to $0.746$ for the $J^P=0^-,1^+,2^-$ family and from $1.039$ to $0.733$ for the $J^P=1^-,2^+,3^-$ family. The increasingly sub-unity values therefore indicate a stronger suppression of curvature associated with the screened long-range interaction.

\begin{enumerate}

\item The radial Regge trajectories of $S$-wave singlet and triplet states (Fig.~\ref{fig:5_Bc_radial_regge}) exhibit pronounced non-linearity under all three potentials, which diminishes for $P$- and $D$-wave trajectories. This behavior reflects the different radial regions sampled by the states. For $S$-waves, the Coulomb and linear contributions remain appreciable over the relevant radial range, leading to stronger curvature. For $P$- and $D$-waves, the centrifugal barrier shifts the wave function toward larger $r$, where the linear confining term becomes more important and the relative Coulomb contribution is reduced, driving the trajectories toward linearity. This trend is consistent with the level-spacing ordering $\Delta E^S > \Delta E^P > \Delta E^D$ from Sec.~\ref{sec:3A_Mass}, reflecting the increasing influence of the centrifugal barrier.

\item The trajectories from Potentials~II and III show systematic shifts relative to Potential~I, reflecting their different long-range confining behavior discussed in Sec.~\ref{sec:3C_ModPot}. Potential~II produces systematically larger radial curvature coefficients, whereas the screened Potential~III produces values below unity. For the parent $nS$ trajectories, for example, $c_{fn_r}$ changes from approximately $1.015$ for Potential~I to $1.144$ for Potential~II and $0.840$ for Potential~III in the singlet sector. The same ordering persists for the daughter radial trajectories. The linear radial slopes show a corresponding sensitivity to the potential modification; for the $n^1S_0$ trajectory, $\beta_{\rm lin}=5.606$, $5.374$, and $4.786$ for Potentials~I, II, and III, respectively. Thus, both the linear slope and the independently fitted non-linear curvature coefficient retain information about the long-range form of the potential through the masses.

\item The orbital Regge trajectories (Fig.~\ref{fig:6_Bc_orbital_regge}) show a characteristic pattern in which the parent trajectories are distinctly non-linear, while the daughter trajectories become comparatively more linear and approximately parallel. Potential~II produces systematically larger curvature coefficients and modified slopes relative to Potential~I, whereas the screened Potential~III produces a substantially stronger reduction of the orbital slope $\alpha_{\rm lin}$ with increasing $n_r$. For the $J^P=0^-,1^+,2^-$ family, for example, $\alpha_{\rm lin}$ decreases from $5.041$ to $2.661$ for Potential~I, from $5.049$ to $2.451$ for Potential~II, and from $5.056$ to $1.776$ for Potential~III as $n_r$ increases from $0$ to $5$. The increasing sensitivity to the potential form with excitation reflects the larger spatial extent of highly excited states and their enhanced sensitivity to the long-range confining interaction.

\item For the highest excitations, some linear trajectory parameters become comparatively weakly constrained, as reflected by the larger \texttt{iminuit} uncertainties for selected daughter trajectories. This is particularly evident for the $D$-wave radial trajectories and for selected orbital slopes in Potentials~II and III. Such behavior is expected when the trajectories become increasingly close to linear and the slope, intercept, and curvature parameters become more strongly correlated. The qualitative trends in the independently fitted curvature coefficients and their evolution with excitation are therefore more robust than the precise values of individual linear-fit parameters at the highest excitations.

\end{enumerate}

The Regge behavior of $B_c$ is broadly intermediate between charmonium and bottomonium~\cite{A:2023bxv}, with the trajectories more closely resembling $b\bar b$ in their degree of non-linearity, although this comparison is phenomenological rather than a unique dynamical correspondence. The increasing approach toward linear behavior also makes the linear parametrization more adequate for the higher daughter trajectories, while the non-linear form of Eq.~\eqref{eqn:23_regge_NL} remains useful for describing the curvature of the low-lying trajectories.

\section{Summary and Conclusions}
\label{sec:4_sum_conclu}

We have presented the first Bayesian MCMC uncertainty-propagated mass spectrum of the $B_c$ system through $6D$, obtained from three confinement models treated on equal statistical footing: the Cornell potential (Potential~I), its logarithmically modified extension (Potential~II), and a screened Cornell form (Potential~III). All three models are constrained using PDG masses and lattice-QCD hyperfine splittings through MCMC posterior inference, providing uncertainty-propagated predictions and a systematic assessment of sensitivity to the long-range interaction. The Bayesian analysis reveals substantial posterior correlations among $\sigma$, $\sigma'$, and $C_0$ for Potential~II, indicating that parameter combinations are more strongly constrained than some individual parameters. A complementary Gaussian-prior analysis motivated by the bottomonium sector gives a largely stable low-lying spectrum and consistent qualitative conclusions, while the WAIC and PSIS-LOO diagnostics provide no strong statistical evidence for preferring one confinement form over another with the present data. Additional experimental or lattice-QCD inputs will therefore be needed to distinguish the models more decisively.

The three potentials give very similar low-lying spectra, while their predictions become increasingly separated for higher-excitation states as the wave functions probe larger distances. The largest inter-potential differences occur in the $D$-wave sector, making the $nD$ multiplets with $n\geq2$ among the more sensitive states to the confinement model. At $6D$, however, the inter-potential separation of approximately $90$~MeV is comparable to the posterior credible interval, for example ${}_{-92}^{+45}$~MeV for the $6^3D_1$ mass in Potential~I, limiting the present discriminating power to at most the $\sim1\sigma$ level. The increasing and asymmetric uncertainties at higher excitation likewise reflect the reduced constraining power of the available data on the long-range parameters.

The spin-dependent decomposition exposes distinct dynamical behavior across the orbital sectors. The $S$-wave hyperfine splitting decreases with radial excitation more slowly than a purely Coulombic expectation, reflecting the interplay between short-distance dynamics and confinement in the contact interaction. In the $P$-wave sector, the symmetric spin-orbit interaction is stabilized by partial cancellation between one-gluon exchange and scalar confinement, while the antisymmetric component governs singlet-triplet mixing. The $D$-wave fine structure is dominated by the antisymmetric spin-orbit term, with a non-standard mass ordering through $4D$ that arises from a level-dependent sign reversal in the symmetric spin-orbit contribution as scalar confinement overtakes one-gluon exchange. The Bayesian treatment provides uncertainty-propagated $P$- and $D$-wave mixing-angle predictions relevant to future radiative-transition and state-identification studies.

The wave-function and velocity diagnostics show that higher-excitation states increasingly probe a regime where relativistic effects become important. The $b$-quark remains non-relativistic throughout the spectrum, whereas the median $\langle v_c^2\rangle$ exceeds the diagnostic benchmark of $0.5$ from approximately the $3S$, $3P$, and $3D$ states onward in Potentials~I and II, and from approximately the $4S$, $4P$, and $4D$ states onward in Potential~III. Screening therefore reduces, but does not eliminate, the growth of the charm-quark velocity. Since the spin-dependent operators arise within the same non-relativistic $1/m_Q^2$ expansion, the higher-excitation predictions should be regarded as indicative rather than precision benchmarks. The same long-range sensitivity appears in the spatial and momentum observables, with Potential~III producing the strongest suppression of the confining interaction, more diffuse higher-excitation states, and lower momentum scales.

The independently fitted Regge trajectories provide a complementary characterization of the excitation dependence. Their curvature coefficients show systematic sensitivity to the potential form without being fixed to a universal value, while the non-linearity weakens with excitation and the higher daughter trajectories become increasingly well described by the linear form. The $B_c$ trajectories are broadly intermediate between the corresponding charmonium and bottomonium patterns, although this comparison remains phenomenological.

For states above the lowest open-flavor threshold, the quoted MCMC credible intervals remain conditional on the adopted single-channel potentials and do not include explicit coupled-channel effects. Existing coupled-channel and screened-potential calculations indicate that such effects can be sizeable and strongly state and model dependent. We therefore do not treat individual coupled-channel mass shifts as universal corrections to the MCMC intervals; instead, their spread provides an additional model-dependent theoretical systematic for the above-threshold spectrum.

The practical utility of the present predictions also depends on whether the corresponding states can be produced and experimentally resolved. The recent LHCb observation of a wide structure in the $B_c^+\gamma$ mass spectrum, consistent with the lowest excited $P$-wave $B_c^+$ states and exceeding $7\sigma$ significance using the full $9~\mathrm{fb}^{-1}$ Run~1--2 dataset, together with the measured relative production cross section $R=0.20\pm0.03\pm0.02\pm0.03$, provides an experimental benchmark for the $1P$ sector~\cite{LHCb:2025ubr,LHCb:2025uce}. The individual $1P$ states remain unresolved, however, so the predicted fine structure and $P_1$--$P_1'$ mixing angle remain relevant for future higher-resolution studies. Higher radial and orbital excitations are expected to be more challenging experimentally, with existing estimates indicating substantially suppressed $D$-wave production~\cite{Berezhnoy:2020gbe,Vagnoni:2025qfv}. Access to these states will therefore depend on increased luminosity and dedicated reconstruction strategies, particularly near or above open-flavor thresholds.

\section*{Acknowledgments}
The author RD gratefully acknowledge the financial support by the Department of Science and Technology (SERB:TAR/2022/000606), New Delhi.

\appendix
\FloatBarrier 

\section{Numerical solution of the radial Schr\"odinger equation}
\label{app:rk}

The radial Schr\"odinger equation in Eq.~\eqref{eqn:2_radial_SE} is solved numerically using a fourth-order Runge-Kutta (RK4) integration scheme. The equation is written in terms of the reduced radial wave function $u_{nl}(r)$, for which the regular solution at the origin satisfies
\begin{equation}
u_{nl}(r)\propto r^{l+1}, \qquad r\rightarrow 0.
\label{eqn:rk_boundary}
\end{equation}
The numerical integration is therefore initiated at the small but finite radius
$r_{\min}=10^{-5}~\mathrm{GeV}^{-1}$, with
\begin{equation}
u_{nl}(r_{\min})=r_{\min}^{\,l+1},
\qquad
u'_{nl}(r_{\min})=(l+1)r_{\min}^{\,l}.
\label{eqn:rk_initial}
\end{equation}
The radial step size used in the calculations is
\begin{equation}
h=10^{-4}~\mathrm{GeV}^{-1}.
\label{eqn:rk_step}
\end{equation}
The integration is continued into the asymptotic region, with the endpoint determined dynamically from the effective potential and the asymptotic behavior of the numerical solution. For the higher radial excitations considered here, the resulting radial domains extend to approximately $13-24~\mathrm{GeV}^{-1}$.

For a given $(n,l)$, the eigenvalue is obtained by iterative bisection of an initial energy bracket. At each midpoint energy, the radial equation is integrated while monitoring the node count and asymptotic behavior of the solution. The bisection was terminated when the energy bracket width reached $10^{-15}$ GeV. The resulting reduced radial wave functions are subsequently normalized according to
\begin{equation}
\int |u_{nl}(r)|^2\,dr=1.
\end{equation}

The numerical stability was checked by repeating the calculation for $h=5\times10^{-4}$, $10^{-4}$, $5\times10^{-5}$, and $10^{-5}~\mathrm{GeV}^{-1}$. The eigenvalues and masses are found to be stable over this range. For example, for the $1S$, $2S$, $5S$, and $6S$ states, changing the step size from $10^{-4}$ to $10^{-5}~\mathrm{GeV}^{-1}$ changes the calculated masses by less than $10^{-5}$~MeV, while $\sqrt{\langle r^2\rangle}$ remains unchanged at the displayed precision.

After normalization, the radial wave function is reconstructed according to
\begin{equation}
R_{nl}(r)=\frac{u_{nl}(r)}{r}.
\end{equation}
For $S$ waves, $|R_{n0}(0)|^2$ is obtained by extrapolating the numerically evaluated $R_{n0}(r)^2$ to $r=0$. For $P$- and $D$-wave states, the first and second derivatives of the reconstructed radial wave function are obtained numerically using \texttt{numpy.gradient}, with the second derivative obtained by differentiating the first-derivative array once more. The corresponding squared quantities, $\left|R'_{n1}(r)\right|^2$ and $\left|R''_{n2}(r)\right|^2$, are then constructed on the finite radial grid and extrapolated to $r=0$ using \texttt{scipy.interpolate.interp1d}. This procedure is consistent with the regular near-origin behavior $R_{nl}(r)\propto r^l$. The numerical stability of the extraction was checked under radial-grid refinement; for the $6D$ state, $\left|R''_{6D}(0)\right|^2$ changes by only approximately $2\times10^{-4}$ in relative terms upon a tenfold refinement of the grid.

As a numerical check, the confining term was removed by setting $\sigma=0$ and $V_c=0$, while choosing the Coulomb coefficient such that
\begin{equation*}
V(r)=-\frac{1}{r}.
\end{equation*}
For equal constituent masses $m_1=m_2=1~\mathrm{GeV}$, the reduced mass is $\mu=0.5~\mathrm{GeV}$, giving the analytic Coulomb spectrum
\begin{equation}
E_n=-\frac{\mu}{2n^2}
=-\frac{1}{4n^2}~\mathrm{GeV}.
\label{eqn:coulomb_spectrum}
\end{equation}
The numerical eigenvalues,
$E_1=-0.250000$,
$E_2=-0.062500$,
$E_5=-0.010000$, and
$E_6=-0.006944~\mathrm{GeV}$,
reproduce the analytic values to the displayed precision. The agreement is obtained for states with different principal quantum numbers and hence provides a check beyond the ground state, including highly excited radial solutions. This provides a check on the implementation of the radial equation, the boundary condition, the eigenvalue search, and the normalization procedure independently of the confining part of the phenomenological potential.

\section{Limitations of Deterministic Parameter Fits}
\label{app:1_min}

Exploratory parameter estimation was first carried out via direct $\chi^2$ minimization,
\begin{equation}
\chi^2 = \sum_k \frac{\left(M_k^{\mathrm{th}}-M_k^{\mathrm{exp}}\right)^2}{\left(\Delta M_k^{\mathrm{exp}}\right)^2},
\end{equation}
using two different sets of experimental inputs. The fitted parameters and the resulting $B_c$ mass spectra are collected in Tables~\ref{tab:min_params} and~\ref{tab:min_massspec}.

Set-I was obtained by minimizing over a combined set of known $B_c$ states and bottomonium masses and hyperfine splittings; Set-II used only the experimentally known $B_c$ masses, mass difference, and LQCD hyperfine splitting~\cite{Mathur:2018epb} (see Table~\ref{tab:min_params}). The resulting $\chi^2$ values,
\begin{equation}
\chi^2_{\mathrm{Set-I}} \approx 1.28\times10^{5}, \qquad \chi^2_{\mathrm{Set-II}} \approx 3.3\times10^{-9},
\end{equation}
reflect the differing levels of constraint rather than differences in physical fidelity. The large Set-I residual arises from the difficulty of simultaneously reproducing the full bottomonium spectrum and the $B_c$ sector within a single parameter set. The near-vanishing Set-II residual simply reflects the fact that a handful of $B_c$ observables leave the multidimensional parameter space almost unconstrained, permitting $M_k^{\mathrm{th}} \approx M_k^{\mathrm{exp}}$ for the calibration states. Consequently, the extremely small $\chi^2$ should not be interpreted as evidence of an exceptionally good fit, but rather as an indication of the under-determined nature of the fit, since the available inputs are insufficient to uniquely constrain the model parameters.

\begin{table}
\centering
\setlength{\tabcolsep}{7pt}
\caption{Cornell potential parameters fitted to two different sets of experimental inputs.}
\label{tab:min_params}
\begin{tabular}{@{}ccc@{}}
\toprule
Parameters              & Set-I\footnote{Inputs: $B_c(1{}^1S_0)$, $B_c(2{}^1S_0)$, $\eta_b(1{}^1S_0)$, $\Upsilon(1{}^3S_1)$, $\eta_b(2{}^1S_0)$, $\Upsilon(2{}^3S_1)$, $\Upsilon(3{}^3S_1)$, $\Upsilon(4{}^3S_1)$, $\Upsilon(5{}^3S_1)$, $\Upsilon(6{}^3S_1)$, $\Upsilon(1{}^3S_1)-\eta_b(1{}^1S_0)$, $\Upsilon(2{}^3S_1)-\eta_b(2{}^1S_0)$, $h_b(1^1P_1)$, $\chi_{b0}(1{}^3P_0)$, $\chi_{b1}(1{}^3P_1)$, $\chi_{b2}(1{}^3P_2)$, $h_b(2^1P_1)$, $\chi_{b0}(2{}^3P_0)$, $\chi_{b1}(2{}^3P_1)$, $\chi_{b2}(2{}^3P_2)$, $\chi_{b1}(3{}^3P_1)$, $\chi_{b2}(3{}^3P_2)$, $\Upsilon_2(1{}^3D_2)$ \cite{ParticleDataGroup:2024cfk}, and $B_c^*(1{}^3S_1)-B_c(1{}^1S_0)$ \cite{Mathur:2018epb}.} & Set-II\footnote{Inputs: $B_c(1{}^1S_0)$, $B_c(2{}^1S_0)$, $B_c(2{}^1S_0) - B_c(1{}^1S_0)$ \cite{ParticleDataGroup:2024cfk}, and $B_c^*(1{}^3S_1)-B_c(1{}^1S_0)$ \cite{Mathur:2018epb}.} \\ \midrule
$\alpha_s$              & 0.373   & 0.48     \\
$\sigma$ (in GeV${}^2$) & 0.182   & 0.15     \\
$\rho$ (in GeV)         & 4       & 1.41     \\
$V_c$ (in GeV)          & $-0.067$  & 0.001    \\ \bottomrule
\end{tabular}
\end{table}

The parameter shifts between the two solutions are significant: $\alpha_s$ increases from $0.373$ to $0.48$, $\sigma$ decreases from $0.182~\text{GeV}^2$ to $0.15~\text{GeV}^2$, $\rho$ drops from $4~\text{GeV}$ to $1.41~\text{GeV}$, and $V_c$ shifts from $-0.067~\text{GeV}$ to near zero. These variations propagate to the mass spectrum in Table~\ref{tab:min_massspec}: the Set-I ground-state mass ($6378~\text{MeV}$) exceeds experiment by more than $100~\text{MeV}$, while Set-II reproduces it exactly by construction. Throughout the excited states, Set-I predictions exceed those of Set-II by roughly $70-250~\text{MeV}$ depending on the state, and the $P$- and $D$-wave mixing angles differ by several degrees, reflecting the sensitivity of spin-dependent interactions to the underlying parameters.

Thus, different choices of calibration observables lead to substantially different parameter solutions and excited-state predictions, illustrating the sensitivity of deterministic fitting to the selected calibration set. Moreover, the limited number of calibration observables provides little redundancy with which to assess the predictive validity of the fitted parameter set. Consequently, while Set-II provides a viable calibration of the model parameters, its predictive robustness cannot be established from the calibration data alone.

\begingroup
\squeezetable
\begin{table}
\centering
\setlength{\tabcolsep}{7pt}
\caption{$B_c$ mass spectra (MeV) for the parameter sets of Table~\ref{tab:min_params}.}
\label{tab:min_massspec}
\begin{tabular}{@{}ccc|ccc|ccc@{}}
\toprule
State        & Set-I & Set-II & State         & Set-I & Set-II & State         & Set-I & Set-II \\ \midrule
${1{}^1S_0}$ & 6378.2  & 6274.47  & ${1{}^3P_0}$  & 6811.71 & 6712.55  & ${1{}^3D_1}$  & 7119.85 & 7021.41  \\
${1{}^3S_1}$ & 6446.91 & 6329.47  & ${1{}P_1}$    & 6841.8  & 6754.98  & ${1{}D_2}$    & 7114.28 & 7024.59  \\
${2{}^1S_0}$ & 6965.44 & 6871.2   & ${1{}P_1'}$   & 6847.68 & 6766.02  & ${1{}D_2'}$   & 7124.63 & 7030.15  \\
${2{}^3S_1}$ & 7005.52 & 6892.47  & ${1{}^3P_2}$  & 6859.6  & 6788.06  & ${1{}^3D_3}$  & 7115.52 & 7029.43  \\
${3{}^1S_0}$ & 7369.2  & 7246.88  & $\theta_{1P}$ & 3.6$^{\circ}$     & 17.35$^{\circ}$    & $\theta_{1D}$ & -52.42$^{\circ}$  & -48.93$^{\circ}$   \\
${3{}^3S_1}$ & 7401.99 & 7261.53  & ${2{}^3P_0}$  & 7227.91 & 7102.1   & ${2{}^3D_1}$  & 7476.85 & 7344.98  \\
${4{}^1S_0}$ & 7707.66 & 7554.83  & ${2{}P_1}$    & 7256.24 & 7139.26  & ${2{}D_2}$    & 7474.81 & 7350.3   \\
${4{}^3S_1}$ & 7736.76 & 7566.45  & ${2{}P_1'}$   & 7262.96 & 7149.54  & ${2{}D_2'}$   & 7482.51 & 7353.96  \\
${5{}^1S_0}$ & 8009.29 & 7826.51  & ${2{}^3P_2}$  & 7275.22 & 7169.94  & ${2{}^3D_3}$  & 7477.08 & 7355.82  \\
${5{}^3S_1}$ & 8036.02 & 7836.31  & $\theta_{2P}$ & 9.86$^{\circ}$    & 19.63$^{\circ}$    & $\theta_{2D}$ & -51.29$^{\circ}$  & -47.2$^{\circ}$    \\
${6{}^1S_0}$ & 8286.19 & 8074.5   & ${3{}^3P_0}$  & 7574.4  & 7418.27  & ${3{}^3D_1}$  & 7791.79 & 7627.92  \\
${6{}^3S_1}$ & 8311.22 & 8083.07  & ${3{}P_1}$    & 7601.93 & 7453.15  & ${3{}D_2}$    & 7791.99 & 7634.59  \\
-            & -       & -        & ${3{}P_1'}$   & 7609.22 & 7463.36  & ${3{}D_2'}$   & 7798.03 & 7637.08  \\
-            & -       & -        & ${3{}^3P_2}$  & 7621.73 & 7482.97  & ${3{}^3D_3}$  & 7794.93 & 7640.57  \\
-            & -       & -        & $\theta_{3P}$ & 12.58$^{\circ}$   & 20.56$^{\circ}$    & $\theta_{3D}$ & -50.06$^{\circ}$  & -45.07$^{\circ}$   \\
-            & -       & -        & ${4{}^3P_0}$  & 7881.85 & 7695.7   & ${4{}^3D_1}$  & 8078.97 & 7884.64  \\
-            & -       & -        & ${4{}P_1}$    & 7908.91 & 7729.29  & ${4{}D_2}$    & 8080.74 & 7892.25  \\
-            & -       & -        & ${4{}P_1'}$   & 7916.59 & 7739.56  & ${4{}D_2'}$   & 8085.63 & 7893.94  \\
-            & -       & -        & ${4{}^3P_2}$  & 7929.29 & 7758.66  & ${4{}^3D_3}$  & 8084.15 & 7898.57  \\
-            & -       & -        & $\theta_{4P}$ & 14.14$^{\circ}$   & 20.99$^{\circ}$    & $\theta_{4D}$ & -48.72$^{\circ}$  & -42.04$^{\circ}$   \\
-            & -       & -        & ${5{}^3P_0}$  & 8163.28 & 7948.1   & ${5{}^3D_1}$  & 8345.98 & 8122.58  \\
-            & -       & -        & ${5{}P_1}$    & 8190.01 & 7980.82  & ${5{}D_2}$    & 8348.92 & 8130.87  \\
-            & -       & -        & ${5{}P_1'}$   & 8197.99 & 7991.2   & ${5{}D_2'}$   & 8352.95 & 8131.99  \\
-            & -       & -        & ${5{}^3P_2}$  & 8210.83 & 8009.92  & ${5{}^3D_3}$  & 8352.7  & 8137.47  \\
-            & -       & -        & $\theta_{5P}$ & 15.19$^{\circ}$   & 21.2$^{\circ}$     & $\theta_{5D}$ & -47.22$^{\circ}$  & -37.1$^{\circ}$    \\
-            & -       & -        & ${6{}^3P_0}$  & 8425.66 & 8182.5   & ${6{}^3D_1}$  & 8597.41 & 8346.14  \\
-            & -       & -        & ${6{}P_1}$    & 8452.16 & 8214.6   & ${6{}D_2}$    & 8601.27 & 8354.96  \\
-            & -       & -        & ${6{}P_1'}$   & 8460.35 & 8225.09  & ${6{}D_2'}$   & 8604.64 & 8355.68  \\
-            & -       & -        & ${6{}^3P_2}$  & 8473.32 & 8243.51  & ${6{}^3D_3}$  & 8605.35 & 8361.81  \\
-            & -       & -        & $\theta_{6P}$ & 15.95$^{\circ}$   & 21.28$^{\circ}$    & $\theta_{6D}$ & -45.51$^{\circ}$  & -27.97$^{\circ}$   \\ \bottomrule
\end{tabular}
\end{table}
\endgroup

\FloatBarrier

\section{MCMC Configuration and Convergence}
\label{app:2_MCMC}

All MCMC calculations employed the affine-invariant ensemble sampler~\cite{Goodman:2010dyf} implemented in \texttt{emcee} (version 3.1.6) \cite{Foreman-Mackey:2012any}, with integrated autocorrelation times ($\tau_{\text{int}}$) computed using the built-in routines. Three independent chains (A-C), initialized from different random seeds and random walker distributions within the prior bounds, were generated for each potential to verify convergence. The corresponding sampler configurations and convergence diagnostics are summarized in Table~\ref{tab:runconfiguniform}. For all three potentials, the production chains satisfy $N_{\rm steps}>50\,\tau_{\rm int}$, with effective sample sizes ranging from $\sim 2 \times 10^3$ to $9 \times 10^3$. Potential II quotes the largest autocorrelation times because of the stronger correlations among $\{\sigma, \sigma', C_0\}$, whereas Potentials I and III converge more rapidly. For Potential II, additional proposal strategies were explored in Chains B and C by employing a smaller StretchMove scale ($a=1.5$) and a mixed StretchMove/DEMove proposal, respectively. These alternatives increased the acceptance fraction relative to Chain A while producing statistically indistinguishable posterior distributions. Increasing the number of walkers provides a further route to improving the acceptance fraction \cite{Foreman-Mackey:2012any}, albeit at higher computational cost.

\begin{table}
\centering
\caption{Configuration and convergence diagnostics for the independent MCMC chains used for Potentials I–III.}
\label{tab:runconfiguniform}
\begin{tabular}{@{}cccccccc@{}}
\toprule
Potential & Chain & Walkers & \makecell{Steps \\per \\walker} & \makecell{$\tau_{\text{int}}$ \\(range)} & \makecell{Flattened \\post-burn-in \\samples} & \makecell{ESS \\(range)}        & \makecell{Mean \\acceptance \\fraction} \\ \midrule
Pot. I    & A     & 32      & 55060          & $\sim 556 - 615$            & $\sim 1.41\times10^6$          & $\sim 2300-2500$   & 0.33                     \\
Pot. I    & B     & 32      & 55000          & $\sim 536 - 565$            & $\sim 1.41\times10^6$          & $\sim 2537 - 2676$ & 0.336                    \\
Pot. I    & C     & 32      & 55000          & $\sim 548 - 563$            & $\sim 1.41\times10^6$          & $\sim 2547 - 2614$ & 0.335                    \\
Pot. II   & A     & 48      & 391951         & $\sim 3460-7270$            & $\sim 1.5\times10^7$           & $\sim 2070-4350$   & 0.088                    \\
Pot. II   & B     & 48      & 390000         & $\sim 3752 - 6452$          & $\sim 1.5\times10^7$           & $\sim 2321 - 3991$ & 0.142                    \\
Pot. II   & C     & 48      & 390000         & $\sim 3672 - 6159$          & $\sim 1.5\times10^7$           & $\sim 2432 - 4079$ & 0.125                    \\
Pot. III  & A     & 40      & 120000         & $\sim 441 - 1153$           & $\sim 3.84\times10^6$          & $\sim 3331 - 8702$ & 0.254                    \\
Pot. III  & B     & 40      & 120000         & $\sim 442 - 1055$           & $\sim 3.84\times10^6$          & $\sim 3639 - 8679$ & 0.257                    \\
Pot. III  & C     & 40      & 120000         & $\sim 435 - 1087$           & $\sim 3.84\times10^6$          & $\sim 3533 - 8831$ & 0.255                    \\ \bottomrule
\end{tabular}
\end{table}

Convergence across the independent chains was further verified using the conventional Gelman-Rubin statistic. Because the post-burn-in chains have slightly different lengths, they were first truncated to a common length before computing $\hat{R}$. For every free parameter in all three potentials, $\hat{R} < 1.001$, indicating no detectable between-chain disagreement according to this diagnostic. Together with the $N_{\rm steps}>50\,\tau_{\rm int}$ criterion and the consistency of posterior summaries across independent chains, this provides evidence that the chains have adequately sampled the stationary posterior. Since the three chains are statistically indistinguishable, all posterior summaries and uncertainty propagation presented in the main text are based on representative Chain A.

The thinning factor was chosen so that each posterior retained approximately $5000$ samples, sufficient for stable percentile estimates while substantially reducing the computational cost of propagating uncertainties to the complete $B_c$ spectrum. The resulting thinned chains contain $5017$ samples (thinning factor $281$) for Potential I, $5001$ samples (thinning factor $3010$) for Potential II, and $5000$ samples (thinning factor $768$) for Potential III. Thus, the retained chains contain approximately 5000 samples, while the corresponding effective sample sizes remain in the range quoted in Table~\ref{tab:runconfiguniform}; these effective sample sizes determine the statistical resolution of the propagated posterior uncertainties. The posterior corner plots, parameter tables, and propagated spectral uncertainties are presented in Sec.~\ref{sec:3_ND}.

Model predictive performance was assessed using WAIC and PSIS-LOO computed from the same thinned posterior samples. Because the experimental splitting $M_{B_c(2S)}-M_{B_c(1S)}$ is not statistically independent of the individual masses, the diagnostics were evaluated both including and excluding this constraint. The results are summarized in Table~\ref{tab:criterianuniform}. The WAIC and PSIS-LOO values are comparable across all three potentials, indicating that the presently available $B_c$ data do not strongly prefer one confinement model over another. Removing the $M_{B_c(2S)}-M_{B_c(1S)}$ constraint changes the absolute values of the information criteria, but leaves the relative comparison essentially unchanged. The effective complexities $p_{\text{WAIC}}\simeq1.1-1.3$ and $p_{\text{LOO}}\simeq1.7-2.4$ remain substantially below the nominal parameter counts, reflecting the limited constraining power of the present experimental dataset. The PSIS diagnostics indicate moderate reliability for Potentials I and II ($k \approx 0.8$) and unreliable LOO estimates for Potential III ($k > 1$).

\begin{table}
\centering
\caption{WAIC and PSIS-LOO diagnostics computed from the thinned posterior samples for Potentials I–III, evaluated both including and excluding the $M_{B_c(2S)}-M_{B_c(1S)}$ constraint.}
\label{tab:criterianuniform}
\begin{tabular}{@{}ccccccc@{}}
\toprule
Potential & Inputs                            & WAIC            & $p_{\text{WAIC}}$ & $\text{ELPD}_{\text{LOO}}$ & $p_{\text{LOO}}$ & max $k$ \\ \midrule
Pot. I    & Full                              & $4.67 \pm 0.94$ & $1.232$           & $-3.01 \pm 1.58$          & $1.911$         & $0.874$ \\
Pot. I    & Without $M_{B_c(2S)}-M_{B_c(1S)}$ & $4 \pm 0.74$    & $1.103$           & $-2.67 \pm 1.27$          & $1.77$          & $0.874$ \\
Pot. II   & Full                              & $4.77 \pm 1$    & $1.266$           & $-2.97 \pm 1.5$           & $1.847$         & $0.805$ \\
Pot. II   & Without $M_{B_c(2S)}-M_{B_c(1S)}$ & $4.11 \pm 0.79$ & $1.138$           & $-2.62 \pm 1.19$          & $1.7$           & $0.805$ \\
Pot. III  & Full                              & $4.75 \pm 1.01$ & $1.27$            & $-3.51 \pm 2.1$           & $2.408$         & $1.152$ \\
Pot. III  & Without $M_{B_c(2S)}-M_{B_c(1S)}$ & $4.1 \pm 0.79$  & $1.15$            & $-3.18 \pm 1.68$          & $2.275$         & $1.152$ \\ \bottomrule
\end{tabular}
\end{table}

\FloatBarrier
\subsection{Prior sensitivity}
\label{sec:prior}

In addition to the uniform-prior analysis, we investigate the complementary Gaussian-prior framework introduced in Sec.~\ref{sec:2c_MCMC}, which propagates prior information from bottomonium-based fit to the $B_c$ sector. The sampler configuration is kept identical to isolate the effect of the prior assumptions. Convergence was again verified using three independent chains for each potential; the corresponding diagnostics are summarized in Table~\ref{tab:runconfigGauss}. As in the uniform-prior analysis, all chains satisfy $N_{\text{steps}}>50~ \tau_{\text{int}}$ and yield $\hat{R} < 1.001$ for every parameter, providing strong evidence that the independent chains have converged to same posterior distribution. Relative to the uniform-prior analysis, the Gaussian priors reduce the autocorrelation times and increase the acceptance fractions, indicating more efficient exploration of the resulting, more localized posterior distributions.

\begin{table}
\centering
\caption{Configuration and convergence diagnostics for the independent MCMC chains used for Potentials I–III with the Gaussian priors described in Sec.~\ref{sec:2c_MCMC}.}
\label{tab:runconfigGauss}
\begin{tabular}{@{}cccccccc@{}}
\toprule
Potential & Chain & Walkers & \makecell{Steps \\per \\walker} & \makecell{$\tau_{\text{int}}$ \\(range)} & \makecell{Flattened \\post-burn-in \\samples} & \makecell{ESS \\(range)}         & \makecell{Mean \\acceptance \\fraction} \\ \midrule
Pot. I    & A     & 32      & 70000          & $\sim 370 - 619$            & $\sim 1.79\times10^6$          & $\sim 2894 - 4845$  & 0.419                    \\
Pot. I    & B     & 32      & 70000          & $\sim 339 - 598$            & $\sim 1.79\times10^6$          & $\sim 2997 - 5287$  & 0.42                     \\
Pot. I    & C     & 32      & 70000          & $\sim 337 - 578$            & $\sim 1.79\times10^6$          & $\sim 3099 - 5314$  & 0.423                    \\
Pot. II   & A     & 48      & 240000         & $\sim 2090 - 3748$          & $\sim 9.2\times10^6$           & $\sim 2459 - 4409$  & 0.181                    \\
Pot. II   & B     & 48      & 240000         & $\sim 1873 - 3877$          & $\sim 9.2\times10^6$           & $\sim 2377 - 4920$  & 0.177                    \\
Pot. II   & C     & 48      & 240000         & $\sim 2080 - 3655$          & $\sim 9.2\times10^6$           & $\sim 2521 - 4432$  & 0.178                    \\
Pot. III  & A     & 40      & 105000         & $\sim 276 - 615$            & $\sim 3.36\times10^6$          & $\sim 5468 - 12189$ & 0.34                     \\
Pot. III  & B     & 40      & 80000          & $\sim 267 - 707$            & $\sim 2.56\times10^6$          & $\sim 3621 - 9590$  & 0.338                    \\
Pot. III  & C     & 40      & 80000          & $\sim 261 - 793$            & $\sim 2.56\times10^6$          & $\sim 3228 - 9824$  & 0.34                     \\ \bottomrule
\end{tabular}
\end{table}

As before, the thinning factor for each potential (Chain A) was chosen such that the retained posterior contained approximately $5000$ samples for further diagnostics and uncertainty propagation. The resulting thinned chains contain $5006$ samples (thinning factor $358$) for Potential I, $5001$ samples (thinning factor $1843$) for Potential II, and $5000$ samples (thinning factor $672$) for Potential III. Model comparison was repeated using WAIC and PSIS-LOO. The corresponding values are listed in Table~\ref{tab:criteriangauss}. The resulting diagnostics remain statistically comparable among the three potentials and are consistent with the conclusions obtained using uniform priors.

\begin{table}
\centering
\caption{WAIC and PSIS-LOO diagnostics computed from the thinned posterior samples for Potentials I–III, with the Gaussian priors described in Sec.~\ref{sec:2c_MCMC}, evaluated both including and excluding the $M_{B_c(2S)}-M_{B_c(1S)}$ constraint.}
\label{tab:criteriangauss}
\begin{tabular}{@{}ccccccc@{}}
\toprule
Potential & Inputs                            & WAIC            & $p_{\text{WAIC}}$ & $\text{ELPD}_{\text{LOO}}$ & $p_{\text{LOO}}$ & max $k$ \\ \midrule
Pot. I    & Full                              & $4.92 \pm 1.06$ & $1.335$           & $-3.22 \pm 1.76$           & $2.091$          & $0.87$  \\
Pot. I    & Without $M_{B_c(2S)}-M_{B_c(1S)}$ & $4.25 \pm 0.84$ & $1.208$           & $-2.87 \pm 1.41$           & $1.949$          & $0.87$  \\
Pot. II   & Full                              & $4.64 \pm 0.94$ & $1.201$           & $-2.82 \pm 1.36$           & $1.702$          & $0.774$ \\
Pot. II   & Without $M_{B_c(2S)}-M_{B_c(1S)}$ & $3.97 \pm 0.74$ & $1.074$           & $-2.47 \pm 1.09$           & $1.555$          & $0.774$ \\
Pot. III  & Full                              & $4.59 \pm 0.91$ & $1.199$           & $-2.8 \pm 1.35$            & $1.7$            & $0.827$ \\
Pot. III  & Without $M_{B_c(2S)}-M_{B_c(1S)}$ & $3.93 \pm 0.72$ & $1.067$           & $-2.45 \pm 1.08$           & $1.551$          & $0.827$ \\ \bottomrule
\end{tabular}
\end{table}

\begin{figure}
    \centering
    \includegraphics[width=0.8\linewidth]{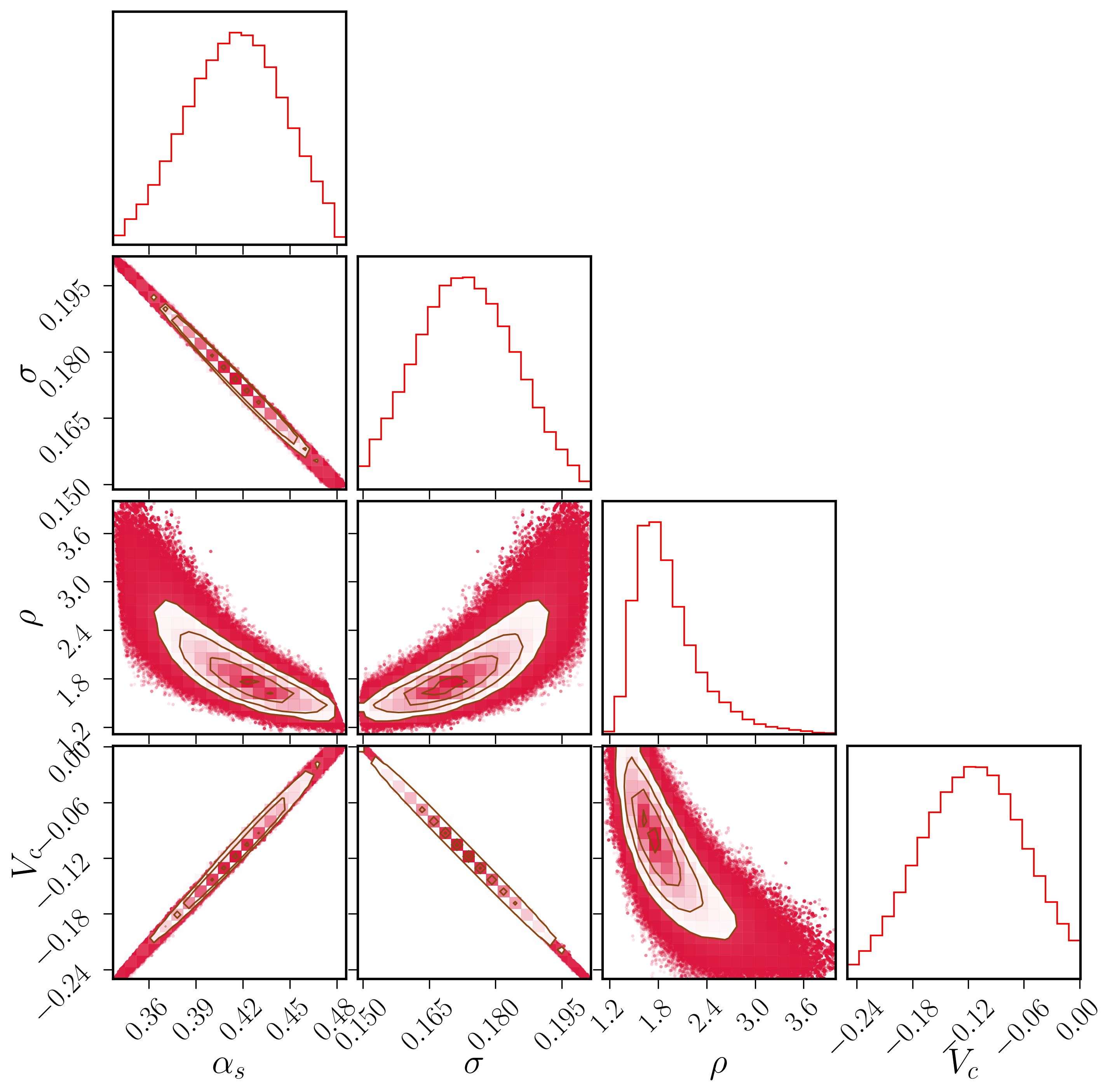}
    \caption{Corner plot for Cornell Potential (Potential I) with the Gaussian priors described in Sec.~\ref{sec:2c_MCMC}.}
    \label{fig:CornerCornelGauss}
\end{figure}

\begin{figure}
    \centering
    \includegraphics[width=\linewidth]{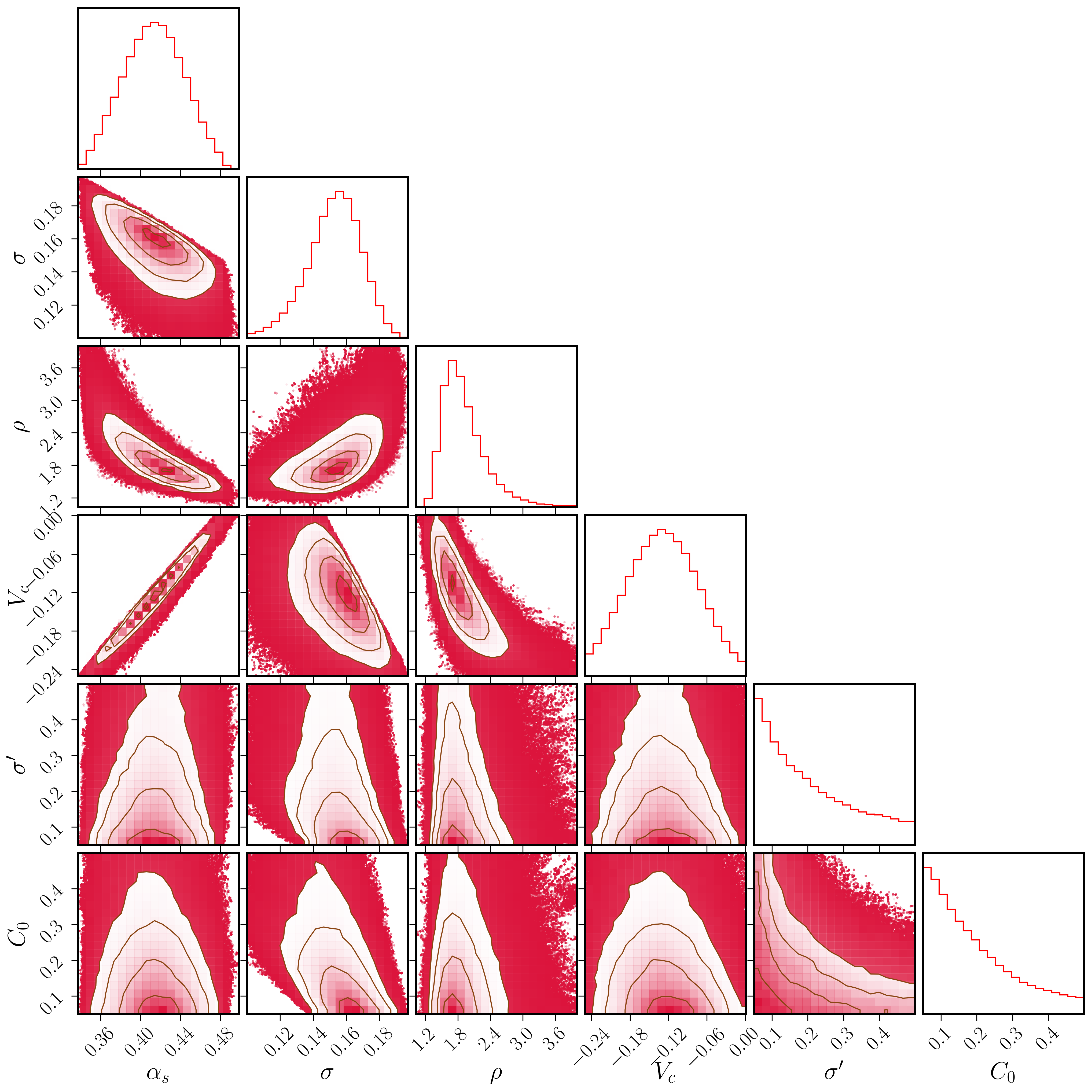}
    \caption{Corner plot for Modified Cornell Potential (Potential II) with the Gaussian priors described in Sec.~\ref{sec:2c_MCMC}.}
    \label{fig:CornerModCornelGauss}
\end{figure}

\begin{figure}
    \centering
    \includegraphics[width=\linewidth]{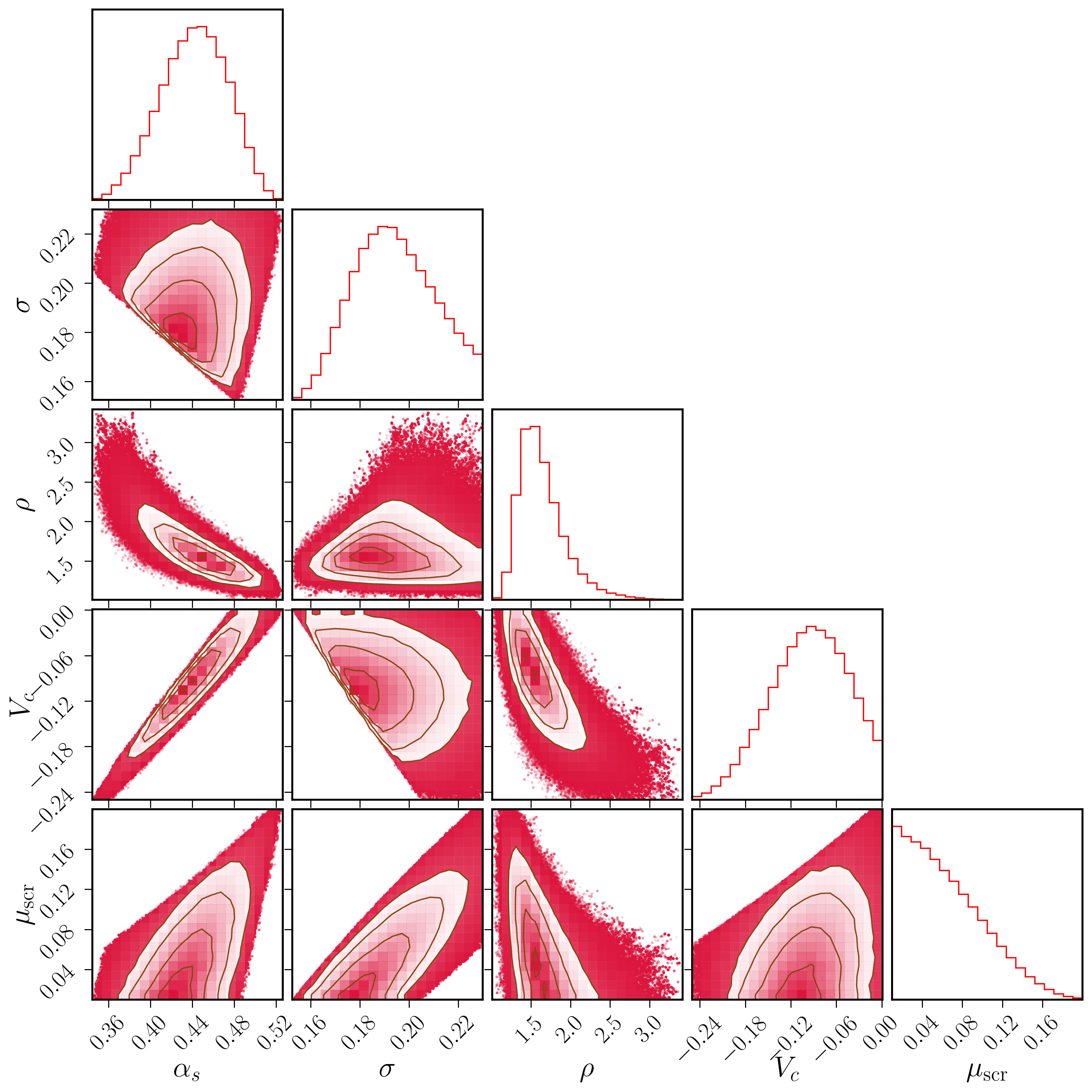}
    \caption{Corner plot for Screened Cornell Potential (Potential III) with the Gaussian priors described in Sec.~\ref{sec:2c_MCMC}.}
    \label{fig:CornerScrCornelGauss}
\end{figure}

The posterior medians and $68\%$ credible intervals are listed in Table~\ref{tab:parametersGauss}. As expected, introducing Gaussian priors on $\alpha_s$, $\sigma$, and $V_c$ shifts the corresponding posterior medians while reducing the associated credible intervals compared to the uniform prior results (Table.~\ref{tab:Bc_parameters}). Similar trends are observed across all three potentials, with slightly larger $\alpha_s$, less negative $V_c$, and reduced values of the unconstrained smearing parameter $\rho$. The latter reflects posterior correlations among the potential parameters rather than a direct prior constraint on $\rho$. The parameter $\sigma$ also shifts under the Gaussian prior, although the direction and magnitude of the shift depend on the potential. Potential I shows the strongest prior dependence because it contains the fewest free parameters. In Potential II, the correlated parameter $\sigma'$ and $C_0$ shift toward smaller values despite retaining uniform priors, reflecting their correlation with $\sigma$. By contrast, the screening parameter $\mu_{\rm scr}$ in Potential III is largely unchanged, indicating weak sensitivity to the adopted Gaussian priors. These trends are evident in the corner plots (Figs.~\ref{fig:CornerCornelGauss}-\ref{fig:CornerScrCornelGauss}), which show more localized, nearly symmetric, and narrower posteriors under the Gaussian-prior analysis, while the parameter correlations become more pronounced.

\begin{table}
\centering
\caption{Potential parameters in $B_c$ system obtained by considering the Gaussian prior defined in Sec.~\ref{sec:2c_MCMC}. Central values denote the posterior median (50th percentile). The asymmetric uncertainties correspond to the 16th and 84th percentiles, which are also given explicitly in brackets.}
\label{tab:parametersGauss}
\begin{tabular}{@{}cccc@{}}
\toprule
Parameters               & Potential - I                                  & Potential - II                                  & Potential – III                                 \\ \midrule
$\alpha_s$               & $0.416_{-0.032}^{+0.03}$ [$0.384$, $0.446$]    & $0.414_{-0.031}^{+0.031}$ [$0.382$, $0.444$]    & $0.443_{-0.034}^{+0.032}$ [$0.409$, $0.475$]    \\
$\sigma$ (in GeV${}^2$)  & $0.173_{-0.011}^{+0.012}$ [$0.162$, $0.185$]   & $0.154_{-0.017}^{+0.015}$ [$0.137$, $0.169$]    & $0.193_{-0.016}^{+0.019}$ [$0.178$, $0.212$]    \\
$\rho$ (in GeV)          & $1.835_{-0.276}^{+0.464}$ [$1.559$, $2.299$]   & $1.838_{-0.28}^{+0.447}$ [$1.558$, $2.286$]     & $1.587_{-0.209}^{+0.323}$ [$1.378$, $1.909$]    \\
$V_c$ (in GeV)           & $-0.115_{-0.056}^{+0.054}$ [$-0.171$, $-0.06$] & $-0.129_{-0.056}^{+0.056}$ [$-0.185$, $-0.073$] & $-0.094_{-0.055}^{+0.051}$ [$-0.149$, $-0.043$] \\
$\sigma'$ (in GeV)       & -                                              & $0.173_{-0.094}^{+0.177}$ [$0.079$, $0.35$]     & -                                               \\
$C_0$ (in GeV)           & -                                              & $0.161_{-0.082}^{+0.154}$ [$0.079$, $0.316$]    & -                                               \\
$\mu_{\rm scr}$ (in GeV) & -                                              & -                                               & $0.057_{-0.033}^{+0.048}$ [$0.024$, $0.105$]    \\ \bottomrule
\end{tabular}
\end{table}

\begin{figure}
    \centering
    \includegraphics[width=\linewidth]{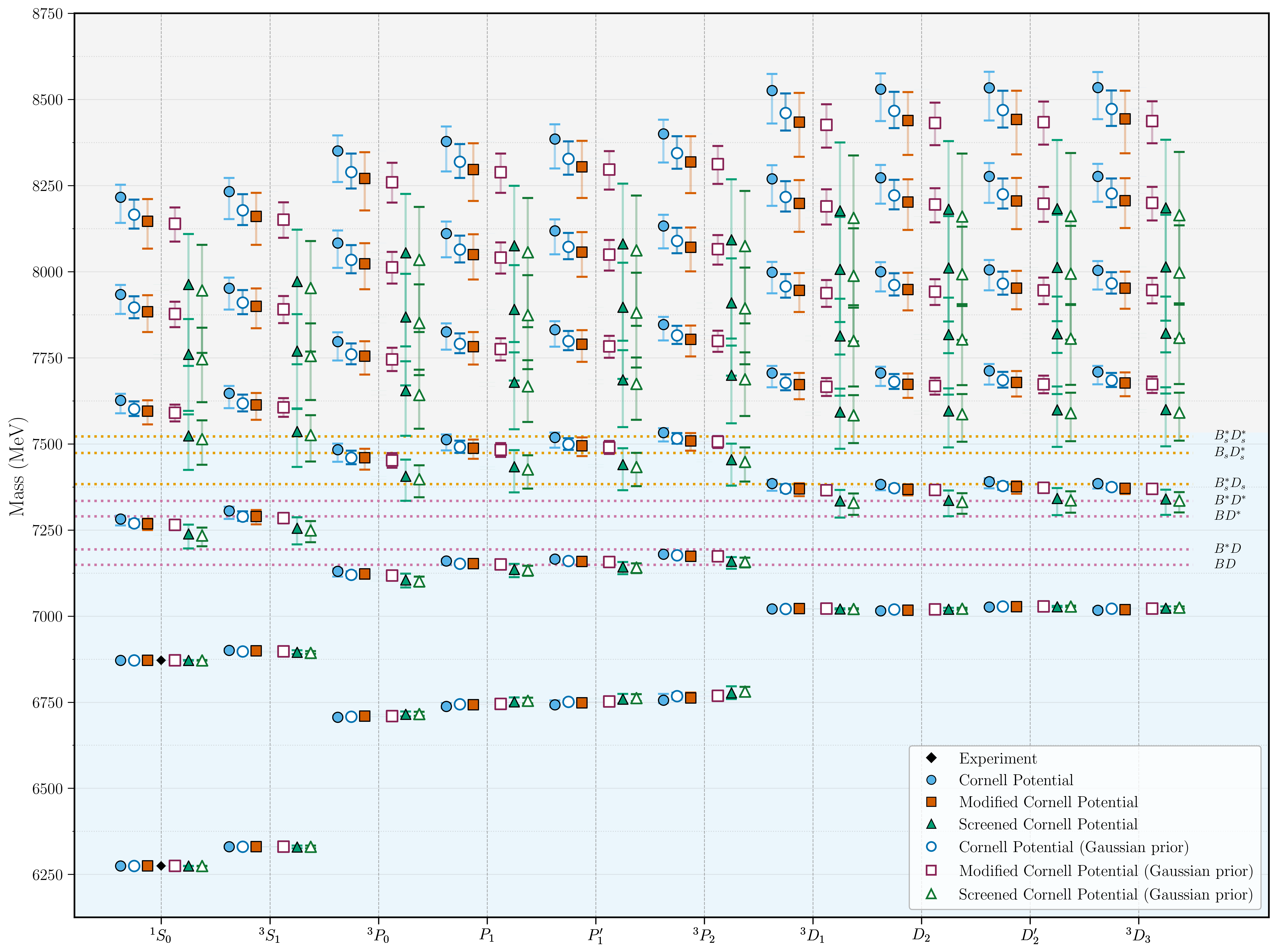}
    \caption{The $B_c$ mass spectrum (in MeV) computed using Cornell (Potential I), modified Cornell (Potential II), and screened Cornell (Potential III) potentials with uniform prior in blue circle, red square, and green triangle, respectively. Corresponding states resulting from Gaussian prior are denoted with hollow shapes. The black diamonds are experimental values \cite{ParticleDataGroup:2024cfk}. The dotted lines indicate the thresholds for open-flavor $B_{(s)}^{(*)} D_{(s)}^{(*)}$ channels.}
    \label{fig:MassSpectraBc_uniandgauss}
\end{figure}

The $B_c$ mass spectra obtained with the Gaussian priors are listed in Table~\ref{tab:masstableGauss}. Despite noticeable shifts in several posterior parameters, the predicted masses remain remarkably stable across all three potentials owing to correlations among the fitted parameters. The ordering of the states and the overall level spacings remain mostly unchanged, while the credible intervals become slightly narrower owing to the additional prior information. The largest deviations occur in the $P$-wave mixing angles, whereas the masses of both low-lying and higher excited states change only marginally. The resulting spectra for all six potential-prior combinations are shown in Fig.~\ref{fig:MassSpectraBc_uniandgauss}.

Overall, the Gaussian-prior analysis leads to the same qualitative physical conclusions as the uniform-prior analysis. Across the six potential–prior combinations, the low-lying 1S and 1D masses remain within a few MeV, whereas the uncertainty bands grow progressively for higher radial excitations. Therefore, the results presented in the main text are based on the uniform-prior analysis, which represents the least informative prior assumption, while the Gaussian-prior analysis serves as a robustness check demonstrating that the principal spectroscopic conclusions are less sensitive to the adopted priors.

\begin{turnpage}
\begingroup
\squeezetable
\begin{table}
\centering
\caption{Mass spectra of $B_c$ mesons (in MeV) resulting from the Gaussian prior defined in Sec.~\ref{sec:2c_MCMC}.}
\label{tab:masstableGauss}
\begin{tabular}{@{}c|ccc|cccc|cccc@{}}
\toprule
State        & Potential - I               & Potential - II              & Potential – III               & State         & Potential - I               & Potential - II              & Potential – III               & State         & Potential - I               & Potential - II              & Potential – III               \\ \midrule
${1{}^1S_0}$ & $6274.47_{-0.31}^{+0.31}$   & $6274.48_{-0.32}^{+0.3}$    & $6274.47_{-0.31}^{+0.31}$     & ${1{}^3P_0}$  & $6708_{-2.9}^{+3.04}$       & $6709.42_{-3.1}^{+3.43}$    & $6715.58_{-5.2}^{+6.46}$      & ${1{}^3D_1}$  & $7021.12_{-1.88}^{+2}$      & $7021.33_{-1.91}^{+1.92}$   & $7021.1_{-1.63}^{+1.71}$      \\
${1{}^3S_1}$ & $6330.18_{-4.03}^{+4.05}$   & $6330.09_{-4.11}^{+4.12}$   & $6330.12_{-4.07}^{+3.86}$     & ${1{}P_1}$    & $6743.75_{-5.54}^{+5.61}$   & $6744.91_{-5.67}^{+5.86}$   & $6753.94_{-7.79}^{+8.88}$     & ${1{}D_2}$    & $7018.97_{-3.5}^{+3.36}$    & $7019.05_{-3.44}^{+3.24}$   & $7021.48_{-3.16}^{+2.5}$      \\
${2{}^1S_0}$ & $6871.17_{-0.72}^{+0.74}$   & $6871.18_{-0.72}^{+0.75}$   & $6871.18_{-0.71}^{+0.73}$     & ${1{}P_1'}$   & $6750.99_{-7.21}^{+7.3}$    & $6751.92_{-7.21}^{+7.56}$   & $6762.38_{-9.48}^{+10.53}$    & ${1{}D_2'}$   & $7027.66_{-2.32}^{+2.3}$    & $7027.73_{-2.25}^{+2.17}$   & $7028.16_{-1.81}^{+1.8}$      \\
${2{}^3S_1}$ & $6897.77_{-3.67}^{+4.36}$   & $6897.44_{-3.71}^{+4.11}$   & $6893.8_{-3.76}^{+4.27}$      & ${1{}^3P_2}$  & $6767.94_{-9.89}^{+9.55}$   & $6768.55_{-9.9}^{+9.91}$    & $6781.36_{-12.12}^{+12.76}$   & ${1{}^3D_3}$  & $7021.68_{-4.4}^{+4.24}$    & $7021.68_{-4.39}^{+4.13}$   & $7025_{-4.07}^{+3.23}$        \\
${3{}^1S_0}$ & $7269.74_{-10.13}^{+10.74}$ & $7264.99_{-12.1}^{+11.21}$  & $7233.82_{-31.04}^{+22.96}$   & $\theta_{1P}$ & $(11.98_{-5.92}^{+3.16})$   & $(11.53_{-6.04}^{+3.44})$   & $(14.61_{-4.04}^{+2.29})$     & $\theta_{1D}$ & $(-51.54_{-0.84}^{+1})$     & $(-51.63_{-0.79}^{+1.05})$  & $(-50.63_{-1.12}^{+1.58})$    \\
${3{}^3S_1}$ & $7289.4_{-12.8}^{+14.53}$   & $7284.17_{-14.62}^{+14.72}$ & $7248.87_{-34.23}^{+26.46}$   & ${2{}^3P_0}$  & $7119.64_{-8.3}^{+9.55}$    & $7117.86_{-8.84}^{+9.21}$   & $7101.14_{-15.66}^{+13.59}$   & ${2{}^3D_1}$  & $7370.36_{-11.54}^{+12.68}$ & $7365.1_{-13.5}^{+13.12}$   & $7329.31_{-35.86}^{+26.49}$   \\
${4{}^1S_0}$ & $7601.24_{-20.64}^{+22.14}$ & $7590_{-25.18}^{+23.45}$    & $7513.76_{-74.18}^{+54.11}$   & ${2{}P_1}$    & $7152.41_{-6.17}^{+6.95}$   & $7149.91_{-6.94}^{+6.96}$   & $7133.18_{-16.66}^{+12.7}$    & ${2{}D_2}$    & $7371.31_{-9.41}^{+10.15}$  & $7365.88_{-11.42}^{+10.77}$ & $7331.85_{-34.72}^{+24.7}$    \\
${4{}^3S_1}$ & $7617.7_{-23.3}^{+25.56}$   & $7605.75_{-27.33}^{+26.76}$ & $7525.63_{-77.31}^{+57.28}$   & ${2{}P_1'}$   & $7159.91_{-5}^{+5.71}$      & $7157.11_{-5.86}^{+5.95}$   & $7140.86_{-16.35}^{+12.15}$   & ${2{}D_2'}$   & $7377.54_{-10.65}^{+11.61}$ & $7372.04_{-12.58}^{+12.16}$ & $7336.33_{-36.02}^{+26.14}$   \\
${5{}^1S_0}$ & $7895.82_{-31.09}^{+32.93}$ & $7876.99_{-38.27}^{+35.4}$  & $7745.88_{-125.2}^{+91.3}$    & ${2{}^3P_2}$  & $7176.65_{-3.26}^{+3.46}$   & $7173.59_{-4.56}^{+4.02}$   & $7157.9_{-16.54}^{+11.4}$     & ${2{}^3D_3}$  & $7374.99_{-8.56}^{+9.14}$   & $7369.47_{-10.64}^{+9.74}$  & $7335.73_{-34.48}^{+24.43}$   \\
${5{}^3S_1}$ & $7909.93_{-33.4}^{+36.5}$   & $7890.71_{-40.43}^{+38.45}$ & $7755.63_{-128.52}^{+94.3}$   & $\theta_{2P}$ & $(16.32_{-3.53}^{+2.02})$   & $(16.09_{-3.53}^{+2.13})$   & $(18_{-2.46}^{+1.34})$        & $\theta_{2D}$ & $(-50.47_{-0.97}^{+1.25})$  & $(-50.58_{-0.95}^{+1.25})$  & $(-49.37_{-1.37}^{+2.14})$    \\
${6{}^1S_0}$ & $8165.64_{-40.98}^{+43.4}$  & $8138.9_{-51.27}^{+47.1}$   & $7945.5_{-181.77}^{+132.32}$  & ${3{}^3P_0}$  & $7459.52_{-18.82}^{+20.87}$ & $7451.85_{-21.39}^{+21.37}$ & $7397.34_{-52.44}^{+40.1}$    & ${3{}^3D_1}$  & $7677.47_{-22.34}^{+24.49}$ & $7665.58_{-26.86}^{+25.71}$ & $7583.02_{-80.8}^{+58.67}$    \\
${6{}^3S_1}$ & $8178.28_{-43.09}^{+46.93}$ & $8151.12_{-53.13}^{+50.1}$  & $7953.02_{-185.09}^{+135.66}$ & ${3{}P_1}$    & $7490.87_{-17.1}^{+18.65}$  & $7482.43_{-20.15}^{+19.43}$ & $7425.74_{-55.82}^{+40.98}$   & ${3{}D_2}$    & $7680.41_{-20.61}^{+22.19}$ & $7668.33_{-25.34}^{+23.62}$ & $7586.56_{-80.51}^{+58.02}$   \\
-            & -                           & -                           & -                             & ${3{}P_1'}$   & $7498.74_{-15.99}^{+17.34}$ & $7489.91_{-19.24}^{+18.63}$ & $7432.91_{-56.13}^{+40.96}$   & ${3{}D_2'}$   & $7685.13_{-21.64}^{+23.56}$ & $7672.96_{-26.37}^{+24.84}$ & $7589.63_{-81.58}^{+59}$      \\
-            & -                           & -                           & -                             & ${3{}^3P_2}$  & $7515.43_{-14.53}^{+15.46}$ & $7506.19_{-18.18}^{+16.73}$ & $7448.07_{-57.61}^{+41.49}$   & ${3{}^3D_3}$  & $7684.72_{-19.77}^{+21.24}$ & $7672.46_{-24.74}^{+22.87}$ & $7590.7_{-81.07}^{+57.69}$    \\
-            & -                           & -                           & -                             & $\theta_{3P}$ & $(18.34_{-2.65}^{+1.41})$   & $(18.16_{-2.6}^{+1.49})$    & $(19.44_{-1.71}^{+0.85})$     & $\theta_{3D}$ & $(-49.37_{-1.16}^{+1.6})$   & $(-49.52_{-1.11}^{+1.58})$  & $(-48_{-1.72}^{+3.05})$       \\
-            & -                           & -                           & -                             & ${4{}^3P_0}$  & $7759.92_{-29.22}^{+31.84}$ & $7745.33_{-34.23}^{+33.37}$ & $7642.33_{-98.81}^{+73.07}$   & ${4{}^3D_1}$  & $7957.18_{-32.71}^{+35.46}$ & $7937.79_{-40.08}^{+37.68}$ & $7799.35_{-132.8}^{+96.26}$   \\
-            & -                           & -                           & -                             & ${4{}P_1}$    & $7790.4_{-27.54}^{+29.97}$  & $7774.87_{-33.23}^{+31.67}$ & $7667.16_{-103.42}^{+75.79}$  & ${4{}D_2}$    & $7961.38_{-30.9}^{+33.48}$  & $7941.84_{-38.9}^{+35.84}$  & $7803.52_{-133.22}^{+96.31}$  \\
-            & -                           & -                           & -                             & ${4{}P_1'}$   & $7798.46_{-26.48}^{+29.02}$ & $7782.83_{-32.86}^{+30.74}$ & $7674.47_{-104.89}^{+75.84}$  & ${4{}D_2'}$   & $7965.07_{-31.93}^{+34.65}$ & $7945.37_{-39.7}^{+36.98}$  & $7805.72_{-134.11}^{+97.18}$  \\
-            & -                           & -                           & -                             & ${4{}^3P_2}$  & $7815.16_{-25.17}^{+27.17}$ & $7798.81_{-31.73}^{+29.37}$ & $7688.07_{-107.16}^{+77.68}$  & ${4{}^3D_3}$  & $7966.13_{-30.14}^{+32.61}$ & $7946.43_{-38.23}^{+35.12}$ & $7807.86_{-134.41}^{+96.18}$  \\
-            & -                           & -                           & -                             & $\theta_{4P}$ & $(19.46_{-2.14}^{+1.01})$   & $(19.31_{-2.07}^{+1.08})$   & $(20.19_{-1.25}^{+0.56})$     & $\theta_{4D}$ & $(-48.17_{-1.44}^{+2.12})$  & $(-48.39_{-1.37}^{+2.06})$  & $(-46.39_{-2.25}^{+4.47})$    \\
-            & -                           & -                           & -                             & ${5{}^3P_0}$  & $8034.27_{-39.09}^{+42.54}$ & $8012.18_{-46.96}^{+44.99}$ & $7851.6_{-151.74}^{+111.44}$  & ${5{}^3D_1}$  & $8216.81_{-42.44}^{+46.05}$ & $8189.63_{-52.96}^{+49.23}$ & $7988.12_{-190.11}^{+137.81}$ \\
-            & -                           & -                           & -                             & ${5{}P_1}$    & $8064.27_{-37.74}^{+40.64}$ & $8041.05_{-46.32}^{+43.45}$ & $7874.42_{-157.95}^{+115.11}$ & ${5{}D_2}$    & $8222.09_{-40.92}^{+44.14}$ & $8194.6_{-51.64}^{+47.61}$  & $7993_{-191.24}^{+137.47}$    \\
-            & -                           & -                           & -                             & ${5{}P_1'}$   & $8072.65_{-36.75}^{+39.68}$ & $8049.12_{-45.86}^{+42.7}$  & $7881.11_{-159.64}^{+115.91}$ & ${5{}D_2'}$   & $8224.99_{-41.79}^{+45.25}$ & $8197.62_{-52.68}^{+48.39}$ & $7994.44_{-191.99}^{+138.54}$ \\
-            & -                           & -                           & -                             & ${5{}^3P_2}$  & $8089.23_{-35.53}^{+38.03}$ & $8065.17_{-45.08}^{+41.21}$ & $7893.97_{-163.05}^{+117.39}$ & ${5{}^3D_3}$  & $8227.23_{-40.2}^{+43.33}$  & $8199.47_{-51.15}^{+46.9}$  & $7997.23_{-192.48}^{+137.39}$ \\
-            & -                           & -                           & -                             & $\theta_{5P}$ & $(20.14_{-1.77}^{+0.71})$   & $(20.02_{-1.72}^{+0.78})$   & $(20.62_{-0.95}^{+0.36})$     & $\theta_{5D}$ & $(-46.8_{-1.83}^{+2.94})$   & $(-47.1_{-1.71}^{+2.81})$   & $(-44.43_{-3.02}^{+6.63})$    \\
-            & -                           & -                           & -                             & ${6{}^3P_0}$  & $8289.57_{-48.5}^{+52.95}$  & $8259.73_{-59.33}^{+56.28}$ & $8034.43_{-209.46}^{+153.75}$ & ${6{}^3D_1}$  & $8461.1_{-51.83}^{+56.43}$  & $8425.92_{-65.49}^{+60.25}$ & $8156.15_{-253.64}^{+180.97}$ \\
-            & -                           & -                           & -                             & ${6{}P_1}$    & $8319.19_{-47.19}^{+51.21}$ & $8288.27_{-59.17}^{+54.75}$ & $8056.18_{-217.28}^{+157.55}$ & ${6{}D_2}$    & $8467.25_{-50.47}^{+54.56}$ & $8431.62_{-64.29}^{+58.8}$  & $8160.7_{-255.15}^{+181.72}$  \\
-            & -                           & -                           & -                             & ${6{}P_1'}$   & $8327.79_{-46.26}^{+50.15}$ & $8296.42_{-58.43}^{+53.84}$ & $8062.3_{-219.29}^{+158.6}$   & ${6{}D_2'}$   & $8469.52_{-51.26}^{+55.6}$  & $8433.88_{-65.08}^{+59.65}$ & $8161.78_{-255.82}^{+182.58}$ \\
-            & -                           & -                           & -                             & ${6{}^3P_2}$  & $8344.41_{-45.31}^{+48.69}$ & $8312.26_{-57.92}^{+52.64}$ & $8074.33_{-224.2}^{+160.31}$  & ${6{}^3D_3}$  & $8472.63_{-49.74}^{+53.7}$  & $8436.95_{-64.12}^{+57.86}$ & $8164.41_{-256.44}^{+182.86}$ \\
-            & -                           & -                           & -                             & $\theta_{6P}$ & $(20.59_{-1.48}^{+0.48})$   & $(20.48_{-1.44}^{+0.54})$   & $(20.87_{-0.71}^{+0.25})$     & $\theta_{6D}$ & $(-45.15_{-2.38}^{+4.27})$  & $(-45.57_{-2.19}^{+4.03})$  & $(-41.96_{-4.11}^{+9.74})$    \\ \bottomrule
\end{tabular}
\end{table}
\endgroup
\end{turnpage}

\FloatBarrier
\newpage 
\bibliography{refs.bib}
\newpage
\FloatBarrier
\begin{table}
\centering
\caption{Potential parameters in $B_c$ system. Central values denote the posterior median (50th percentile). The asymmetric uncertainties correspond to the 16th and 84th percentiles, which are also given explicitly in brackets.}
\label{tab:Bc_parameters}
\begin{tabular}{@{}cccc@{}}
\toprule
Parameters              & Potential I & Potential II & Potential III \\ \midrule
$\alpha_s$              & $0.378_{-0.028}^{+0.056}$ [$0.350$, $0.434$]    &     $0.388_{-0.029}^{+0.051}$ [$0.359$, $0.439$] & $0.42_{-0.051}^{+0.056}$ [$0.369$, $0.476$] \\
$\sigma$ (in GeV${}^2$) & $0.187_{-0.020}^{+0.010}$ [$0.167$, $0.197$]     &     $0.150_{-0.028}^{+0.024}$ [$0.122$, $0.174$] & $0.206_{-0.02}^{+0.016}$ [$0.186$, $0.221$] \\
$\rho$ (in GeV)         & $2.284_{-0.618}^{+0.795}$ [$1.666$, $3.079$]    &     $2.085_{-0.477}^{+0.686}$ [$1.608$, $2.771$] & $1.742_{-0.358}^{+0.686}$ [$1.383$, $2.428$] \\
$V_c$ (in GeV)          & $-0.182_{-0.050}^{+0.099}$ [$-0.232$, $-0.083$]    &     $-0.185_{-0.049}^{+0.095}$ [$-0.234$, $-0.090$] & $-0.138_{-0.081}^{+0.093}$ [$-0.219$, $-0.045$] \\ 
$\sigma'$ (in GeV) & -                         & $0.220_{-0.120}^{+0.166}$ [$0.100$, $0.386$] & - \\
$C_0$ (in GeV)          & -                          & $0.225_{-0.121}^{+0.159}$ [$0.104$, $0.384$] & - \\
$\mu_{\rm scr}$ (in GeV)  & - & - & $0.058_{-0.034}^{+0.054}$ [$0.024$, $0.112$] \\
\bottomrule
\end{tabular}
\end{table}

\begin{turnpage}
\begingroup
\squeezetable
\begin{table}
\centering
\caption{Mass spectra of $S$-wave $B_c$ mesons in MeV. States above $n=3$ exceed $\langle v_c^2\rangle\sim0.5$.}
\label{tab:Bcmass_S}
\begin{tabular}{@{}cccccccccccccccccccc@{}}
\toprule
State    & Potential I & Potential II  &Potential III& \makecell{PDG \\ \cite{ParticleDataGroup:2024cfk}} & \makecell{LQCD \\ \cite{Mathur:2018epb}} & \makecell{LQCD \\ \cite{Dowdall:2012ab}} & \makecell{GI \\ \cite{Godfrey:2004ya}} & \makecell{EFG \\ \cite{Ebert:2002pp, Ebert:2011jc}} & \makecell{LLWL \\ \cite{Li:2023wgq}} & \makecell{WWLC \\ \cite{Wang:2022cxy}} & \makecell{MBK \\ \cite{Monteiro:2016rzi}}  & \makecell{BB \\ \cite{Bokade:2025lmn}}&\makecell{PLCR \\ \cite{Patel:2026prg}}& \makecell{EQ \\ \cite{Eichten:2019gig}} & \makecell{LZ \\ \cite{Li:2019tbn}} & \makecell{AAMS \\ \cite{Asghar:2019qjl}} & \makecell{DKR \\ \cite{Devlani:2014nda}} & \makecell{LTFWP \\ \cite{Li:2022bre}}  &\makecell{HZ \\ \cite{Hao:2024nqb}}\\ \midrule
$1^1S_0$ & $6274.48_{-0.3}^{+0.31}$      &      $6274.47_{-0.31}^{+0.31}$  &$6274.47_{-0.31}^{+0.32}$& $6274.47 (0.32)$\footnotemark[1]                      & $6276(3)(6)$              & $6278(9)$                   & 6271                     & 6272                                  & 6271                   & 6277                     & 6275                         & 6275.8&6275& 6275                       & 6271                     & 6318                       & 6278                       & 6269                     &6272\\
$1^3S_1$ & $6330.19_{-3.99}^{+3.72}$      &      $6330.12_{-3.93}^{+3.92}$  &$6329.74_{-3.9}^{+4.04}$& -                                    & $6331(4)(6)$              & $6332(9)$                   & 6338                     & 6333                                  & 6338                   & 6332                     & 6314                         & 6272&6332& 6329                       & 6326                     & 6338                       & 6331                       & 6322                     &6335\\
$2^1S_0$ & $6871.2_{-0.74}^{+0.71}$      &      $6871.17_{-0.72}^{+0.75}$  &$6871.17_{-0.71}^{+0.75}$& $6871.2 (1.0)$\footnotemark[1]                       & -                         & $6894(19)(8)$               & 6855                     & 6842                                  & 6855                   & 6867                     & 6838                         & 6869.8&6858& 6866.5                     & 6871                     & 6741                       & 6863                       & 6886                     &6874\\
$2^3S_1$ & $6900.82_{-5.19}^{+4.35}$      &      $6898.85_{-4.63}^{+4.78}$  &$6894.77_{-5.14}^{+5.57}$& -                                    & -                         & $6922(19)(8)$               & 6887                     & 6882                                  & 6886                   & 6911                     & 6850                         & 6905.6&6868& 6897.5                     & 6890                     & 6747                       & 6873                       & 6907                     &6897\\
$3^1S_0$ & $7282.2_{-18.74}^{+8.83}$      &      $7268.11_{-18.64}^{+14.54}$  &$7238.66_{-42.23}^{+27.14}$& -                                    & -                         & -                           & 7250                     & 7226                                  & 7220                   & 7228                     & -                            & 7252.1&7223& 7253.5                     & 7239                     & 7014                       & 7244                       & 7261                     &7252\\
$3^3S_1$ & $7305.87_{-24.04}^{+11.85}$      &      $7289.32_{-23.23}^{+18.37}$  &$7254.91_{-46.85}^{+31.74}$& -                                    & -                         & -                           & 7272                     & 7258                                  & 7240                   & 7272                     & -                            & 7278.4&7227& 7279.5                     & 7252                     & 7018                       & 7249                       & 7275                     &7281\\
$4^1S_0$ & $7626.76_{-38}^{+18.35}$      &      $7595.35_{-39.17}^{+31.03}$  &$7523.22_{-98.6}^{+62}$& -                                    & -                         & -                           & -                        & 7585                                  & 7496                   & -                        & -                            & 7545.1&7514& 7572.5                     & 7540                     & 7239                       & 7564                       & 7551                     &-\\
$4^3S_1$ & $7647.08_{-43.33}^{+21.32}$      &      $7612.72_{-42.72}^{+34.93}$  &$7535.7_{-102.9}^{+67.05}$& -                                    & -                         & -                           & -                        & 7609                                  & 7512                   & -                        & -                            & 7565.9&7517& 7595.5                     & 7550                     & 7242                       & 7568                       & 7561                     &-\\
$5^1S_0$ & $7933.99_{-56.71}^{+27.47}$      &      $7883.8_{-59.34}^{+47.79}$  &$7759.9_{-163.87}^{+102.26}$& -                                    & -                         & -                           & -                        & 7928                                  & 7722                   & -                        & -                            & 7785.4&7765& -                          & 7805                     & -                          & 7852                       & 7790                     &-\\
$5^3S_1$ & $7952.09_{-62.19}^{+30.69}$      &      $7899.24_{-63.39}^{+51.58}$  &$7769.72_{-168.28}^{+106.54}$& -                                    & -                         & -                           & -                        & 7947                                  & 7735                   & -                        & -                            & 7802.6&7767& -                          & 7813                     & -                          & 7855                       & 7798                     &-\\
$6^1S_0$ & $8216.02_{-74.89}^{+36.22}$      &      $8146.37_{-79.36}^{+64.31}$  &$7963.04_{-236.73}^{+146.58}$& -                                    & -                         & -                           & -                        & -                                     & -                      & -                        & -                            & -&7988& -                          & 8046                     & -                          & 8120                       & 7994                     &-\\
$6^3S_1$ & $8232.47_{-80.35}^{+39.46}$      &      $8160.12_{-82.67}^{+68.33}$  &$7971.76_{-241.11}^{+150.42}$& -                                    & -                         & -                           & -                        & -                                     & -                      & -                        & -                            & -&7990& -                          & 8054                     & -                          & 8122                       & 8001                     &-\\ \bottomrule
\end{tabular}
\footnotetext[1]{Used as input, along with $M_{B_c^*(1S)}-M_{B_c(1S)} = 55(3)$ MeV from LQCD~\cite{Mathur:2018epb}.}
\footnotetext[2]{GI: relativized quark model; EFG: relativistic quark model; LLWL: screened GI model; WWLC: full Salpeter equation; MBK: relativistic model with coupled-channel effects; BB: relativistic screened potential model; PLCR: semi-relativistic framework with screened potential; EQ, LZ, AAMS, DKR, LTFWP, and HZ: non-relativistic potential models.}
\end{table}
\endgroup
\end{turnpage}

\begin{turnpage}
\begingroup
\squeezetable
\begin{table}
\centering
\caption{Mass spectra of $P$-wave $B_c$ mesons in MeV. States above $n=3$ exceed $\langle v_c^2\rangle\sim0.5$.}
\label{tab:Bcmass_P}
\scalebox{0.94}{
\begin{tabular}{@{}ccccccccccccccccccc@{}}
\toprule
State         & Potential I   & Potential II        &Potential III& \makecell{LQCD \\ \cite{Mathur:2018epb}} & \makecell{LQCD \\ \cite{Dowdall:2012ab, Davies:1996gi}}                                    & \makecell{GI \\ \cite{Godfrey:2004ya}} & \makecell{EFG \\ \cite{Ebert:2002pp, Ebert:2011jc}} & \makecell{LLWL \\ \cite{Li:2023wgq}} & \makecell{WWLC \\ \cite{Wang:2022cxy}} & \makecell{MBK \\ \cite{Monteiro:2016rzi}}  & \makecell{BB \\ \cite{Bokade:2025lmn}}&\makecell{PLCR \\ \cite{Patel:2026prg}}& \makecell{EQ \\ \cite{Eichten:2019gig}} & \makecell{LZ \\ \cite{Li:2019tbn}} & \makecell{AAMS \\ \cite{Asghar:2019qjl}} & \makecell{DKR \\ \cite{Devlani:2014nda}} & \makecell{LTFWP \\ \cite{Li:2022bre}}  &\makecell{HZ \\ \cite{Hao:2024nqb}}\\ \midrule
$1^3P_0$      & $6706.33_{-2.93}^{+3.6}$      &      $6709.63_{-3.81}^{+4.31}$          &$6714.64_{-5.74}^{+8.31}$& $6712(18)(7)$              & $6707(14)(8)$      & 6706                     & 6699                                  & 6701                   & 6705                     & 6672                         & 6703.9&6742& 6692.5                     & 6714                     & 6631                       & 6748                       & 6712                     &6706\\
$1P_1$        & $6737.64_{-5.41}^{+9.55}$      &      $6742.31_{-6.7}^{+8.82}$          &$6751.48_{-11.02}^{+12.6}$& $6736(17)(7)$              & $6743(30)$          & 6741                     & 6743                                  & 6745                   & 6739                     & 6766                         & 6751.9&6758& 6730.5                     & 6757                     & 6650                       & 6767                       & 6761                     &6750\\
$1P_1'$       & $6742.93_{-6.63}^{+12.47}$      &      $6747.8_{-7.92}^{+11.72}$          &$6759.06_{-13.63}^{+15.21}$& -                          & $6765(30)$          & 6750                     & 6750                                  & 6754                   & 6748                     & 6828                         & 6744.8&6774& 6738.5                     & 6776                     & 6656                       & 6769                       & 6770                     &6785\\
$1^3P_2$      & $6756.28_{-9.12}^{+17.34}$      &      $6762.19_{-10.44}^{+16.1}$          &$6776.86_{-17.98}^{+18.79}$& -                          & $6783(30)$          & 6768                     & 6761                                  & 6773                   & 6762                     & 6776                         & 6772.6&6782& 6750.5                     & 6787                     & 6665                       & 6775                       & 6783                     &6802\\
$\theta_{1P}$ & $(4.82_{-10.01}^{+9.19})$$^{\circ}$      &      $(6.62_{-8.89}^{+7.83})$$^{\circ}$  &$(12.85_{-9.87}^{+4.32})$$^{\circ}$& -                          & $33.4(1.5)^{\circ}$ & 22.4$^{\circ}$           & 20.4$^{\circ}$                        & 35.2$^{\circ}$         & 32.2$^{\circ}$           & 0.4$^{\circ}$                & 11.8$^{\circ}$&-& 18.7$^{\circ}$             & 35.5$^{\circ}$           & 35.3$^{\circ}$             & 20.57$^{\circ}$            & $-24.3^{\circ}$          &$-35.0^{\circ}$\\
$2^3P_0$      & $7130.32_{-15.49}^{+7.89}$      &      $7122.31_{-14.01}^{+10.86}$         &$7104.79_{-22.1}^{+18.15}$& -                          & -                                        & 7122                     & 7094                                  & 7097                   & 7112                     & 6914                         & 7111.4&7126& 7104.5                     & 7107                     & 6915                       & 7139                       & 7118                     &7099\\
$2P_1$        & $7159.81_{-11.02}^{+5.58}$      &      $7152.1_{-10.87}^{+8.61}$         &$7136.04_{-22.93}^{+15.06}$& -                          & -                                        & 7145                     & 7134                                  & 7125                   & 7144                     & 7259                         & 7157.9&7138& 7135.5                     & 7134                     & 6930                       & 7155                       & 7156                     &7128\\
$2P_1'$       & $7165.57_{-8.64}^{+4.94}$      &      $7158.21_{-9.29}^{+7.67}$         &$7143.19_{-21.99}^{+13.67}$& -                          & -                                        & 7150                     & 7147                                  & 7133                   & 7149                     & 7322                         & 7147.8&7150& 7143.5                     & 7150                     & 6939                       & 7156                       & 7164                     &7165\\
$2^3P_2$      & $7179.47_{-4.69}^{+3.1}$      &      $7173.09_{-7.38}^{+5.62}$         &$7158.95_{-21.52}^{+12.04}$& -                          & -                                        & 7164                     & 7157                                  & 7148                   & 7163                     & 7232                         & 7178.4&7156& 7154.5                     & 7160                     & 6946                       & 7162                       & 7175                     &7172\\
$\theta_{2P}$ & $(12.12_{-5.2}^{+5.51})$$^{\circ}$      &      $(13.36_{-4.88}^{+4.54})$$^{\circ}$  &$(16.97_{-5.52}^{+2.51})$$^{\circ}$& -                          & -                                        & 18.9$^{\circ}$           & 23.2$^{\circ}$                        & 26.5$^{\circ}$         & 30.9$^{\circ}$           & 0.05$^{\circ}$               & 25.5$^{\circ}$&-& 21.2$^{\circ}$             & 38$^{\circ}$             & 35.3$^{\circ}$             & 19.94$^{\circ}$            & $-28.4^{\circ}$          &$-36.3^{\circ}$\\
$3^3P_0$      & $7483.48_{-35.27}^{+17.43}$      &      $7459.36_{-33.9}^{+26.57}$         &$7406.1_{-71.73}^{+48.48}$& -                          & -                                        & -                        & 7474                                  & 7393                   & 7408                     & -                            & 7423.2&7427& 7436.5                     & 7420                     & 7147                       & 7463                       & 7427                     &-\\
$3P_1$        & $7512.29_{-31.68}^{+15.2}$      &      $7487.58_{-31.25}^{+25.21}$         &$7433.42_{-74.64}^{+47.94}$& -                          & -                                        & -                        & 7500                                  & 7414                   & 7440                     & -                            & 7467.7&7437& 7464.5                     & 7441                     & 7162                       & 7479                       & 7458                     &-\\
$3P_1'$       & $7518.83_{-29.67}^{+14.24}$      &      $7494.16_{-30.05}^{+24.39}$         &$7439.96_{-74.82}^{+47.4}$& -                          & -                                        & -                        & 7510                                  & 7421                   & 7442                     & -                            & 7456.1&7447& 7473.5                     & 7458                     & 7168                       & 7479                       & 7466                     &-\\
$3^3P_2$      & $7533.16_{-26.52}^{+12.25}$      &      $7508.81_{-28.28}^{+22.77}$         &$7454.45_{-76.07}^{+46.41}$& -                          & -                                        & -                        & 7524                                  & 7434                   & 7456                     & -                            & 7487.2&7452& 7482.5                     & 7464                     & 7176                       & 7485                       & 7476                     &-\\
$\theta_{3P}$ & $(15.33_{-3.79}^{+3.91})$$^{\circ}$      &      $(16.25_{-3.59}^{+3.17})$$^{\circ}$  &$(18.76_{-3.92}^{+1.61})$$^{\circ}$& -                          & -                                        & -                        & -                                     & 23.6$^{\circ}$         & 29.9$^{\circ}$           & -                            & 29.2$^{\circ}$&-& -                          & 39.7$^{\circ}$           & 35.3$^{\circ}$             & 17.68$^{\circ}$            & $-30.2^{\circ}$          &-\\
$4^3P_0$      & $7796.83_{-54.47}^{+26.87}$      &      $7754.28_{-53.16}^{+43.52}$          &$7654.73_{-131.07}^{+85.26}$& -                          & -                                        & -                        & 7817                                  & 7633                   & -                        & -                            & 7678.9&7684& -                          & 7693                     & 7350                       & -                          & 7682                     &-\\
$4P_1$        & $7824.99_{-51.28}^{+24.92}$      &      $7782.19_{-52.36}^{+42.08}$         &$7679.23_{-136.68}^{+86.25}$& -                          & -                                        & -                        & 7844                                  & 7650                   & -                        & -                            & 7721.4&7693& -                          & 7710                     & 7364                       & -                          & 7708                     &-\\
$4P_1'$       & $7831.9_{-49.49}^{+24.1}$      &      $7788.83_{-50.62}^{+41.53}$         &$7685.83_{-137.43}^{+85.72}$& -                          & -                                        & -                        & 7853                                  & 7656                   & -                        & -                            & 7709&7702& -                          & 7727                     & 7373                       & -                          & 7715                     &-\\
$4^3P_2$      & $7846.53_{-46.7}^{+22.17}$      &      $7803.63_{-49.63}^{+40.03}$         &$7699.76_{-140.43}^{+85.68}$& -                          & -                                        & -                        & 7867                                  & 7667                   & -                        & -                            & 7739.6&7707& -                          & 7732                     & 7379                       & -                          & 7724                     &-\\
$\theta_{4P}$ & $(17.07_{-3.11}^{+3.05})$$^{\circ}$      &      $(17.83_{-2.95}^{+2.4})$$^{\circ}$  &$(19.68_{-3}^{+1.11})$$^{\circ}$& -                          & -                                        & -                        & -                                     & 22.2$^{\circ}$         & -                        & -                            & 30.8$^{\circ}$&-& -                          & 39.7$^{\circ}$           & 35.3$^{\circ}$             & -                          & $-31^{\circ}$            &-\\
$5^3P_0$      & $8083.36_{-72.79}^{+36.01}$      &      $8022.41_{-73.49}^{+59.79}$         &$7868.25_{-198.83}^{+125.64}$& -                          & -                                        & -                        & -                                     & -                      & -                        & -                            & 7896.4&7875& -                          & -                        & -                          & -                          & 7899                     &-\\
$5P_1$        & $8111.21_{-69.91}^{+34.23}$      &      $8049.26_{-71.84}^{+58.99}$         &$7890.79_{-206.68}^{+128.33}$& -                          & -                                        & -                        & -                                     & -                      & -                        & -                            & 7936.6&7885& -                          & -                        & -                          & -                          & 7921                     &-\\
$5P_1'$       & $8118.52_{-68.29}^{+33.42}$      &      $8056.34_{-71.18}^{+58.42}$         &$7896.88_{-208}^{+128.72}$& -                          & -                                        & -                        & -                                     & -                      & -                        & -                            & 7924.1&7896& -                          & -                        & -                          & -                          & 7927                     &-\\
$5^3P_2$      & $8133.07_{-65.43}^{+31.75}$      &      $8070.89_{-70.08}^{+57.38}$         &$7909.54_{-211.7}^{+129.01}$& -                          & -                                        & -                        & -                                     & -                      & -                        & -                            & 7953.7&7901& -                          & -                        & -                          & -                          & 7936                     &-\\
$\theta_{5P}$ & $(18.21_{-2.75}^{+2.42})$$^{\circ}$      &      $(18.84_{-2.57}^{+1.84})$$^{\circ}$  &$(20.23_{-2.41}^{+0.76})$$^{\circ}$& -                          & -                                        & -                        & -                                     & -                      & -                        & -                            & 31.7$^{\circ}$&-& -                          & -                        & -                          & -                          & $-31.6^{\circ}$          &-\\
$6^3P_0$      & $8350.51_{-90.25}^{+44.9}$      &      $8270.54_{-92.55}^{+76.56}$         &$8055.37_{-272.36}^{+170.36}$& -                          & -                                        & -                        & -                                     & -                      & -                        & -                            & -&-& -                          & -                        & -                          & -                          & -                        &-\\
$6P_1$        & $8378.16_{-87.5}^{+43.09}$      &      $8296.78_{-91.48}^{+76.03}$          &$8075.44_{-279.75}^{+173.66}$& -                          & -                                        & -                        & -                                     & -                      & -                        & -                            & -&-& -                          & -                        & -                          & -                          & -                        &-\\
$6P_1'$       & $8385.59_{-85.84}^{+42.39}$      &      $8303.93_{-90.27}^{+75.56}$         &$8081.15_{-281.55}^{+174.13}$& -                          & -                                        & -                        & -                                     & -                      & -                        & -                            & -&-& -                          & -                        & -                          & -                          & -                        &-\\
$6^3P_2$      & $8400.57_{-83.52}^{+40.6}$      &      $8318.44_{-90.29}^{+74.42}$         &$8092.89_{-286.8}^{+175.08}$& -                          & -                                        & -                        & -                                     & -                      & -                        & -                            & -&-& -                          & -                        & -                          & -                          & -                        &-\\
$\theta_{6P}$ & $(18.99_{-2.48}^{+1.93})$$^{\circ}$      &      $(19.53_{-2.28}^{+1.41})$$^{\circ}$  &$(20.58_{-2}^{+0.53})$$^{\circ}$& -                          & -                                        & -                        & -                                     & -                      & -                        & -                            & -&-& -                          & -                        & -                          & -                          & -                        &-\\ \bottomrule
\end{tabular}
\footnotetext{GI: relativized quark model; EFG: relativistic quark model; LLWL: screened GI model; WWLC: full Salpeter equation; MBK: relativistic model with coupled-channel effects; BB: relativistic screened potential model; PLCR: semi-relativistic framework with screened potential; EQ, LZ, AAMS, DKR, LTFWP, and HZ: non-relativistic potential models.}
}
\end{table}
\endgroup
\end{turnpage}

\begin{turnpage}
\begingroup
\squeezetable
\begin{table}
\centering
\caption{Mass spectra of $D$-wave $B_c$ mesons in MeV. States above $n=3$ exceed $\langle v_c^2\rangle\sim0.5$.}
\label{tab:Bcmass_D}
\scalebox{0.94}{
\begin{tabular}{@{}ccccccccccccccccc@{}}
\toprule
State         & Potential I   & Potential II          &Potential III& \makecell{GI \\ \cite{Godfrey:2004ya}} & \makecell{EFG \\ \cite{Ebert:2002pp, Ebert:2011jc}} & \makecell{LLWL \\ \cite{Li:2023wgq}} & \makecell{WWLC \\ \cite{Wang:2022cxy}} & \makecell{MBK \\ \cite{Monteiro:2016rzi}}  & \makecell{BB \\ \cite{Bokade:2025lmn}}&\makecell{PLCR \\ \cite{Patel:2026prg}}& \makecell{EQ \\ \cite{Eichten:2019gig}} & \makecell{LZ \\ \cite{Li:2019tbn}} & \makecell{AAMS \\ \cite{Asghar:2019qjl}} & \makecell{DKR \\ \cite{Devlani:2014nda}} & \makecell{LTFWP \\ \cite{Li:2022bre}}  &\makecell{HZ \\ \cite{Hao:2024nqb}}\\ \midrule
$1^3D_1$      & $7021.03_{-2.08}^{+1.96}$      &      $7021.48_{-2.06}^{+1.94}$          &$7020.86_{-1.66}^{+1.89}$& 7028                     & 7021                                  & 7023                   & 7014                     & 7078                         & 7038.3&7030& 7006.5                     & 7020                     & 6841                       & 7030                       & 7037                     &7023\\
$1D_2$        & $7015.51_{-3.72}^{+5.53}$      &      $7017.16_{-3.78}^{+4.73}$          &$7020.17_{-4.87}^{+3.53}$& 7041                     & 7025                                  & 7032                   & 7025                     & 7009                         & 7053&7033& 7005.5                     & 7024                     & 6845                       & 7025                       & 7046                     &7033\\
$1D_2'$       & $7026.33_{-2.66}^{+2.78}$      &      $7027.11_{-2.63}^{+2.46}$          &$7027.41_{-2.16}^{+2.03}$& 7036                     & 7026                                  & 7039                   & 7029                     & 7154                         & 7031.9&7033& 7015.5                     & 7032                     & 6845                       & 7035                       & 7037                     &7041\\
$1^3D_3$      & $7016.85_{-4.63}^{+7.46}$      &      $7018.82_{-4.68}^{+6.47}$          &$7023.23_{-6.71}^{+4.81}$& 7045                     & 7029                                  & 7042                   & 7035                     & 6980                         & 7041.4&7036& 7010.5                     & 7030                     & 6847                       & 7026                       & 7042                     &7043\\
$\theta_{1D}$ & $(-52.52_{-0.6}^{+1.53})$$^{\circ}$      &      $(-52.29_{-0.64}^{+1.54})$$^{\circ}$  &$(-51.22_{-1.38}^{+2.32})$$^{\circ}$& 44.5$^{\circ}$           & $-35.9$$^{\circ}$                       & $-53.4$$^{\circ}$        & 31.8$^{\circ}$           & 0.05$^{\circ}$               & 41.8$^{\circ}$&-& $-49.2$$^{\circ}$            & 45$^{\circ}$             & 39.2$^{\circ}$             & $-2.49$$^{\circ}$            & $-41.7$$^{\circ}$          &$-42.8^{\circ}$\\
$2^3D_1$      & $7384.77_{-21.34}^{+10.28}$      &      $7369.14_{-21.22}^{+16.56}$          &$7334.76_{-48.52}^{+30.98}$& -                        & 7392                                  & 7327                   & 7335                     & -                            & 7361.8&7361& 7346.5                     & 7336                     & 7080                       & 7365                       & 7357                     &7023\\
$2D_2$        & $7382.69_{-17}^{+7.94}$      &      $7368.05_{-17.98}^{+14.24}$          &$7336.22_{-46.35}^{+28.06}$& -                        & 7399                                  & 7335                   & 7345                     & -                            & 7375.9&7366& 7348.5                     & 7343                     & 7084                       & 7361                       & 7365                     &7033\\
$2D_2'$       & $7390.66_{-19.48}^{+9.25}$      &      $7375.22_{-19.95}^{+15.68}$          &$7341.39_{-48.56}^{+30.05}$& -                        & 7400                                  & 7340                   & 7349                     & -                            & 7361.9&7365& 7355.5                     & 7347                     & 7084                       & 7370                       & 7360                     &7041\\
$2^3D_3$      & $7385.11_{-15.12}^{+7.04}$      &      $7370.93_{-16.86}^{+13.24}$          &$7339.8_{-46.22}^{+27.01}$& -                        & 7405                                  & 7344                   & 7355                     & -                            & 7371.4&7365& 7353.5                     & 7348                     & 7087                       & 7363                       & 7364                     &7043\\
$\theta_{2D}$ & $(-51.59_{-0.72}^{+1.83})$$^{\circ}$      &      $(-51.37_{-0.75}^{+1.82})$$^{\circ}$  &$(-50.09_{-1.63}^{+3.03})$$^{\circ}$& -                        & -                                     & $-48.4$$^{\circ}$        & 32.6$^{\circ}$           & -                            & 42.8$^{\circ}$&-& $-40.3$$^{\circ}$            & 45$^{\circ}$             & 39.2$^{\circ}$             & $-2.8$$^{\circ}$             & $-42.6$$^{\circ}$          &$-43.9^{\circ}$\\
$3^3D_1$      & $7705.69_{-41.64}^{+20.24}$      &      $7671.8_{-42.07}^{+33.74}$          &$7592.92_{-107.23}^{+67.56}$& -                        & 7732                                  & 7573                   & -                        & -                            & 7625.7&7615& -                          & 7611                     & 7289                       & -                          & 7619                     &-\\
$3D_2$        & $7705.84_{-37.79}^{+18.12}$      &      $7672.98_{-39.54}^{+31.59}$          &$7595.68_{-106.44}^{+65.42}$& -                        & 7741                                  & 7581                   & -                        & -                            & 7639.2&7613& -                          & 7620                     & 7293                       & -                          & 7627                     &-\\
$3D_2'$       & $7712.15_{-40}^{+19.31}$      &      $7678.42_{-41.08}^{+32.96}$          &$7599.24_{-108.05}^{+67.21}$& -                        & 7743                                  & 7584                   & -                        & -                            & 7629.5&7614& -                          & 7623                     & 7293                       & -                          & 7623                     &-\\
$3^3D_3$      & $7709.03_{-36.18}^{+17.15}$      &      $7676.36_{-38.28}^{+30.91}$          &$7599.41_{-106.82}^{+64.62}$& -                        & 7750                                  & 7589                   & -                        & -                            & 7638.8&7615& -                          & 7625                     & 7296                       & -                          & 7627                     &-\\
$\theta_{3D}$ & $(-50.67_{-0.84}^{+2.19})$$^{\circ}$      &      $(-50.46_{-0.85}^{+2.24})$$^{\circ}$  &$(-48.92_{-1.94}^{+4.21})$$^{\circ}$& -                        & -                                     & $-42.9$$^{\circ}$        & -                        & -                            & 43.9$^{\circ}$&-& -                          & 45$^{\circ}$             & 39.2$^{\circ}$             & -                          & $-43.6$$^{\circ}$           &-\\
$4^3D_1$      & $7998.21_{-60.96}^{+29.7}$      &      $7945.39_{-62.31}^{+50.95}$          &$7813.93_{-174.28}^{+107.48}$& -                        & -                                     & -                      & -                        & -                            & 7849.3&7845& -                          & -                        & 7478                       & -                          & 7842                     &-\\
$4D_2$        & $8000.06_{-57.48}^{+27.7}$      &      $7947.91_{-60.26}^{+49.03}$          &$7817.51_{-174.54}^{+106.6}$& -                        & -                                     & -                      & -                        & -                            & 7862.1&7847& -                          & -                        & 7482                       & -                          & 7849                     &-\\
$4D_2'$       & $8005.21_{-59.57}^{+28.68}$      &      $7952.21_{-61.34}^{+50.34}$          &$7820.19_{-176}^{+107.76}$& -                        & -                                     & -                      & -                        & -                            & 7855.3&7848& -                          & -                        & 7483                       & -                          & 7846                     &-\\
$4^3D_3$      & $8003.76_{-55.96}^{+26.86}$      &      $7951.81_{-59.45}^{+48.31}$          &$7821.35_{-175.52}^{+106.46}$& -                        & -                                     & -                      & -                        & -                            & 7864.4&7849& -                          & -                        & 7489                       & -                          & 7850                     &-\\
$\theta_{4D}$ & $(-49.73_{-0.96}^{+2.7})$$^{\circ}$      &      $(-49.52_{-0.98}^{+2.78})$$^{\circ}$  &$(-47.65_{-2.34}^{+6.09})$$^{\circ}$& -                        & -                                     & -                      & -                        & -                            & $-44.9$$^{\circ}$&-& -                          & -                        & 39.2$^{\circ}$             & -                          & $-44.6$$^{\circ}$          &-\\
$5^3D_1$      & $8269.86_{-78.83}^{+38.93}$      &      $8198_{-82.27}^{+67.65}$          &$8006.71_{-247.24}^{+152.21}$& -                        & -                                     & -                      & -                        & -                            & 8043.3&8057& -                          & -                        & -                          & -                          & -                        &-\\
$5D_2$        & $8272.94_{-75.7}^{+37.05}$      &      $8201.67_{-80.29}^{+66.07}$          &$8010.82_{-247.62}^{+150.62}$& -                        & -                                     & -                      & -                        & -                            & 8055.4&8058& -                          & -                        & -                          & -                          & -                        &-\\
$5D_2'$       & $8277.09_{-77.48}^{+38.11}$      &      $8204.96_{-81.49}^{+67.25}$          &$8012.79_{-249.09}^{+152.15}$& -                        & -                                     & -                      & -                        & -                            & 8050.6&8059& -                          & -                        & -                          & -                          & -                        &-\\
$5^3D_3$      & $8277.11_{-74.24}^{+36.18}$      &      $8205.68_{-79.43}^{+65.58}$          &$8014.1_{-248.39}^{+151.01}$& -                        & -                                     & -                      & -                        & -                            & 8059.4&8060& -                          & -                        & -                          & -                          & -                        &-\\
$\theta_{5D}$ & $(-48.72_{-1.11}^{+3.46})$$^{\circ}$      &      $(-48.51_{-1.16}^{+3.59})$$^{\circ}$  &$(-46.14_{-2.94}^{+8.91})$$^{\circ}$& -                        & -                                     & -                      & -                        & -                            & $-43.4$$^{\circ}$&-& -                          & -                        & -                          & -                          & -                        &-\\
$6^3D_1$      & $8526.08_{-96.15}^{+47.63}$      &      $8434.34_{-100.89}^{+84.67}$          &$8176.91_{-323.61}^{+198.45}$& -                        & -                                     & -                      & -                        & -                            & -&-& -                          & -                        & -                          & -                          & -                        &-\\
$6D_2$        & $8530.02_{-93.16}^{+45.89}$      &      $8438.75_{-100.08}^{+82.95}$          &$8181.66_{-326.37}^{+197.64}$& -                        & -                                     & -                      & -                        & -                            & -&-& -                          & -                        & -                          & -                          & -                        &-\\
$6D_2'$       & $8533.5_{-94.85}^{+46.84}$      &      $8441.62_{-100.91}^{+83.97}$          &$8182.88_{-326.98}^{+199.09}$& -                        & -                                     & -                      & -                        & -                            & -&-& -                          & -                        & -                          & -                          & -                        &-\\
$6^3D_3$      & $8534.51_{-91.78}^{+45.03}$      &      $8443.43_{-100.07}^{+82.13}$          &$8185.28_{-327.98}^{+197.65}$& -                        & -                                     & -                      & -                        & -                            & -&-& -                          & -                        & -                          & -                          & -                        &-\\
$\theta_{6D}$ & $(-47.58_{-1.35}^{+4.64})$$^{\circ}$      &      $(-47.38_{-1.37}^{+4.89})$$^{\circ}$  &$(-44.35_{-3.73}^{+13.42})$$^{\circ}$& -                        & -                                     & -                      & -                        & -                            & -&-& -                          & -                        & -                          & -                          & -                        &-\\ \bottomrule
\end{tabular}
\footnotetext{GI: relativized quark model; EFG: relativistic quark model; LLWL: screened GI model; WWLC: full Salpeter equation; MBK: relativistic model with coupled-channel effects; BB: relativistic screened potential model; PLCR: semi-relativistic framework with screened potential; EQ, LZ, AAMS, DKR, LTFWP, and HZ: non-relativistic potential models.}
}
\end{table}
\endgroup
\end{turnpage}

\begin{table}
\centering
\caption{Energy eigenvalue $E_{nl}$ and spin-dependent contributions (in MeV) of $S$-wave $B_c$ mesons.}
\label{tab:Bcmass_S_contri}
\begin{tabular}{@{}c|cc|cc|cc@{}}
\toprule
& \multicolumn{2}{c|}{\textbf{Potential I}}
& \multicolumn{2}{c|}{\textbf{Potential II}}
& \multicolumn{2}{c}{\textbf{Potential III}} \\
State        & $E_{nl}$                    & $\langle V_{SS} \rangle$ & $E_{nl}$                    & $\langle V_{SS} \rangle$ & $E_{nl}$& $\langle V_{SS} \rangle$ \\ \midrule
${1{}^1S_0}$ & $16.27_{-3.01}^{+2.78}$     & $-41.78_{-2.72}^{+2.97}$ & $16.21_{-2.96}^{+2.95}$     & $-41.74_{-2.92}^{+2.94}$ & $15.91_{-2.92}^{+3.04}$& $-41.46_{-2.99}^{+2.93}$ \\
${1{}^3S_1}$ & $16.27_{-3.01}^{+2.78}$     & $13.93_{-0.99}^{+0.91}$  & $16.21_{-2.96}^{+2.95}$     & $13.91_{-0.98}^{+0.97}$  & $15.91_{-2.92}^{+3.04}$& $13.82_{-0.98}^{+1}$ \\
${2{}^1S_0}$ & $593.4_{-3.87}^{+3.3}$      & $-22.25_{-3.28}^{+3.92}$ & $591.95_{-3.52}^{+3.59}$    & $-20.82_{-3.48}^{+3.51}$ & $588.89_{-3.88}^{+4.17}$& $-17.7_{-4.12}^{+3.81}$ \\
${2{}^3S_1}$ & $593.4_{-3.87}^{+3.3}$      & $7.42_{-1.31}^{+1.09}$   & $591.95_{-3.52}^{+3.59}$    & $6.94_{-1.17}^{+1.16}$   & $588.89_{-3.88}^{+4.17}$& $5.9_{-1.27}^{+1.38}$ \\
${3{}^1S_0}$ & $1000.04_{-22.78}^{+10.95}$ & $-17.31_{-3.35}^{+4.02}$ & $984.04_{-22.23}^{+17.42}$  & $-15.59_{-3.63}^{+3.4}$  & $950.88_{-45.88}^{+30.51}$& $-12.02_{-4.2}^{+3.62}$ \\
${3{}^3S_1}$ & $1000.04_{-22.78}^{+10.95}$ & $5.77_{-1.34}^{+1.12}$   & $984.04_{-22.23}^{+17.42}$  & $5.2_{-1.13}^{+1.21}$    & $950.88_{-45.88}^{+30.51}$& $4.01_{-1.21}^{+1.4}$ \\
${4{}^1S_0}$ & $1342.12_{-42.13}^{+20.5}$  & $-14.76_{-3.38}^{+3.94}$ & $1308.35_{-41.81}^{+33.97}$ & $-12.98_{-3.6}^{+3.24}$  & $1232.42_{-101.69}^{+66.11}$& $-9.21_{-4.11}^{+3.42}$ \\
${4{}^3S_1}$ & $1342.12_{-42.13}^{+20.5}$  & $4.92_{-1.31}^{+1.13}$   & $1308.35_{-41.81}^{+33.97}$ & $4.33_{-1.08}^{+1.2}$    & $1232.42_{-101.69}^{+66.11}$& $3.07_{-1.14}^{+1.37}$ \\
${5{}^1S_0}$ & $1647.63_{-61.12}^{+29.69}$ & $-13.11_{-3.38}^{+3.83}$ & $1595.09_{-62.14}^{+50.97}$ & $-11.33_{-3.53}^{+3.06}$ & $1467.15_{-166.97}^{+105.44}$& $-7.49_{-3.92}^{+3.26}$ \\
${5{}^3S_1}$ & $1647.63_{-61.12}^{+29.69}$ & $4.37_{-1.28}^{+1.13}$   & $1595.09_{-62.14}^{+50.97}$ & $3.78_{-1.02}^{+1.18}$   & $1467.15_{-166.97}^{+105.44}$& $2.5_{-1.09}^{+1.31}$ \\
${6{}^1S_0}$ & $1928.09_{-78.79}^{+38.86}$ & $-11.91_{-3.38}^{+3.68}$ & $1856.72_{-81.84}^{+67.42}$ & $-10.16_{-3.48}^{+2.9}$  & $1669.83_{-240.13}^{+149.35}$& $-6.29_{-3.77}^{+3.11}$ \\
${6{}^3S_1}$ & $1928.09_{-78.79}^{+38.86}$ & $3.97_{-1.23}^{+1.13}$   & $1856.72_{-81.84}^{+67.42}$ & $3.39_{-0.97}^{+1.16}$   & $1669.83_{-240.13}^{+149.35}$& $2.1_{-1.04}^{+1.26}$ \\  \bottomrule
\end{tabular}
\end{table}

\begin{turnpage}
\begingroup
\squeezetable
\begin{table}
\centering
\caption{Energy eigenvalue $E_{nl}$ and spin-dependent contributions (in MeV) of $P$-wave\footnote{The spin dependent contributions pertaining to ${}^3P_1$ and ${}^1P_1$ states are given here. These states mix to form the physical states presented in Table \ref{tab:Bcmass_P}.} $B_c$ mesons.}
\label{tab:Bcmass_P_contri}
\scalebox{0.94}{
\begin{tabular}{@{}c|ccccc|ccccc|ccccc@{}}
\toprule
& \multicolumn{5}{c|}{\textbf{Potential I}}
& \multicolumn{5}{c|}{\textbf{Potential II}}
& \multicolumn{5}{c}{\textbf{Potential III}} \\
State        & $E_{nl}$                    & $\langle V_{SS} \rangle$ & $\langle \lsp \rangle$ & $\langle \lsm \rangle$ & $\langle V_{T} \rangle$  & $E_{nl}$                    & $\langle V_{SS} \rangle$ & $\langle \lsp \rangle$ & $\langle \lsm \rangle$ & $\langle V_{T} \rangle$   & $E_{nl}$& $\langle V_{SS} \rangle$& $\langle \lsp \rangle$& $\langle \lsm \rangle$&$\langle V_{T} \rangle$\\ \midrule
${1{}^3P_0}$ & $444.1_{-6.94}^{+13.01}$    & $0.45_{-0.18}^{+0.32}$   & $-26.2_{-8.45}^{+4.54}$    & $0$               & $-12.35_{-1.62}^{+0.88}$ & $449.28_{-8.31}^{+12.01}$   & $0.53_{-0.2}^{+0.28}$    & $-27.42_{-7.81}^{+4.51}$   & $0$               & $-12.64_{-1.47}^{+0.9}$   & $460.84_{-14.06}^{+15.57}$& $0.74_{-0.29}^{+0.26}$& $-32.97_{-7.32}^{+7.61}$& $0$&$-13.73_{-1.36}^{+1.47}$\\
${1{}^3P_1}$   & $444.1_{-6.94}^{+13.01}$    & $0.45_{-0.18}^{+0.32}$   & $-13.1_{-4.23}^{+2.27}$    & $0.44_{-0.81}^{+1.51}$     & $6.17_{-0.44}^{+0.81}$   & $449.28_{-8.31}^{+12.01}$   & $0.53_{-0.2}^{+0.28}$    & $-13.71_{-3.91}^{+2.25}$   & $0.63_{-0.8}^{+1.41}$      & $6.32_{-0.45}^{+0.74}$    & $460.84_{-14.06}^{+15.57}$& $0.74_{-0.29}^{+0.26}$& $-16.48_{-3.66}^{+3.81}$& $1.6_{-1.35}^{+1.34}$&$6.87_{-0.74}^{+0.68}$\\
${1{}^1P_1}$  & $444.1_{-6.94}^{+13.01}$    & $-1.34_{-0.95}^{+0.54}$  & $0$               & $0.44_{-0.81}^{+1.51}$     & $0$             & $449.28_{-8.31}^{+12.01}$   & $-1.58_{-0.83}^{+0.61}$  & $0$               & $0.63_{-0.8}^{+1.41}$      & $0$              & $460.84_{-14.06}^{+15.57}$& $-2.22_{-0.79}^{+0.88}$& $0$& $1.6_{-1.35}^{+1.34}$&$0$\\
${1{}^3P_2}$ & $444.1_{-6.94}^{+13.01}$    & $0.45_{-0.18}^{+0.32}$   & $13.1_{-2.27}^{+4.23}$     & $0$               & $-1.24_{-0.16}^{+0.09}$  & $449.28_{-8.31}^{+12.01}$   & $0.53_{-0.2}^{+0.28}$    & $13.71_{-2.25}^{+3.91}$    & $0$               & $-1.26_{-0.15}^{+0.09}$   & $460.84_{-14.06}^{+15.57}$& $0.74_{-0.29}^{+0.26}$& $16.48_{-3.81}^{+3.66}$& $0$&$-1.37_{-0.14}^{+0.15}$\\
${2{}^3P_0}$ & $867.03_{-8.24}^{+4.46}$    & $0.57_{-0.21}^{+0.31}$   & $-26.45_{-6.34}^{+3.57}$   & $0$               & $-11.13_{-1.17}^{+0.68}$ & $859.82_{-8.99}^{+7.28}$    & $0.63_{-0.21}^{+0.25}$   & $-26.6_{-5.74}^{+3.27}$    & $0$               & $-11.03_{-1.04}^{+0.62}$  & $844.93_{-22.05}^{+13.42}$& $0.76_{-0.23}^{+0.13}$& $-28.26_{-3.62}^{+4.06}$& $0$&$-10.74_{-0.89}^{+0.52}$\\
${2{}^3P_1}$   & $867.03_{-8.24}^{+4.46}$    & $0.57_{-0.21}^{+0.31}$   & $-13.23_{-3.17}^{+1.78}$   & $1.22_{-0.64}^{+1.16}$     & $5.56_{-0.34}^{+0.58}$   & $859.82_{-8.99}^{+7.28}$    & $0.63_{-0.21}^{+0.25}$   & $-13.3_{-2.87}^{+1.63}$    & $1.33_{-0.6}^{+1.06}$      & $5.51_{-0.31}^{+0.52}$    & $844.93_{-22.05}^{+13.42}$& $0.76_{-0.23}^{+0.13}$& $-14.13_{-1.81}^{+2.03}$& $1.93_{-0.92}^{+0.74}$&$5.37_{-0.26}^{+0.44}$\\
${2{}^1P_1}$  & $867.03_{-8.24}^{+4.46}$    & $-1.72_{-0.92}^{+0.63}$  & $0$               & $1.22_{-0.64}^{+1.16}$     & $0$             & $859.82_{-8.99}^{+7.28}$    & $-1.9_{-0.76}^{+0.64}$   & $0$               & $1.33_{-0.6}^{+1.06}$      & $0$              & $844.93_{-22.05}^{+13.42}$& $-2.29_{-0.4}^{+0.69}$& $0$& $1.93_{-0.92}^{+0.74}$&$0$\\
${2{}^3P_2}$ & $867.03_{-8.24}^{+4.46}$    & $0.57_{-0.21}^{+0.31}$   & $13.23_{-1.78}^{+3.17}$    & $0$               & $-1.11_{-0.12}^{+0.07}$  & $859.82_{-8.99}^{+7.28}$    & $0.63_{-0.21}^{+0.25}$   & $13.3_{-1.63}^{+2.87}$     & $0$               & $-1.1_{-0.1}^{+0.06}$     & $844.93_{-22.05}^{+13.42}$& $0.76_{-0.23}^{+0.13}$& $14.13_{-2.03}^{+1.81}$& $0$&$-1.07_{-0.09}^{+0.05}$\\
${3{}^3P_0}$ & $1220.26_{-29.35}^{+13.92}$ & $0.64_{-0.22}^{+0.28}$   & $-26.77_{-5.37}^{+3.1}$    & $0$               & $-10.55_{-0.97}^{+0.59}$ & $1195.7_{-29.9}^{+24.11}$   & $0.68_{-0.21}^{+0.22}$   & $-26.31_{-4.78}^{+2.77}$   & $0$               & $-10.27_{-0.87}^{+0.52}$  & $1141.53_{-75.06}^{+47.33}$& $0.68_{-0.13}^{+0.15}$& $-24.52_{-4.06}^{+2.42}$& $0$&$-9.33_{-0.98}^{+1.06}$\\
${3{}^3P_1}$   & $1220.26_{-29.35}^{+13.92}$ & $0.64_{-0.22}^{+0.28}$   & $-13.39_{-2.68}^{+1.55}$   & $1.65_{-0.56}^{+0.99}$     & $5.27_{-0.29}^{+0.49}$   & $1195.7_{-29.9}^{+24.11}$   & $0.68_{-0.21}^{+0.22}$   & $-13.16_{-2.39}^{+1.38}$   & $1.69_{-0.51}^{+0.88}$     & $5.13_{-0.26}^{+0.44}$    & $1141.53_{-75.06}^{+47.33}$& $0.68_{-0.13}^{+0.15}$& $-12.26_{-2.03}^{+1.21}$& $1.91_{-0.57}^{+0.54}$&$4.66_{-0.53}^{+0.49}$\\
${3{}^1P_1}$  & $1220.26_{-29.35}^{+13.92}$ & $-1.93_{-0.83}^{+0.66}$  & $0$               & $1.65_{-0.56}^{+0.99}$     & $0$             & $1195.7_{-29.9}^{+24.11}$   & $-2.05_{-0.65}^{+0.63}$  & $0$               & $1.69_{-0.51}^{+0.88}$     & $0$              & $1141.53_{-75.06}^{+47.33}$& $-2.04_{-0.44}^{+0.4}$& $0$& $1.91_{-0.57}^{+0.54}$&$0$\\
${3{}^3P_2}$ & $1220.26_{-29.35}^{+13.92}$ & $0.64_{-0.22}^{+0.28}$   & $13.39_{-1.55}^{+2.68}$    & $0$               & $-1.06_{-0.1}^{+0.06}$   & $1195.7_{-29.9}^{+24.11}$   & $0.68_{-0.21}^{+0.22}$   & $13.16_{-1.38}^{+2.39}$    & $0$               & $-1.03_{-0.09}^{+0.05}$   & $1141.53_{-75.06}^{+47.33}$& $0.68_{-0.13}^{+0.15}$& $12.26_{-1.21}^{+2.03}$& $0$&$-0.93_{-0.1}^{+0.11}$\\
${4{}^3P_0}$ & $1533.33_{-49.19}^{+23.74}$ & $0.69_{-0.22}^{+0.25}$   & $-27.03_{-4.78}^{+2.82}$   & $0$               & $-10.2_{-0.87}^{+0.54}$  & $1490.33_{-50.46}^{+41.26}$ & $0.71_{-0.2}^{+0.18}$    & $-26.15_{-4.24}^{+2.45}$   & $0$               & $-9.81_{-0.78}^{+0.49}$   & $1387.22_{-137.76}^{+85.7}$& $0.62_{-0.13}^{+0.16}$& $-22.62_{-4.21}^{+3.58}$& $0$&$-8.34_{-1.2}^{+1.76}$\\
${4{}^3P_1}$   & $1533.33_{-49.19}^{+23.74}$ & $0.69_{-0.22}^{+0.25}$   & $-13.52_{-2.39}^{+1.41}$   & $1.93_{-0.51}^{+0.89}$     & $5.1_{-0.27}^{+0.43}$    & $1490.33_{-50.46}^{+41.26}$ & $0.71_{-0.2}^{+0.18}$    & $-13.08_{-2.12}^{+1.23}$   & $1.92_{-0.45}^{+0.78}$     & $4.9_{-0.24}^{+0.39}$     & $1387.22_{-137.76}^{+85.7}$& $0.62_{-0.13}^{+0.16}$& $-11.31_{-2.1}^{+1.79}$& $1.78_{-0.35}^{+0.59}$&$4.17_{-0.88}^{+0.6}$\\
${4{}^1P_1}$  & $1533.33_{-49.19}^{+23.74}$ & $-2.06_{-0.74}^{+0.67}$  & $0$               & $1.93_{-0.51}^{+0.89}$     & $0$             & $1490.33_{-50.46}^{+41.26}$ & $-2.14_{-0.55}^{+0.6}$   & $0$               & $1.92_{-0.45}^{+0.78}$     & $0$              & $1387.22_{-137.76}^{+85.7}$& $-1.85_{-0.48}^{+0.4}$& $0$& $1.78_{-0.35}^{+0.59}$&$0$\\
${4{}^3P_2}$ & $1533.33_{-49.19}^{+23.74}$ & $0.69_{-0.22}^{+0.25}$   & $13.52_{-1.41}^{+2.39}$    & $0$               & $-1.02_{-0.09}^{+0.05}$  & $1490.33_{-50.46}^{+41.26}$ & $0.71_{-0.2}^{+0.18}$    & $13.08_{-1.23}^{+2.12}$    & $0$               & $-0.98_{-0.08}^{+0.05}$   & $1387.22_{-137.76}^{+85.7}$& $0.62_{-0.13}^{+0.16}$& $11.31_{-1.79}^{+2.1}$& $0$&$-0.83_{-0.12}^{+0.18}$\\
${5{}^3P_0}$ & $1819.89_{-67.94}^{+33.08}$ & $0.72_{-0.22}^{+0.22}$   & $-27.25_{-4.38}^{+2.62}$   & $0$               & $-9.96_{-0.8}^{+0.5}$    & $1757.8_{-71.1}^{+58.09}$   & $0.73_{-0.19}^{+0.15}$   & $-26.06_{-3.87}^{+2.27}$   & $0$               & $-9.5_{-0.71}^{+0.51}$    & $1598.08_{-208.42}^{+129}$& $0.57_{-0.15}^{+0.17}$& $-21.07_{-4.55}^{+5.53}$& $0$&$-7.57_{-1.46}^{+2.32}$\\
${5{}^3P_1}$   & $1819.89_{-67.94}^{+33.08}$ & $0.72_{-0.22}^{+0.22}$   & $-13.62_{-2.19}^{+1.31}$   & $2.13_{-0.47}^{+0.81}$     & $4.98_{-0.25}^{+0.4}$    & $1757.8_{-71.1}^{+58.09}$   & $0.73_{-0.19}^{+0.15}$   & $-13.03_{-1.93}^{+1.14}$   & $2.07_{-0.41}^{+0.71}$     & $4.75_{-0.25}^{+0.36}$    & $1598.08_{-208.42}^{+129}$& $0.57_{-0.15}^{+0.17}$& $-10.54_{-2.27}^{+2.77}$& $1.72_{-0.35}^{+0.6}$&$3.78_{-1.16}^{+0.73}$\\
${5{}^1P_1}$  & $1819.89_{-67.94}^{+33.08}$ & $-2.16_{-0.65}^{+0.66}$  & $0$               & $2.13_{-0.47}^{+0.81}$     & $0$             & $1757.8_{-71.1}^{+58.09}$   & $-2.19_{-0.46}^{+0.57}$  & $0$               & $2.07_{-0.41}^{+0.71}$     & $0$              & $1598.08_{-208.42}^{+129}$& $-1.72_{-0.51}^{+0.45}$& $0$& $1.72_{-0.35}^{+0.6}$&$0$\\
${5{}^3P_2}$ & $1819.89_{-67.94}^{+33.08}$ & $0.72_{-0.22}^{+0.22}$   & $13.62_{-1.31}^{+2.19}$    & $0$               & $-1_{-0.08}^{+0.05}$     & $1757.8_{-71.1}^{+58.09}$   & $0.73_{-0.19}^{+0.15}$   & $13.03_{-1.14}^{+1.93}$    & $0$               & $-0.95_{-0.07}^{+0.05}$   & $1598.08_{-208.42}^{+129}$& $0.57_{-0.15}^{+0.17}$& $10.54_{-2.77}^{+2.27}$& $0$&$-0.76_{-0.15}^{+0.23}$\\
${6{}^3P_0}$ & $2087.07_{-85.73}^{+41.97}$ & $0.75_{-0.22}^{+0.18}$   & $-27.42_{-4.07}^{+2.47}$   & $0$               & $-9.78_{-0.75}^{+0.48}$  & $2005.42_{-90.27}^{+75.39}$ & $0.74_{-0.18}^{+0.13}$   & $-26.01_{-3.62}^{+2.13}$   & $0$               & $-9.27_{-0.67}^{+0.52}$   & $1782.18_{-281.99}^{+174.41}$& $0.53_{-0.18}^{+0.18}$& $-19.63_{-5.08}^{+7.26}$& $0$&$-6.87_{-1.74}^{+2.73}$\\
${6{}^3P_1}$   & $2087.07_{-85.73}^{+41.97}$ & $0.75_{-0.22}^{+0.18}$   & $-13.71_{-2.04}^{+1.24}$   & $2.28_{-0.44}^{+0.75}$     & $4.89_{-0.24}^{+0.37}$   & $2005.42_{-90.27}^{+75.39}$ & $0.74_{-0.18}^{+0.13}$   & $-13.01_{-1.81}^{+1.07}$   & $2.2_{-0.38}^{+0.66}$      & $4.63_{-0.26}^{+0.34}$    & $1782.18_{-281.99}^{+174.41}$& $0.53_{-0.18}^{+0.18}$& $-9.81_{-2.54}^{+3.63}$& $1.67_{-0.48}^{+0.61}$&$3.44_{-1.37}^{+0.87}$\\
${6{}^1P_1}$  & $2087.07_{-85.73}^{+41.97}$ & $-2.24_{-0.55}^{+0.65}$  & $0$               & $2.28_{-0.44}^{+0.75}$     & $0$             & $2005.42_{-90.27}^{+75.39}$ & $-2.21_{-0.38}^{+0.53}$  & $0$               & $2.2_{-0.38}^{+0.66}$      & $0$              & $1782.18_{-281.99}^{+174.41}$& $-1.6_{-0.54}^{+0.55}$& $0$& $1.67_{-0.48}^{+0.61}$&$0$\\
${6{}^3P_2}$ & $2087.07_{-85.73}^{+41.97}$ & $0.75_{-0.22}^{+0.18}$   & $13.71_{-1.24}^{+2.04}$    & $0$               & $-0.98_{-0.08}^{+0.05}$  & $2005.42_{-90.27}^{+75.39}$ & $0.74_{-0.18}^{+0.13}$   & $13.01_{-1.07}^{+1.81}$    & $0$               & $-0.93_{-0.07}^{+0.05}$   & $1782.18_{-281.99}^{+174.41}$& $0.53_{-0.18}^{+0.18}$& $9.81_{-3.63}^{+2.54}$& $0$&$-0.69_{-0.17}^{+0.27}$\\ \bottomrule
\end{tabular}%
}
\end{table}
\endgroup
\end{turnpage}

\begin{turnpage}
\begingroup
\squeezetable
\begin{table}
\centering
\caption{Energy eigenvalue $E_{nl}$ and spin-dependent contributions (in MeV) of $D$-wave\footnote{The spin dependent contributions pertaining to ${}^3D_2$ and ${}^1D_2$ states are given here. These states mix to form the physical states presented in Table \ref{tab:Bcmass_D}.} $B_c$ mesons.}
\label{tab:Bcmass_D_contri}
\scalebox{0.94}{
\begin{tabular}{@{}c|ccccc|ccccc|ccccc@{}}
\toprule
& \multicolumn{5}{c|}{\textbf{Potential I}}
& \multicolumn{5}{c|}{\textbf{Potential II}}
& \multicolumn{5}{c}{\textbf{Potential III}} \\
State        & $E_{nl}$                    & $\langle V_{SS} \rangle$ & $\langle \lsp \rangle$ & $\langle \lsm \rangle$ & $\langle V_{T} \rangle$ & $E_{nl}$                     & $\langle V_{SS} \rangle$ & $\langle \lsp \rangle$ & $\langle \lsm \rangle$ & $\langle V_{T} \rangle$  & $E_{nl}$& $\langle V_{SS} \rangle$& $\langle \lsp \rangle$& $\langle \lsm \rangle$&$\langle V_{T} \rangle$\\ \midrule
${1{}^3D_1}$ & $719.56_{-3.37}^{+4.72}$    & $0.02_{-0.01}^{+0.03}$   & $3.24_{-4.18}^{+2.36}$     & $0$               & $-1.63_{-0.13}^{+0.08}$ & $720.95_{-3.44}^{+4.03}$     & $0.02_{-0.01}^{+0.03}$   & $2.41_{-3.78}^{+2.37}$     & $0$               & $-1.61_{-0.11}^{+0.07}$  & $723.21_{-4.05}^{+2.93}$& $0.04_{-0.02}^{+0.03}$& $-0.41_{-3.36}^{+3.8}$& $0$&$-1.61_{-0.09}^{+0.06}$\\
${1{}^3D_2}$   & $719.56_{-3.37}^{+4.72}$    & $0.02_{-0.01}^{+0.03}$   & $1.08_{-1.39}^{+0.79}$     & $-5.16_{-0.74}^{+1.36}$    & $1.63_{-0.08}^{+0.13}$  & $720.95_{-3.44}^{+4.03}$     & $0.02_{-0.01}^{+0.03}$   & $0.8_{-1.26}^{+0.79}$      & $-4.8_{-0.79}^{+1.23}$     & $1.61_{-0.07}^{+0.11}$   & $723.21_{-4.05}^{+2.93}$& $0.04_{-0.02}^{+0.03}$& $-0.14_{-1.12}^{+1.27}$& $-3.74_{-1.33}^{+1.27}$&$1.61_{-0.06}^{+0.09}$\\
${1{}^1D_2}$  & $719.56_{-3.37}^{+4.72}$    & $-0.04_{-0.08}^{+0.03}$  & $0$               & $-5.16_{-0.74}^{+1.36}$    & $0$            & $720.95_{-3.44}^{+4.03}$     & $-0.06_{-0.08}^{+0.04}$  & $0$               & $-4.8_{-0.79}^{+1.23}$     & $0$             & $723.21_{-4.05}^{+2.93}$& $-0.11_{-0.09}^{+0.07}$& $0$& $-3.74_{-1.33}^{+1.27}$&$0$\\
${1{}^3D_3}$ & $719.56_{-3.37}^{+4.72}$    & $0.02_{-0.01}^{+0.03}$   & $-2.16_{-1.57}^{+2.79}$    & $0$               & $-0.46_{-0.04}^{+0.02}$ & $720.95_{-3.44}^{+4.03}$     & $0.02_{-0.01}^{+0.03}$   & $-1.61_{-1.58}^{+2.52}$    & $0$               & $-0.46_{-0.03}^{+0.02}$  & $723.21_{-4.05}^{+2.93}$& $0.04_{-0.02}^{+0.03}$& $0.27_{-2.53}^{+2.24}$& $0$&$-0.46_{-0.03}^{+0.02}$\\
${2{}^3D_1}$ & $1085.79_{-17.57}^{+8.35}$  & $0.03_{-0.02}^{+0.05}$   & $0.38_{-3.58}^{+2.06}$     & $0$               & $-1.56_{-0.11}^{+0.07}$ & $1070.95_{-18.47}^{+14.65}$  & $0.04_{-0.02}^{+0.04}$   & $-0.24_{-3.16}^{+2}$       & $0$               & $-1.52_{-0.1}^{+0.06}$   & $1038.54_{-47.11}^{+28.71}$& $0.06_{-0.03}^{+0.03}$& $-2.35_{-2.12}^{+2.94}$& $0$&$-1.41_{-0.12}^{+0.14}$\\
${2{}^3D_2}$   & $1085.79_{-17.57}^{+8.35}$  & $0.03_{-0.02}^{+0.05}$   & $0.13_{-1.19}^{+0.69}$     & $-3.87_{-0.64}^{+1.16}$    & $1.56_{-0.07}^{+0.11}$  & $1070.95_{-18.47}^{+14.65}$  & $0.04_{-0.02}^{+0.04}$   & $-0.08_{-1.05}^{+0.67}$    & $-3.51_{-0.7}^{+1.03}$     & $1.52_{-0.06}^{+0.1}$    & $1038.54_{-47.11}^{+28.71}$& $0.06_{-0.03}^{+0.03}$& $-0.78_{-0.71}^{+0.98}$& $-2.47_{-1.17}^{+1.07}$&$1.41_{-0.14}^{+0.12}$\\
${2{}^1D_2}$  & $1085.79_{-17.57}^{+8.35}$  & $-0.09_{-0.15}^{+0.06}$  & $0$               & $-3.87_{-0.64}^{+1.16}$    & $0$            & $1070.95_{-18.47}^{+14.65}$  & $-0.11_{-0.13}^{+0.07}$  & $0$               & $-3.51_{-0.7}^{+1.03}$     & $0$             & $1038.54_{-47.11}^{+28.71}$& $-0.17_{-0.1}^{+0.1}$& $0$& $-2.47_{-1.17}^{+1.07}$&$0$\\
${2{}^3D_3}$ & $1085.79_{-17.57}^{+8.35}$  & $0.03_{-0.02}^{+0.05}$   & $-0.26_{-1.37}^{+2.38}$    & $0$               & $-0.45_{-0.03}^{+0.02}$ & $1070.95_{-18.47}^{+14.65}$  & $0.04_{-0.02}^{+0.04}$   & $0.16_{-1.33}^{+2.1}$      & $0$               & $-0.43_{-0.03}^{+0.02}$  & $1038.54_{-47.11}^{+28.71}$& $0.06_{-0.03}^{+0.03}$& $1.57_{-1.96}^{+1.41}$& $0$&$-0.4_{-0.03}^{+0.04}$\\
${3{}^3D_1}$ & $1408.49_{-38.36}^{+18.46}$ & $0.04_{-0.03}^{+0.06}$   & $-1.43_{-3.2}^{+1.86}$     & $0$               & $-1.52_{-0.1}^{+0.07}$  & $1375.39_{-39.93}^{+31.95}$  & $0.05_{-0.03}^{+0.06}$   & $-1.84_{-2.79}^{+1.74}$    & $0$               & $-1.46_{-0.09}^{+0.06}$  & $1297.41_{-107}^{+65.98}$& $0.07_{-0.03}^{+0.04}$& $-3.15_{-1.45}^{+2.15}$& $0$&$-1.27_{-0.16}^{+0.25}$\\
${3{}^3D_2}$   & $1408.49_{-38.36}^{+18.46}$ & $0.04_{-0.03}^{+0.06}$   & $-0.48_{-1.07}^{+0.62}$    & $-3.06_{-0.58}^{+1.04}$    & $1.52_{-0.07}^{+0.1}$   & $1375.39_{-39.93}^{+31.95}$  & $0.05_{-0.03}^{+0.06}$   & $-0.61_{-0.93}^{+0.58}$    & $-2.73_{-0.63}^{+0.91}$    & $1.46_{-0.06}^{+0.09}$   & $1297.41_{-107}^{+65.98}$& $0.07_{-0.03}^{+0.04}$& $-1.05_{-0.48}^{+0.72}$& $-1.73_{-1.04}^{+0.9}$&$1.27_{-0.25}^{+0.16}$\\
${3{}^1D_2}$  & $1408.49_{-38.36}^{+18.46}$ & $-0.13_{-0.19}^{+0.08}$  & $0$               & $-3.06_{-0.58}^{+1.04}$    & $0$            & $1375.39_{-39.93}^{+31.95}$  & $-0.16_{-0.17}^{+0.09}$  & $0$               & $-2.73_{-0.63}^{+0.91}$    & $0$             & $1297.41_{-107}^{+65.98}$& $-0.2_{-0.11}^{+0.1}$& $0$& $-1.73_{-1.04}^{+0.9}$&$0$\\
${3{}^3D_3}$ & $1408.49_{-38.36}^{+18.46}$ & $0.04_{-0.03}^{+0.06}$   & $0.95_{-1.24}^{+2.13}$     & $0$               & $-0.44_{-0.03}^{+0.02}$ & $1375.39_{-39.93}^{+31.95}$  & $0.05_{-0.03}^{+0.06}$   & $1.23_{-1.16}^{+1.86}$     & $0$               & $-0.42_{-0.03}^{+0.02}$  & $1297.41_{-107}^{+65.98}$& $0.07_{-0.03}^{+0.04}$& $2.1_{-1.43}^{+0.96}$& $0$&$-0.36_{-0.05}^{+0.07}$\\
${4{}^3D_1}$ & $1702.41_{-58.03}^{+27.92}$ & $0.06_{-0.04}^{+0.08}$   & $-2.71_{-2.93}^{+1.72}$    & $0$               & $-1.5_{-0.09}^{+0.06}$  & $1650.01_{-60.57}^{+49.29}$  & $0.07_{-0.04}^{+0.06}$   & $-2.94_{-2.54}^{+1.55}$    & $0$               & $-1.43_{-0.08}^{+0.07}$  & $1518.97_{-175.19}^{+107.09}$& $0.07_{-0.03}^{+0.04}$& $-3.16_{-1.63}^{+1.25}$& $0$&$-1.15_{-0.21}^{+0.33}$\\
${4{}^3D_2}$   & $1702.41_{-58.03}^{+27.92}$ & $0.06_{-0.04}^{+0.08}$   & $-0.9_{-0.98}^{+0.58}$     & $-2.49_{-0.53}^{+0.95}$    & $1.5_{-0.06}^{+0.09}$   & $1650.01_{-60.57}^{+49.29}$  & $0.07_{-0.04}^{+0.06}$   & $-0.98_{-0.85}^{+0.52}$    & $-2.19_{-0.57}^{+0.82}$    & $1.43_{-0.07}^{+0.08}$   & $1518.97_{-175.19}^{+107.09}$& $0.07_{-0.03}^{+0.04}$& $-1.05_{-0.54}^{+0.42}$& $-1.24_{-0.94}^{+0.75}$&$1.15_{-0.33}^{+0.21}$\\
${4{}^1D_2}$  & $1702.41_{-58.03}^{+27.92}$ & $-0.17_{-0.23}^{+0.1}$   & $0$               & $-2.49_{-0.53}^{+0.95}$    & $0$            & $1650.01_{-60.57}^{+49.29}$  & $-0.2_{-0.19}^{+0.11}$   & $0$               & $-2.19_{-0.57}^{+0.82}$    & $0$             & $1518.97_{-175.19}^{+107.09}$& $-0.2_{-0.13}^{+0.09}$& $0$& $-1.24_{-0.94}^{+0.75}$&$0$\\
${4{}^3D_3}$ & $1702.41_{-58.03}^{+27.92}$ & $0.06_{-0.04}^{+0.08}$   & $1.81_{-1.15}^{+1.96}$     & $0$               & $-0.43_{-0.03}^{+0.02}$ & $1650.01_{-60.57}^{+49.29}$  & $0.07_{-0.04}^{+0.06}$   & $1.96_{-1.03}^{+1.69}$     & $0$               & $-0.41_{-0.02}^{+0.02}$  & $1518.97_{-175.19}^{+107.09}$& $0.07_{-0.03}^{+0.04}$& $2.11_{-0.83}^{+1.09}$& $0$&$-0.33_{-0.06}^{+0.1}$\\
${5{}^3D_1}$ & $1974.98_{-76.14}^{+37.28}$ & $0.07_{-0.04}^{+0.09}$   & $-3.68_{-2.73}^{+1.62}$    & $0$               & $-1.48_{-0.09}^{+0.06}$ & $1903.33_{-80.52}^{+66.48}$  & $0.08_{-0.04}^{+0.07}$   & $-3.77_{-2.36}^{+1.41}$    & $0$               & $-1.4_{-0.08}^{+0.08}$   & $1712.08_{-248.3}^{+151.07}$& $0.07_{-0.03}^{+0.05}$& $-3.19_{-1.74}^{+1.01}$& $0$&$-1.06_{-0.25}^{+0.4}$\\
${5{}^3D_2}$   & $1974.98_{-76.14}^{+37.28}$ & $0.07_{-0.04}^{+0.09}$   & $-1.23_{-0.91}^{+0.54}$    & $-2.06_{-0.5}^{+0.88}$     & $1.48_{-0.06}^{+0.09}$  & $1903.33_{-80.52}^{+66.48}$  & $0.08_{-0.04}^{+0.07}$   & $-1.26_{-0.79}^{+0.47}$    & $-1.79_{-0.53}^{+0.75}$    & $1.4_{-0.08}^{+0.08}$    & $1712.08_{-248.3}^{+151.07}$& $0.07_{-0.03}^{+0.05}$& $-1.07_{-0.58}^{+0.34}$& $-0.91_{-0.84}^{+0.62}$&$1.06_{-0.4}^{+0.25}$\\
${5{}^1D_2}$  & $1974.98_{-76.14}^{+37.28}$ & $-0.2_{-0.26}^{+0.12}$   & $0$               & $-2.06_{-0.5}^{+0.88}$     & $0$            & $1903.33_{-80.52}^{+66.48}$  & $-0.24_{-0.21}^{+0.13}$  & $0$               & $-1.79_{-0.53}^{+0.75}$    & $0$             & $1712.08_{-248.3}^{+151.07}$& $-0.2_{-0.15}^{+0.09}$& $0$& $-0.91_{-0.84}^{+0.62}$&$0$\\
${5{}^3D_3}$ & $1974.98_{-76.14}^{+37.28}$ & $0.07_{-0.04}^{+0.09}$   & $2.45_{-1.08}^{+1.82}$     & $0$               & $-0.42_{-0.03}^{+0.02}$ & $1903.33_{-80.52}^{+66.48}$  & $0.08_{-0.04}^{+0.07}$   & $2.51_{-0.94}^{+1.57}$     & $0$               & $-0.4_{-0.02}^{+0.02}$   & $1712.08_{-248.3}^{+151.07}$& $0.07_{-0.03}^{+0.05}$& $2.13_{-0.67}^{+1.16}$& $0$&$-0.3_{-0.07}^{+0.12}$\\
${6{}^3D_1}$ & $2231.87_{-93.54}^{+46.09}$ & $0.08_{-0.05}^{+0.09}$   & $-4.44_{-2.57}^{+1.54}$    & $0$               & $-1.46_{-0.08}^{+0.06}$ & $2140.38_{-100.42}^{+83.18}$ & $0.09_{-0.05}^{+0.08}$   & $-4.41_{-2.2}^{+1.32}$     & $0$               & $-1.38_{-0.08}^{+0.08}$  & $1882.67_{-326.92}^{+198.11}$& $0.07_{-0.03}^{+0.05}$& $-3.26_{-1.79}^{+1}$& $0$&$-0.97_{-0.29}^{+0.46}$\\
${6{}^3D_2}$   & $2231.87_{-93.54}^{+46.09}$ & $0.08_{-0.05}^{+0.09}$   & $-1.48_{-0.86}^{+0.51}$    & $-1.72_{-0.47}^{+0.82}$    & $1.46_{-0.06}^{+0.08}$  & $2140.38_{-100.42}^{+83.18}$ & $0.09_{-0.05}^{+0.08}$   & $-1.47_{-0.73}^{+0.44}$    & $-1.48_{-0.49}^{+0.7}$     & $1.38_{-0.08}^{+0.08}$   & $1882.67_{-326.92}^{+198.11}$& $0.07_{-0.03}^{+0.05}$& $-1.09_{-0.6}^{+0.33}$& $-0.67_{-0.75}^{+0.51}$&$0.97_{-0.46}^{+0.29}$\\
${6{}^1D_2}$  & $2231.87_{-93.54}^{+46.09}$ & $-0.24_{-0.28}^{+0.14}$  & $0$               & $-1.72_{-0.47}^{+0.82}$    & $0$            & $2140.38_{-100.42}^{+83.18}$ & $-0.27_{-0.23}^{+0.15}$  & $0$               & $-1.48_{-0.49}^{+0.7}$     & $0$             & $1882.67_{-326.92}^{+198.11}$& $-0.2_{-0.16}^{+0.1}$& $0$& $-0.67_{-0.75}^{+0.51}$&$0$\\
${6{}^3D_3}$ & $2231.87_{-93.54}^{+46.09}$ & $0.08_{-0.05}^{+0.09}$   & $2.96_{-1.02}^{+1.71}$     & $0$               & $-0.42_{-0.02}^{+0.02}$ & $2140.38_{-100.42}^{+83.18}$ & $0.09_{-0.05}^{+0.08}$   & $2.94_{-0.88}^{+1.47}$     & $0$               & $-0.39_{-0.02}^{+0.02}$  & $1882.67_{-326.92}^{+198.11}$& $0.07_{-0.03}^{+0.05}$& $2.18_{-0.67}^{+1.2}$& $0$&$-0.28_{-0.08}^{+0.13}$\\ \bottomrule
\end{tabular}%
}
\end{table}
\endgroup
\end{turnpage}

\begin{turnpage}
\begingroup
\squeezetable
\begin{table}
\centering
\caption{Square of radial wave function at the origin $|R_{nl}^{(l)}(0)|^2$, RMS radii $\sqrt{\langle r^2 \rangle}$, expectation values of kinetic energy $\langle T \rangle$, squared momentum $\langle p^2 \rangle$, and heavy quark velocities $\langle v_c^2 \rangle$ and $\langle v_b^2 \rangle$ for the $B_c$ system. Results obtained are shown together with the corresponding uncertainties propagated from the MCMC parameter distributions.}
\label{tab:Bc_otherobs}
\scalebox{0.84}{
\begin{tabular}{@{}c|cccccc|cccccc|cccccc@{}}
\toprule
& \multicolumn{6}{c|}{\textbf{Potential I}}
& \multicolumn{6}{c|}{\textbf{Potential II}}
& \multicolumn{6}{c}{\textbf{Potential III}} \\
State & \makecell{$|R_{nl}^{(l)}(0)|^2$ \\ (GeV${}^{3+2l}$)}    & \makecell{$\sqrt{\langle r^2 \rangle}$ \\ (GeV${}^{-1}$)} & \makecell{$\langle T \rangle$ \\ (GeV)}      & \makecell{$\langle p^2 \rangle$ \\ (GeV${}^2$)}     & $\langle v_c^2 \rangle$   & $\langle v_b^2 \rangle$   & \makecell{$|R_{nl}^{(l)}(0)|^2$ \\ (GeV${}^{3+2l}$)}    & \makecell{$\sqrt{\langle r^2 \rangle}$ \\ (GeV${}^{-1}$)} & \makecell{$\langle T \rangle$ \\ (GeV)}      & \makecell{$\langle p^2 \rangle$ \\ (GeV${}^2$)}     & $\langle v_c^2 \rangle$   & $\langle v_b^2 \rangle$    & \makecell{$|R_{nl}^{(l)}(0)|^2$ \\ (GeV${}^{3+2l}$)}    & \makecell{$\sqrt{\langle r^2 \rangle}$ \\ (GeV${}^{-1}$)} & \makecell{$\langle T \rangle$ \\ (GeV)}      & \makecell{$\langle p^2 \rangle$ \\ (GeV${}^2$)}     & $\langle v_c^2 \rangle$   &$\langle v_b^2 \rangle$    \\ \midrule
$1S$  & $2.155_{-0.175}^{+0.399}$ & $1.706_{-0.037}^{+0.018}$    & $0.372_{-0.011}^{+0.025}$ & $0.85_{-0.026}^{+0.058}$  & $0.378_{-0.012}^{+0.026}$ & $0.037_{-0.001}^{+0.003}$ & $2.271_{-0.212}^{+0.36}$  & $1.691_{-0.033}^{+0.023}$    & $0.381_{-0.014}^{+0.023}$ & $0.871_{-0.033}^{+0.052}$ & $0.387_{-0.015}^{+0.023}$ & $0.038_{-0.001}^{+0.002}$  & $2.57_{-0.385}^{+0.46}$& $1.659_{-0.043}^{+0.039}$& $0.402_{-0.026}^{+0.03}$& $0.919_{-0.059}^{+0.069}$& $0.408_{-0.026}^{+0.031}$&$0.04_{-0.003}^{+0.003}$\\
$2S$  & $1.304_{-0.047}^{+0.09}$  & $3.511_{-0.02}^{+0.045}$     & $0.429_{-0.012}^{+0.005}$ & $0.981_{-0.027}^{+0.012}$ & $0.436_{-0.012}^{+0.005}$ & $0.043_{-0.001}^{+0.001}$ & $1.302_{-0.044}^{+0.081}$ & $3.546_{-0.035}^{+0.047}$    & $0.419_{-0.012}^{+0.009}$ & $0.959_{-0.028}^{+0.021}$ & $0.426_{-0.013}^{+0.01}$  & $0.042_{-0.001}^{+0.001}$  & $1.314_{-0.051}^{+0.058}$& $3.615_{-0.067}^{+0.108}$& $0.402_{-0.027}^{+0.017}$& $0.918_{-0.061}^{+0.04}$& $0.408_{-0.027}^{+0.018}$&$0.04_{-0.003}^{+0.002}$\\
$3S$  & $1.095_{-0.025}^{+0.042}$ & $4.972_{-0.053}^{+0.116}$    & $0.524_{-0.025}^{+0.012}$ & $1.199_{-0.058}^{+0.027}$ & $0.533_{-0.026}^{+0.012}$ & $0.052_{-0.002}^{+0.001}$ & $1.071_{-0.029}^{+0.038}$ & $5.076_{-0.096}^{+0.132}$    & $0.501_{-0.027}^{+0.021}$ & $1.146_{-0.062}^{+0.048}$ & $0.509_{-0.027}^{+0.021}$ & $0.05_{-0.003}^{+0.002}$   & $0.995_{-0.091}^{+0.064}$& $5.332_{-0.221}^{+0.419}$& $0.45_{-0.071}^{+0.044}$& $1.028_{-0.161}^{+0.101}$& $0.457_{-0.072}^{+0.045}$&$0.045_{-0.007}^{+0.004}$\\
$4S$  & $0.993_{-0.015}^{+0.023}$ & $6.246_{-0.079}^{+0.177}$    & $0.618_{-0.035}^{+0.017}$ & $1.412_{-0.08}^{+0.038}$  & $0.627_{-0.035}^{+0.017}$ & $0.061_{-0.003}^{+0.002}$ & $0.961_{-0.038}^{+0.024}$ & $6.42_{-0.154}^{+0.211}$     & $0.583_{-0.039}^{+0.031}$ & $1.332_{-0.088}^{+0.07}$  & $0.592_{-0.039}^{+0.031}$ & $0.058_{-0.004}^{+0.003}$  & $0.833_{-0.154}^{+0.095}$& $6.933_{-0.418}^{+0.897}$& $0.492_{-0.117}^{+0.071}$& $1.124_{-0.268}^{+0.163}$& $0.5_{-0.119}^{+0.073}$&$0.049_{-0.012}^{+0.007}$\\
$5S$  & $0.93_{-0.009}^{+0.012}$  & $7.401_{-0.102}^{+0.228}$    & $0.706_{-0.043}^{+0.021}$ & $1.614_{-0.098}^{+0.047}$ & $0.717_{-0.044}^{+0.021}$ & $0.07_{-0.004}^{+0.002}$  & $0.892_{-0.045}^{+0.024}$ & $7.643_{-0.208}^{+0.282}$    & $0.66_{-0.048}^{+0.039}$  & $1.508_{-0.111}^{+0.09}$  & $0.67_{-0.049}^{+0.04}$   & $0.065_{-0.005}^{+0.004}$  & $0.723_{-0.199}^{+0.122}$& $8.484_{-0.658}^{+1.561}$& $0.526_{-0.166}^{+0.1}$& $1.201_{-0.38}^{+0.23}$& $0.534_{-0.169}^{+0.102}$&$0.052_{-0.016}^{+0.01}$\\
$6S$  & $0.886_{-0.006}^{+0.005}$ & $8.472_{-0.123}^{+0.276}$    & $0.79_{-0.05}^{+0.024}$   & $1.806_{-0.115}^{+0.055}$ & $0.803_{-0.051}^{+0.025}$ & $0.078_{-0.005}^{+0.002}$ & $0.843_{-0.049}^{+0.026}$ & $8.78_{-0.259}^{+0.35}$      & $0.733_{-0.058}^{+0.047}$ & $1.676_{-0.132}^{+0.108}$ & $0.745_{-0.059}^{+0.048}$ & $0.073_{-0.006}^{+0.005}$  & $0.639_{-0.231}^{+0.144}$& $10.009_{-0.941}^{+2.514}$& $0.551_{-0.218}^{+0.132}$& $1.26_{-0.499}^{+0.302}$& $0.56_{-0.222}^{+0.134}$&$0.055_{-0.022}^{+0.013}$\\ \midrule
$1P$  & $0.25_{-0.005}^{+0.009}$  & $2.799_{-0.018}^{+0.04}$     & $0.357_{-0.009}^{+0.004}$ & $0.817_{-0.019}^{+0.009}$ & $0.363_{-0.009}^{+0.004}$ & $0.035_{-0.001}^{+0}$     & $0.254_{-0.006}^{+0.008}$ & $2.817_{-0.025}^{+0.037}$    & $0.354_{-0.008}^{+0.005}$ & $0.809_{-0.018}^{+0.012}$ & $0.36_{-0.008}^{+0.005}$  & $0.035_{-0.001}^{+0.001}$  & $0.261_{-0.009}^{+0.008}$& $2.857_{-0.047}^{+0.062}$& $0.346_{-0.011}^{+0.009}$& $0.792_{-0.026}^{+0.021}$& $0.352_{-0.012}^{+0.009}$&$0.034_{-0.001}^{+0.001}$\\
$2P$  & $0.348_{-0.002}^{+0.002}$ & $4.377_{-0.05}^{+0.111}$     & $0.463_{-0.022}^{+0.01}$  & $1.058_{-0.05}^{+0.024}$  & $0.47_{-0.022}^{+0.011}$  & $0.046_{-0.002}^{+0.001}$ & $0.34_{-0.01}^{+0.005}$   & $4.461_{-0.085}^{+0.117}$    & $0.446_{-0.022}^{+0.017}$ & $1.018_{-0.051}^{+0.04}$  & $0.453_{-0.022}^{+0.018}$ & $0.044_{-0.002}^{+0.002}$  & $0.317_{-0.034}^{+0.018}$& $4.653_{-0.179}^{+0.324}$& $0.409_{-0.052}^{+0.033}$& $0.935_{-0.119}^{+0.077}$& $0.416_{-0.053}^{+0.034}$&$0.041_{-0.005}^{+0.003}$\\
$3P$  & $0.42_{-0.01}^{+0.004}$   & $5.713_{-0.076}^{+0.17}$     & $0.562_{-0.032}^{+0.015}$ & $1.285_{-0.073}^{+0.035}$ & $0.571_{-0.032}^{+0.015}$ & $0.056_{-0.003}^{+0.002}$ & $0.399_{-0.022}^{+0.014}$ & $5.865_{-0.141}^{+0.195}$    & $0.533_{-0.034}^{+0.027}$ & $1.218_{-0.078}^{+0.061}$ & $0.541_{-0.035}^{+0.027}$ & $0.053_{-0.003}^{+0.003}$  & $0.34_{-0.075}^{+0.045}$& $6.286_{-0.355}^{+0.739}$& $0.461_{-0.097}^{+0.059}$& $1.053_{-0.221}^{+0.135}$& $0.468_{-0.098}^{+0.06}$&$0.046_{-0.01}^{+0.006}$\\
$4P$  & $0.479_{-0.018}^{+0.007}$ & $6.91_{-0.099}^{+0.222}$     & $0.655_{-0.04}^{+0.019}$  & $1.497_{-0.092}^{+0.044}$ & $0.665_{-0.041}^{+0.02}$  & $0.065_{-0.004}^{+0.002}$ & $0.447_{-0.034}^{+0.023}$ & $7.129_{-0.194}^{+0.269}$    & $0.614_{-0.044}^{+0.035}$ & $1.404_{-0.101}^{+0.081}$ & $0.624_{-0.045}^{+0.036}$ & $0.061_{-0.004}^{+0.004}$  & $0.349_{-0.112}^{+0.072}$& $7.85_{-0.574}^{+1.337}$& $0.501_{-0.144}^{+0.087}$& $1.146_{-0.329}^{+0.198}$& $0.509_{-0.146}^{+0.088}$&$0.05_{-0.014}^{+0.009}$\\
$5P$  & $0.531_{-0.025}^{+0.011}$ & $8.011_{-0.121}^{+0.269}$    & $0.742_{-0.048}^{+0.023}$ & $1.697_{-0.109}^{+0.053}$ & $0.754_{-0.048}^{+0.023}$ & $0.074_{-0.005}^{+0.002}$ & $0.488_{-0.044}^{+0.03}$  & $8.295_{-0.244}^{+0.336}$    & $0.691_{-0.053}^{+0.044}$ & $1.579_{-0.122}^{+0.1}$   & $0.702_{-0.054}^{+0.044}$ & $0.069_{-0.005}^{+0.004}$  & $0.349_{-0.146}^{+0.099}$& $9.381_{-0.839}^{+2.167}$& $0.532_{-0.195}^{+0.117}$& $1.216_{-0.445}^{+0.267}$& $0.54_{-0.198}^{+0.119}$&$0.053_{-0.019}^{+0.012}$\\
$6P$  & $0.578_{-0.032}^{+0.014}$ & $9.04_{-0.14}^{+0.312}$      & $0.825_{-0.055}^{+0.026}$ & $1.887_{-0.125}^{+0.06}$  & $0.839_{-0.055}^{+0.027}$ & $0.082_{-0.005}^{+0.003}$ & $0.525_{-0.053}^{+0.038}$ & $9.39_{-0.294}^{+0.399}$     & $0.763_{-0.062}^{+0.051}$ & $1.745_{-0.142}^{+0.117}$ & $0.775_{-0.063}^{+0.052}$ & $0.076_{-0.006}^{+0.005}$  & $0.343_{-0.175}^{+0.126}$& $10.903_{-1.155}^{+3.316}$& $0.554_{-0.247}^{+0.15}$& $1.265_{-0.563}^{+0.344}$& $0.562_{-0.25}^{+0.153}$&$0.055_{-0.024}^{+0.015}$\\ \midrule
$1D$  & $0.066_{-0.005}^{+0.002}$ & $3.653_{-0.04}^{+0.09}$      & $0.406_{-0.019}^{+0.009}$ & $0.929_{-0.043}^{+0.02}$  & $0.413_{-0.019}^{+0.009}$ & $0.04_{-0.002}^{+0.001}$  & $0.063_{-0.004}^{+0.003}$ & $3.707_{-0.062}^{+0.089}$    & $0.395_{-0.018}^{+0.013}$ & $0.903_{-0.04}^{+0.03}$   & $0.402_{-0.018}^{+0.013}$ & $0.039_{-0.002}^{+0.001}$  & $0.058_{-0.008}^{+0.006}$& $3.826_{-0.122}^{+0.2}$& $0.373_{-0.033}^{+0.024}$& $0.852_{-0.076}^{+0.054}$& $0.379_{-0.034}^{+0.024}$&$0.037_{-0.003}^{+0.002}$\\
$2D$  & $0.141_{-0.013}^{+0.006}$ & $5.096_{-0.068}^{+0.153}$    & $0.511_{-0.029}^{+0.014}$ & $1.168_{-0.066}^{+0.031}$ & $0.519_{-0.029}^{+0.014}$ & $0.051_{-0.003}^{+0.001}$ & $0.13_{-0.013}^{+0.01}$   & $5.22_{-0.121}^{+0.168}$     & $0.487_{-0.03}^{+0.023}$  & $1.114_{-0.067}^{+0.053}$ & $0.495_{-0.03}^{+0.023}$  & $0.048_{-0.003}^{+0.002}$  & $0.107_{-0.029}^{+0.02}$& $5.538_{-0.28}^{+0.556}$& $0.433_{-0.075}^{+0.047}$& $0.989_{-0.172}^{+0.108}$& $0.44_{-0.076}^{+0.048}$&$0.043_{-0.007}^{+0.005}$\\
$3D$  & $0.221_{-0.023}^{+0.011}$ & $6.355_{-0.092}^{+0.206}$    & $0.608_{-0.037}^{+0.018}$ & $1.389_{-0.085}^{+0.041}$ & $0.617_{-0.038}^{+0.018}$ & $0.06_{-0.004}^{+0.002}$  & $0.199_{-0.024}^{+0.02}$  & $6.546_{-0.174}^{+0.241}$    & $0.572_{-0.04}^{+0.032}$  & $1.308_{-0.092}^{+0.073}$ & $0.581_{-0.041}^{+0.032}$ & $0.057_{-0.004}^{+0.003}$  & $0.147_{-0.057}^{+0.041}$& $7.132_{-0.48}^{+1.082}$& $0.48_{-0.122}^{+0.073}$& $1.097_{-0.279}^{+0.168}$& $0.487_{-0.124}^{+0.075}$&$0.048_{-0.012}^{+0.007}$\\
$4D$  & $0.305_{-0.034}^{+0.017}$ & $7.498_{-0.114}^{+0.254}$    & $0.698_{-0.045}^{+0.022}$ & $1.595_{-0.103}^{+0.05}$  & $0.709_{-0.046}^{+0.022}$ & $0.069_{-0.004}^{+0.002}$ & $0.268_{-0.036}^{+0.031}$ & $7.756_{-0.226}^{+0.311}$    & $0.652_{-0.05}^{+0.04}$   & $1.489_{-0.114}^{+0.091}$ & $0.662_{-0.05}^{+0.041}$  & $0.065_{-0.005}^{+0.004}$  & $0.179_{-0.087}^{+0.067}$& $8.678_{-0.726}^{+1.81}$& $0.515_{-0.171}^{+0.103}$& $1.177_{-0.39}^{+0.235}$& $0.523_{-0.173}^{+0.104}$&$0.051_{-0.017}^{+0.01}$\\
$5D$  & $0.392_{-0.047}^{+0.023}$ & $8.559_{-0.134}^{+0.299}$    & $0.783_{-0.052}^{+0.025}$ & $1.79_{-0.119}^{+0.057}$  & $0.796_{-0.053}^{+0.025}$ & $0.078_{-0.005}^{+0.002}$ & $0.339_{-0.051}^{+0.044}$ & $8.882_{-0.275}^{+0.376}$    & $0.726_{-0.058}^{+0.048}$ & $1.66_{-0.133}^{+0.109}$  & $0.738_{-0.059}^{+0.048}$ & $0.072_{-0.006}^{+0.005}$  & $0.204_{-0.118}^{+0.097}$& $10.207_{-1.018}^{+2.836}$& $0.542_{-0.224}^{+0.134}$& $1.239_{-0.511}^{+0.306}$& $0.551_{-0.227}^{+0.136}$&$0.054_{-0.022}^{+0.013}$\\
$6D$  & $0.482_{-0.06}^{+0.03}$   & $9.557_{-0.152}^{+0.34}$     & $0.864_{-0.058}^{+0.028}$ & $1.975_{-0.134}^{+0.065}$ & $0.878_{-0.059}^{+0.029}$ & $0.086_{-0.006}^{+0.003}$ & $0.411_{-0.066}^{+0.057}$ & $9.944_{-0.323}^{+0.438}$    & $0.797_{-0.067}^{+0.055}$ & $1.822_{-0.153}^{+0.126}$ & $0.81_{-0.068}^{+0.056}$  & $0.079_{-0.007}^{+0.005}$  & $0.222_{-0.147}^{+0.13}$& $11.746_{-1.371}^{+4.23}$& $0.559_{-0.275}^{+0.169}$& $1.278_{-0.629}^{+0.385}$& $0.568_{-0.28}^{+0.171}$&$0.055_{-0.027}^{+0.017}$\\ \bottomrule
\end{tabular}
}
\end{table}
\endgroup
\end{turnpage}

\begingroup
\squeezetable
\begin{table}[]
\centering
\caption{Numerical values of linear and non-linear Radial Regge parameters.}
\label{tab:Regge_radial}
\begin{tabular}{@{}ccccccc@{}}
\toprule
Potential & $l$ & $\alpha_{lin}$     & $\beta_{lin}$     & $C_{lin}$           & $c_{fn_r}$        & $c_{0n_r}$        \\ \midrule
\multicolumn{7}{l}{For $n^1S_0$, $nP_1'$, and $nD_2'$ states}                                                          \\
Pot. I    & 0   & -                  & $5.606 \pm 0.035$ & $40.67 \pm 0.096$   & $1.015 \pm 0.004$ & $0.103 \pm 0.01$  \\
          & 1   & $23.029 \pm 0.837$ & $4.97 \pm 0.036$  & $23.029 \pm 0.837$  & $1.002 \pm 0.005$ & $0.842 \pm 0.022$ \\
          & 2   & $12.438 \pm 0.43$  & $4.692 \pm 0.037$ & $24.875 \pm 0.86$   & $0.994 \pm 0.005$ & $1.493 \pm 0.03$  \\
Pot. II   & 0   & -                  & $5.374 \pm 0.035$ & $40.809 \pm 0.096$  & $1.144 \pm 0.004$ & $0.121 \pm 0.011$ \\
          & 1   & $23.1 \pm 0.834$   & $4.682 \pm 0.036$ & $23.1 \pm 0.834$    & $1.113 \pm 0.006$ & $0.931 \pm 0.024$ \\
          & 2   & $12.454 \pm 0.428$ & $4.377 \pm 0.037$ & $24.907 \pm 0.855$  & $1.096 \pm 0.006$ & $1.633 \pm 0.033$ \\
Pot. III  & 0   & -                  & $4.786 \pm 0.034$ & $41.198 \pm 0.096$  & $0.84 \pm 0.003$  & $0.107 \pm 0.011$ \\
          & 1   & $23.321 \pm 0.825$ & $3.926 \pm 0.035$ & $23.321 \pm 0.825$  & $0.786 \pm 0.005$ & $1.113 \pm 0.03$  \\
          & 2   & $12.522 \pm 5.759$ & $3.52 \pm 0.036$  & $25.043 \pm 11.517$ & $0.752 \pm 0.005$ & $2.011 \pm 0.044$ \\ \midrule
\multicolumn{7}{l}{For $n^3S_1$, $n^3P_2$, and $n^3D_3$ states}                                                        \\
Pot. I    & 0   & -                  & $5.524 \pm 0.035$ & $41.268 \pm 0.097$  & $1.015 \pm 0.004$ & $0.164 \pm 0.011$ \\
          & 1   & $23.121 \pm 0.839$ & $4.984 \pm 0.036$ & $23.121 \pm 0.839$  & $1.006 \pm 0.005$ & $0.865 \pm 0.022$ \\
          & 2   & $12.408 \pm 0.43$  & $4.722 \pm 0.037$ & $24.815 \pm 0.859$  & $0.998 \pm 0.005$ & $1.462 \pm 0.029$ \\
Pot. II   & 0   & -                  & $5.283 \pm 0.035$ & $41.397 \pm 0.097$  & $1.142 \pm 0.004$ & $0.188 \pm 0.012$ \\
          & 1   & $23.2 \pm 0.835$   & $4.691 \pm 0.036$ & $23.2 \pm 0.835$    & $1.117 \pm 0.006$ & $0.959 \pm 0.024$ \\
          & 2   & $12.428 \pm 0.427$ & $4.406 \pm 0.037$ & $24.856 \pm 0.855$  & $1.1 \pm 0.006$   & $1.603 \pm 0.032$ \\
Pot. III  & 0   & -                  & $4.678 \pm 0.034$ & $41.763 \pm 0.097$  & $0.835 \pm 0.004$ & $0.182 \pm 0.014$ \\
          & 1   & $23.44 \pm 0.826$  & $3.916 \pm 0.035$ & $23.44 \pm 0.826$   & $0.787 \pm 0.005$ & $1.16 \pm 0.03$   \\
          & 2   & $12.51 \pm 10.215$ & $3.539 \pm 0.036$ & $25.019 \pm 20.429$ & $0.754 \pm 0.005$ & $1.989 \pm 0.044$ \\ \bottomrule
\end{tabular}
\end{table}
\endgroup

\begingroup
\squeezetable
\begin{table}[]
\centering
\caption{Numerical values of linear and non-linear Orbital Regge parameters.}
\label{tab:Regge_orbital}
\begin{tabular}{@{}ccccccc@{}}
\toprule
Potential & $n_r$ & $\alpha_{lin}$    & $\beta_{lin}$       & $C_{lin}$           & $c_{fl}$          & $c_{0l}$           \\ \midrule
\multicolumn{7}{l}{For states with $J^P = 0^-, 1^+$, and $2^-$}                                                            \\
Pot. I    & 0     & $5.041 \pm 0.094$ & -                   & $39.689 \pm 0.116$  & $1.173 \pm 0.014$ & $0.158 \pm 0.016$  \\
          & 1     & $3.715 \pm 0.1$   & $23.672 \pm 1.126$  & $23.672 \pm 1.126$  & $1.144 \pm 0.021$ & $1.679 \pm 0.071$  \\
          & 2     & $3.229 \pm 0.106$ & $13.279 \pm 0.591$  & $26.559 \pm 1.183$  & $1.117 \pm 0.024$ & $3.327 \pm 0.141$  \\
          & 3     & $2.961 \pm 0.11$  & $9.706 \pm 0.411$   & $29.117 \pm 1.232$  & $1.099 \pm 0.027$ & $5.019 \pm 0.224$  \\
          & 4     & $2.784 \pm 0.114$ & $7.876 \pm 0.319$   & $31.503 \pm 1.278$  & $1.085 \pm 0.03$  & $6.754 \pm 0.319$  \\
          & 5     & $2.661 \pm 0.118$ & $6.755 \pm 0.264$   & $33.777 \pm 1.32$   & $1.075 \pm 0.032$ & $8.503 \pm 0.424$  \\
Pot. II   & 0     & $5.049 \pm 0.094$ & -                   & $39.706 \pm 0.116$  & $1.366 \pm 0.016$ & $0.163 \pm 0.016$  \\
          & 1     & $3.601 \pm 0.101$ & $23.674 \pm 27.178$ & $23.674 \pm 27.178$ & $1.298 \pm 0.024$ & $1.745 \pm 0.075$  \\
          & 2     & $3.071 \pm 0.105$ & $13.228 \pm 0.589$  & $26.455 \pm 1.179$  & $1.248 \pm 0.029$ & $3.463 \pm 0.152$  \\
          & 3     & $2.777 \pm 0.109$ & $9.626 \pm 0.409$   & $28.877 \pm 1.226$  & $1.215 \pm 0.032$ & $5.234 \pm 0.246$  \\
          & 4     & $2.586 \pm 0.113$ & $7.776 \pm 0.317$   & $31.104 \pm 1.268$  & $1.191 \pm 0.035$ & $7.039 \pm 0.354$  \\
          & 5     & $2.451 \pm 0.117$ & $6.641 \pm 0.261$   & $33.205 \pm 1.307$  & $1.172 \pm 0.038$ & $8.865 \pm 0.473$  \\
Pot. III  & 0     & $5.056 \pm 0.094$ & -                   & $39.748 \pm 0.116$  & $1.072 \pm 0.012$ & $0.103 \pm 0.014$  \\
          & 1     & $3.352 \pm 0.1$   & $23.679 \pm 20.123$ & $23.679 \pm 20.123$ & $0.982 \pm 0.019$ & $1.757 \pm 0.081$  \\
          & 2     & $2.68 \pm 0.104$  & $13.122 \pm 0.585$  & $26.243 \pm 1.17$   & $0.904 \pm 0.024$ & $3.697 \pm 0.183$  \\
          & 3     & $2.281 \pm 0.108$ & $9.444 \pm 0.403$   & $28.331 \pm 1.209$  & $0.843 \pm 0.027$ & $5.813 \pm 0.324$  \\
          & 4     & $1.996 \pm 0.111$ & $7.533 \pm 0.311$   & $30.132 \pm 1.243$  & $0.792 \pm 0.03$  & $8.119 \pm 0.51$   \\
          & 5     & $1.776 \pm 0.114$ & $6.345 \pm 7.734$   & $31.724 \pm 38.669$ & $0.746 \pm 0.032$ & $10.613 \pm 0.749$ \\ \midrule
\multicolumn{7}{l}{For states with $J^P = 1^-, 2^+$, and $3^-$}                                                            \\
Pot. I    & 0     & $4.616 \pm 0.094$ & -                   & $40.364 \pm 0.117$  & $1.126 \pm 0.015$ & $0.266 \pm 0.021$  \\
          & 1     & $3.47 \pm 0.101$  & $23.882 \pm 1.128$  & $23.882 \pm 1.128$  & $1.098 \pm 0.021$ & $1.898 \pm 0.083$  \\
          & 2     & $3.033 \pm 0.106$ & $13.371 \pm 0.592$  & $26.742 \pm 1.184$  & $1.074 \pm 0.025$ & $3.646 \pm 0.161$  \\
          & 3     & $2.796 \pm 0.11$  & $9.762 \pm 0.411$   & $29.286 \pm 1.234$  & $1.06 \pm 0.028$  & $5.422 \pm 0.254$  \\
          & 4     & $2.641 \pm 0.114$ & $7.915 \pm 0.32$    & $31.662 \pm 1.279$  & $1.049 \pm 0.03$  & $7.224 \pm 0.357$  \\
          & 5     & $2.535 \pm 0.118$ & $6.786 \pm 0.264$   & $33.929 \pm 1.321$  & $1.042 \pm 0.033$ & $9.03 \pm 0.47$    \\
Pot. II   & 0     & $4.632 \pm 0.094$ & -                   & $40.383 \pm 0.117$  & $1.312 \pm 0.017$ & $0.27 \pm 0.021$   \\
          & 1     & $3.379 \pm 0.1$   & $23.873 \pm 1.127$  & $23.873 \pm 1.127$  & $1.249 \pm 0.025$ & $1.956 \pm 0.086$  \\
          & 2     & $2.902 \pm 0.105$ & $13.311 \pm 0.59$   & $26.622 \pm 1.181$  & $1.205 \pm 0.029$ & $3.762 \pm 0.172$  \\
          & 3     & $2.643 \pm 0.11$  & $9.675 \pm 0.409$   & $29.025 \pm 1.227$  & $1.177 \pm 0.033$ & $5.595 \pm 0.274$  \\
          & 4     & $2.471 \pm 0.114$ & $7.811 \pm 9.64$    & $31.243 \pm 38.559$ & $1.157 \pm 0.036$ & $7.463 \pm 0.39$   \\
          & 5     & $2.355 \pm 0.117$ & $6.667 \pm 0.262$   & $33.335 \pm 1.308$  & $1.143 \pm 0.038$ & $9.321 \pm 0.516$  \\
Pot. III  & 0     & $4.672 \pm 0.094$ & -                   & $40.426 \pm 0.117$  & $1.039 \pm 0.013$ & $0.196 \pm 0.019$  \\
          & 1     & $3.179 \pm 0.1$   & $23.854 \pm 1.124$  & $23.854 \pm 1.124$  & $0.952 \pm 0.02$  & $1.941 \pm 0.091$  \\
          & 2     & $2.564 \pm 0.105$ & $13.188 \pm 0.586$  & $26.376 \pm 1.172$  & $0.88 \pm 0.024$  & $3.948 \pm 0.201$  \\
          & 3     & $2.197 \pm 0.108$ & $9.481 \pm 0.404$   & $28.442 \pm 1.211$  & $0.824 \pm 0.027$ & $6.119 \pm 0.351$  \\
          & 4     & $1.931 \pm 0.111$ & $7.557 \pm 0.311$   & $30.227 \pm 1.244$  & $0.775 \pm 0.03$  & $8.472 \pm 0.548$  \\
          & 5     & $1.727 \pm 0.114$ & $6.362 \pm 0.255$   & $31.81 \pm 1.273$   & $0.733 \pm 0.032$ & $10.997 \pm 0.797$ \\ \bottomrule
\end{tabular}
\end{table}
\endgroup

\end{document}